\documentclass[]{aa} 

\usepackage{natbib,twoopt}
\usepackage[varg]{txfonts}
\usepackage{hyperref} 
\usepackage{multicol}
\usepackage{textcomp} 
\usepackage{pifont}
\usepackage{booktabs}
\usepackage{amssymb}
\usepackage{amsmath}
\graphicspath{{images/}{../images/}{../../images/}}
\usepackage{graphicx}
\usepackage{geometry}
\usepackage{physics}
\usepackage[math]{cellspace}
\usepackage{tabu}
\usepackage{gensymb}
\usepackage{svg}

\usepackage{bm}

\hypersetup{
    colorlinks=true,
    urlcolor=blue,
    linkcolor=blue,
    citecolor=blue,
}

\bibpunct{(}{)}{;}{a}{}{,}             
\title{Detecting axisymmetric magnetic fields using gravity modes in intermediate-mass stars}
\subtitle{}
\authorrunning{J.V.B. et al.}
\titlerunning{Detect axisymmetric magnetic fields with $g$ modes in intermediate-mass stars}
\author{J.~Van~Beeck \inst{\ref{kul} \href{https://orcid.org/0000-0002-5082-3887}{\includegraphics[width=3mm]{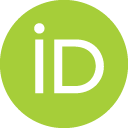}}} \and V.~Prat \inst{\ref{cea} \href{https://orcid.org/0000-0002-5335-4991}{\includegraphics[width=3mm]{Images/labels/ORCIDiD_icon128x128.png}}} \and T.~Van~Reeth \inst{\ref{kul} \href{https://orcid.org/0000-0003-2771-1745}{\includegraphics[width=3mm]{Images/labels/ORCIDiD_icon128x128.png}}} \and S.~Mathis \inst{\ref{cea},\ref{lesia} \href{https://orcid.org/0000-0001-9491-8012}{\includegraphics[width=3mm]{Images/labels/ORCIDiD_icon128x128.png}}} \and D.~M.~Bowman \inst{\ref{kul} \href{https://orcid.org/0000-0001-7402-3852}{\includegraphics[width=3mm]{Images/labels/ORCIDiD_icon128x128.png}}} \and C.~Neiner \inst{\ref{lesia} \href{https://orcid.org/0000-0003-1978-9809+}{\includegraphics[width=3mm]{Images/labels/ORCIDiD_icon128x128.png}}} \and C.~Aerts \inst{\ref{kul},\ref{nijm},\ref{MPIA}  \href{https://orcid.org/0000-0003-1822-7126}{\includegraphics[width=3mm]{Images/labels/ORCIDiD_icon128x128.png}}}}
\institute{Instituut voor Sterrenkunde, KU Leuven, Celestijnenlaan 200D, 3001 Leuven, Belgium \label{kul}\\ e-mail:~\href{mailto:jordan.vanbeeck@kuleuven.be}{\color{black}\texttt{jordan.vanbeeck@kuleuven.be}} \and AIM, CEA, CNRS, Universit\'e Paris-Saclay, Universit\'e Paris Diderot, Sorbonne Paris Cit\'e, F-91191 Gif-sur-Yvette, France \label{cea} \and LESIA, Observatoire de Paris, PSL University, CNRS, Sorbonne Universit\'e, Univ. Paris Diderot, Sorbonne Paris Cit\'e, 5 place Jules
Janssen, F-92195 Meudon, France \label{lesia} \and Dept. of Astrophysics, IMAPP, Radboud University Nijmegen, 6500 GL, Nijmegen, The Netherlands \label{nijm} \and Max Planck Institute for Astronomy, Koenigstuhl 17, 69117 Heidelberg, Germany \label{MPIA}}

\date{Received / Accepted}
\abstract {Angular momentum (AM) transport models of stellar interiors require improvements to explain the strong extraction of AM from stellar cores that is observed with asteroseismology. One of the often invoked mediators of AM transport are internal magnetic fields, even though their properties, observational signatures and influence on stellar evolution are largely unknown.}
{We study how a fossil, axisymmetric internal magnetic field affects period spacing patterns of dipolar gravity mode oscillations in main-sequence stars with masses of 1.3, 2.0 and 3.0 M$_\sun$. We assess the influence of fundamental stellar parameters on the magnitude of pulsation mode frequency shifts.} 
{We compute dipolar gravity mode frequency shifts due to a fossil, axisymmetric poloidal-toroidal internal magnetic field for a grid of stellar evolution models, varying stellar fundamental parameters. Rigid rotation is taken into account using the traditional approximation of rotation and the influence of the magnetic field is computed using a perturbative approach.} 
{We find magnetic signatures for dipolar gravity mode oscillations in terminal-age main-sequence stars that are measurable for a near-core field strength larger than $10^{\,5}$ G. The predicted signatures differ appreciably from those due to rotation.} 
{Our formalism demonstrates the potential for the future detection and characterization of strong fossil, axisymmetric internal magnetic fields in gravity-mode pulsators near the end of core-hydrogen burning from {\it Kepler\/} photometry, if such fields exist.}

\begin{document}

\maketitle
\section{Introduction}\label{sec:1}
Recent developments in asteroseismology broadened our knowledge of stellar interiors, allowing us to characterize internal stellar rotation and chemical mixing from non-radial oscillations \cite[see reviews by][]{2013_Chaplin,2017_Hekker,Aerts2019_ARAA}. Thanks to these efforts, we learned that current models severely underestimate angular momentum (AM) transport throughout stellar evolution \citep{Aerts2019_ARAA,2019_Ouazzani,2019_Eggenberger,2020_Hartogh}, and thus fail to explain the observed (quasi-)\,rigid rotation in most intermediate-mass main sequence (MS) stars \citep{2014_Kurtz,2015_Saio,2016_Murphy,2017_Aerts, Van_Reeth_2018_differential_rotation,2019_Ouazzani}.
Two physical processes are often invoked in the literature to explain the missing transport of angular momentum: internal gravity waves \cite[e.g.][]{2005_Talon,2007_Pantillon,Mathis_2009,2013_Rogers,2015_Rogers} and stable or unstable magnetic fields \cite[e.g.][]{1999_Spruit,2005_Mathis,2019_Fuller}. We limit our focus to magnetic fields, and build upon the recently developed formalism described in \citeauthor{Prat_formalism} (\citeyear{Prat_formalism}, hereafter referred to as P+19), which considers interior axisymmetric stellar magnetic fields.

The number of stars for which surface magnetic fields have been detected using the Zeeman effect \cite[with techniques described in e.g.][]{2009_ARAA_Donati_Landstreet} has significantly increased thanks to improvements in instrumentation and the increasing number of stars observed in spectropolarimetric campaigns. This led to the detection of large-scale surface magnetic fields in approximately 7\% of O- and B-type stars from surveys such as Magnetism in Massive Stars \cite[MiMeS;][]{2016_MIMES_WADE} and B fields in OB stars \cite[BOB;][]{2014_Morel_survey}.

Most detected large-scale surface magnetic fields have a simple geometry \cite[usually an inclined dipole, see e.g.][]{1990_Moss,2007_auriere,2012_Walder,2018_schultz} and have polar field strengths ranging from a few hundred G to a few tens of kG. Weak magnetic fields of a few G have also been detected in Vega \citep{2009_lignieres,2010_Petit,2011_Petit,2014_Petit} and several Am stars \cite[e.g.][]{2011_Petit,2016_Blazere_b,2016_Blazere_a}.
The origin of these weak fields remains elusive, but seems to be associated with sharp vertical gradients in velocity in the stellar surface layers \citep{2016_Blazere_a} or with configurations that cannot reach an equilibrium state \citep{2013_braitwaite,2019_braitwaite}.
The properties of the stronger large-scale surface magnetic fields do not scale with stellar parameters or rotation, and their incidence rate throughout the MS is approximately constant \citep{2018_buysschaert}. Therefore, they are thought to be of \mbox{fossil origin \citep{2015_Neiner,2015emeriau}}, even though a convective dynamo might still be present in the core of these stars, as indicated by the simulations of \citet{2005_Brun} and \citet{Augustson_dynamos}. The stability of fossil fields depends critically on their configuration: to be stable, they should extend deep within the stellar radiative envelope and have a mixed configuration with both poloidal and toroidal components \citep{1973_Markey,1973_Tayler,1980_Tayler,2007_Braithwaite_stab_pol,2009_Braithwaite}.

The characterization of internal magnetic fields is virtually unexplored, because spectropolarimetric techniques do not probe the stellar interior. A promising way to constrain internal magnetic fields is considering their effect on stellar oscillation modes, in particular, gravity ($g$) modes, which probe near-core regions \cite[e.g.][]{Aerts_book}.
Purely poloidal dipolar stellar magnetic fields have been constrained in the past by inspecting their influence on $g$-mode frequencies. These magnetic fields cause splitting of $g$-mode frequencies of the same angular degree but different azimuthal order \cite[e.g.][]{2005_Hasan,Buysschaert_2018_magnetic_forward_model}. For slowly rotating stars that have weak near-core magnetic field strengths, both the rotational and magnetic influence can be considered as perturbations \cite[e.g][]{1957_Ledoux}. 
However, P+19 stressed that one cannot describe rotation as a perturbation for rapidly-rotating stars, such as $\gamma$~Doradus ($\gamma$~Dor) or slowly pulsating B-type (SPB) stars.
Instead, the influence of rotation on g modes is commonly described within the traditional approximation (TAR), where one neglects the horizontal component of the rotation vector and assumes spherical symmetry \cite[e.g.][]{1960_TAR_eckart,Lee_1997,2003_Townsend}.
The latter assumption breaks down if the star rotates close to critical, due to centrifugal deformation. 
However, recent efforts have made it possible to use the TAR in centrifugally slightly deformed stars \citep{2019_mathis_TAR_centri}, and magnetic stars \cite[][P+19]{Mathis_debrye_2011}.

One of the main tools to interpret $g$-mode oscillations are period spacing patterns, based on modes of consecutive radial order and the same angular degree and azimuthal order.
The period spacings of chemically homogeneous, non-rotating and non-magnetic stars are asymptotically constant. An oscillatory component in the period spacing pattern appears if there is a strong chemical gradient near the core that is left behind by a receding convective core during MS evolution \cite[e.g.][]{2008_Miglio,2010_Degroote}. Observed period spacing patterns of pulsating, rotating stars also display a slope which is a function of the rotation rate \citep{2013_Bouabid,2015_Van_Reeth_BIS,2015_Van_Reeth,Ouazzani_2017}. Some modeled stars display features in their period spacing patterns that cannot be explained by the input physics of the oscillation models \cite[e.g.][]{2015_Moravveji,2016_Moravveji}. In these cases taking the effect of a magnetic field into account could aid asteroseismic modeling efforts and can influence derived chemical mixing and convective core overshooting levels \cite[e.g.][]{2012_Briquet,Buysschaert_2018_magnetic_forward_model}.

An example of a modeled rotating, magnetic, B3.5\,V SPB star is HD~43317, whose magnetic characterization is discussed in \citet{2013_Briquet_mag_char} and \citet{2017_Buysschaert_mag_char}. This SPB star displays $g$-mode pulsations \citep{2012_Papics} and magnetic splittings obtained with the formalism of \citet{2005_Hasan} were shown to be negligible compared to the rotational splittings for this star \citep{Buysschaert_2018_magnetic_forward_model}. However, the assumed purely poloidal field is inherently unstable. 
Using the best-fitting stellar model of \citet{Buysschaert_2018_magnetic_forward_model}, a 5.8-M$_\sun$ model with an age of 28.4 Myr, a radius of 3.39 R$_\sun$, an effective temperature of 17\,800~K, a central hydrogen mass fraction of 0.54 and solar-like metallicity, P+19 concluded that magnetic frequency shifts significantly change if rotation is taken into account using the TAR, and that a magnetic field with a near-core field strength of 150~kG would be detectable in period spacing patterns of high-radial-order $g$ modes.
Even though the magnetic influence is found to be significant for higher radial order modes, it remains perturbative, and the magnetic frequency shifts are too small to be detected using 150~d of CoRoT photometry (P+19). However, P+19 only considered one stellar model and a small range of magnetic field strengths. A natural extension to their work consists of investigating the influence of stellar parameters and large variations of magnetic field strength on the derived period spacing patterns.

In this work, we investigate the parameter space of stars lower in mass than the single model considered by P+19. We compute pulsation mode frequencies for a stellar model grid to derive correlations between stellar parameters and simulated period spacing patterns.

\section{Checking the consistency of the perturbative magnetic frequency shift}\label{sec:2}

Following the setup in P+19, we determine the parameter domain in which the axisymmetric magnetic field exerts a perturbative influence on the frequency, by comparing the angular Alfv\'en frequency $\omega_{\rm A}$  with the computed angular pulsation frequency in the co-rotating frame $\omega_{\rm co}$. It is assumed that
\begin{equation}                                                                                                                                 
    \dfrac{\omega_{\rm A}}{\omega_{\rm co}} \ll 1 \label{eq:criterion}
\end{equation}
holds if the Lorentz force is acting perturbatively. If this does not hold, the Lorentz force is dynamically significant, and non-perturbative treatments of the Lorentz force are necessary to evaluate the influence of the internal magnetic field on $g$-mode frequencies. For dynamically significant Lorentz forces, conversion of $g$ modes into (pure) Alfv\'en waves is possible, as was proposed by \citet{2015_Fuller_magnetic_greenhouse} and \citet{2017_Lecoanet}. Those studies do not consider any rotational influence, while this changes the wave vector and therefore also changes the regime in which mode conversion is expected. 

The characteristic frequencies such as the Brunt-V\"ais\"al\"a frequency, Lamb frequency, and Coriolis frequency define pulsation mode cavities \cite[e.g.][]{Aerts_book,2010_BOOK_Smeyers_VanHoolst}. The angular Alfv\'en frequency $\omega_{\rm A}$ is another characteristic frequency that further modifies the pulsation mode cavity. For $\omega_{\rm A} \sim \omega_{\rm co}$, the Lorentz force becomes a significant restoring force \cite[e.g.][]{Aerts2019_ARAA}.
P+19 considered high-radial-order modes, obtaining the following estimate of the angular Alfv\'en frequency:
\begin{align}
    \omega_{\rm A} = \bm{v}_A \bm{\cdot} \bm{k} = \dfrac{\bm{B}\bm{\cdot}\bm{k}}{\sqrt{\ \mu_0\ \rho}} \approx \dfrac{b_r\left(r\right) k_r \cos\theta}{\sqrt{\ \mu_0\ \rho\left(r\right)}}~, \label{eq:alfven_vel_prat}
\end{align}
where $\bm{v}_A$ is the Alfv\'en velocity vector, $\mu_0$ is the vacuum permeability, $\bm{k}$ is the wave vector, $\theta$ is the colatitude, and the subscript $r$ denotes the radial component.
As noted by P+19, these modes are typically subinertial ($\omega_{\rm co} < 2\, \Omega$), and are trapped in an equatorial waveguide defined by $\left|\,\cos\theta\,\right| \leq \omega_{\rm co}\, /\, 2\,\Omega = 1 / \left|\,s\,\right|$, where $\Omega$ is the angular rotation rate of the star and $s$ is the spin parameter \cite[e.g.][]{2003_Townsend}. The maximal angular Alfv\'en frequency estimate to be compared with the angular pulsation frequency, based on Eq. (\ref{eq:alfven_vel_prat}), is thus
\begin{equation}
    \omega_{\rm \,A,max} = \dfrac{b_r\left(r\right) k_r\ \omega_{\rm co}}{2\, \Omega \sqrt{\ \mu_0\ \rho\left(r\right)}} = \dfrac{b_r\left(r\right) k_r}{\left|\,s\,\right| \sqrt{\ \mu_0\ \rho\left(r\right)}} ~. \label{eq:wamax_1}
\end{equation}

Adopting a rough estimate for $k_r$ led P+19 to consider the following consistency check: 
\begin{align}
    \dfrac{0.17 B_0 \left|\,n\,\right|}{\Omega\ R \sqrt{\ \mu_0\ \rho_c}} \ll 1~,
\end{align}
which is equivalent to Eq. (\ref{eq:criterion}) when estimating the Alfv\'en frequency in the following form, similar to Eq. (\ref{eq:wamax_1}):
\begin{equation}
    \omega_{\rm A} = \dfrac{0.34 B_0 \left|\,n\,\right|}{\left|\,s\,\right| R\sqrt{\ \mu_0\ \rho_c}}~, \label{eq:omega_a_prat}
\end{equation}
where $\rho_c$ is the central density, $n$ is the mode radial order, $R$ is the stellar radius, and $B_0$ is the magnetic field amplitude scaling factor.

The Alfv\'en frequency estimate defined in Eq. (\ref{eq:omega_a_prat}) does not take into account the spatial variability of the radial component of the magnetic field vector, nor of the density profile. The radial magnetic field strength estimate is moreover biased towards the stellar model used in P+19. We therefore develop two extensions of this $\omega_{\rm A}$ estimate, to improve the consistency check of the perturbative magnetic frequency shifts.

The first extension explicitly evaluates the radial magnetic field $b_r\left(r\right)$ and density profile $\rho\left(r\right)$ but estimates $k_r$ in the same way as P+19. The Alfv\'en frequency profile is therefore equal to:
\begin{align}
    \omega_{\rm A}\left(r,\theta_{\rm T}\right) \sim \dfrac{B\left(r,\theta\right) \left|\,n\,\right|}{R\sqrt{\ \mu_0\ \rho\left(r\right)}} \equiv \dfrac{b_r\left(r\right) \left|\,n\,\right| \cos\theta_{\rm T}}{R\sqrt{\ \mu_0\ \rho\left(r\right)}}~, \label{eq:omega_a_first_ext}
\end{align}
where the density profile does not depend on $\theta$ for our 1D stellar models and where $\theta_{\rm T}$ is defined as:
\begin{equation} \label{eq:omega_T}
    \theta_{\rm T} = \left\{
  \begin{array}{lcr}
    \cos^{\,-1} \left(\dfrac{1}{\left|\,s\,\right|}\right) & \text{if } & \omega_{\rm co} < 2\, \Omega~,\\
    0\degree & \text{if } & \omega_{\rm co} \geq 2\, \Omega~.
  \end{array}
\right.
\end{equation}

The second extension uses the dispersion relation for gravito-inertial waves within the TAR \citep{Mathis_2009} to estimate the radial wave vector as:
\begin{align}
    k_r\left(r\right) = \dfrac{N\left(r\right)}{\omega_{\rm co}} \dfrac{\sqrt{\lambda_{l,m}\left(s\right)}}{r}~, \label{eq:radial_mathis2009}
\end{align}
where $\lambda_{l,m}\left(s\right)$ is the eigenvalue of the Laplace tidal equations which need to be solved in the TAR, and where $N(r)$ is the Brunt-V\"ais\"al\"a frequency profile. We verified numerically that the radial wave vector estimation based upon \citet{Mathis_2009} is a good approximation, fulfilling the condition $\left(\bm{k}\bm{\cdot}\bm{\xi}\right) \approx 0$, where $\bm{\xi}$ is the Lagrangian displacement, for all models in the near-core overshoot zone. We then obtain the Alfv\'en frequency profile at $\theta_{\rm T}$ from Eq. (\ref{eq:radial_mathis2009}) by computing:
\begin{align}
    \omega_{\rm A}\left(r,\theta_{\rm T}\right) \sim \dfrac{B\left(r,\theta_{\rm T}\right)N\left(r\right)\sqrt{\lambda_{l,m}\left(s\right)}}{\omega_{\rm co}\ r\sqrt{\ \mu_0\ \rho\left(r\right)}}~. \label{eq:omega_a_second_ext}
\end{align}

It is expected that modes undergo interactions reminiscent of avoided crossings when their frequencies change throughout stellar evolution and become nearly the same, a phenomenon called mode bumping \cite[e.g.][]{1977_aizenman_smeyers,1979_roth_weigert,1981_JCD,1992_Gautschy,2010_BOOK_Smeyers_VanHoolst}. Such interactions are not included in the perturbative criterion (Eq. (\ref{eq:criterion})) and lead to negative magneto-rotationally modified period spacings. This is discussed further in Sect. \ref{subsect:mode_bump}.

\section{Computational setup}\label{sec:3}

We used the Modules for Experiments in Stellar Astrophysics (MESA) one-dimensional stellar structure and evolution code \cite[][version 10396]{MESA_2011,MESA_2013,MESA_2015,MESA_2015_erratum,MESA_2018} to calculate non-rotating, non-magnetic stellar models for a range of input parameters. The linear, adiabatic eigenmodes were computed using the stellar oscillation code GYRE \cite[][ version 5.2]{GYRE_2013,GYRE_2018}, including the effects of rotation within the TAR. The link to the example inlists is provided in Appendix \ref{sec:C1}. 

\begin{table}[t!]\centering
\caption{Values of the parameters varied in the MESA grid.}
\label{tab:MESA_param_table}
\begin{tabular}{@{}Sl l Sl Sl Sl@{}}
\hline \hline
    MESA parameter && \multicolumn{3}{c}{Explicit values}   \\ \cmidrule{1-1} \cmidrule{3-5}
    $f_{\rm ov}$ && $0.004$ & $0.014$ & $0.024$ \\
    $X_{\rm c}$ && $0.005$ & $0.340$ &$0.675$\\
    $D_{\rm mix}$ (cm$^{\,2}$ s$^{\,-1}$)  && $0.1$ & $1.0$ & $10.0$\\
    $Z_{\rm ini}$ && $0.010$ & $0.014$ & $0.018$\\
    $M_{\rm ini}$ (M$_\sun$) && $1.3$ & $2.0$ & $3.0$ \\
    $\alpha_{\rm MLT}$ && $1.5$ & $1.8$ & $2.0$\\
    \hline
\end{tabular}
\end{table}

\subsection{MESA setup}\label{sec:3.1}
Our grid parameter ranges are selected based on typical values used for $\gamma$~Dor and SPB asteroseismic modeling \citep{2012_Papics,2016_Van_Reeth,Van_Reeth_2018_differential_rotation,Buysschaert_2018_magnetic_forward_model,Mombarg_2019_gamma_dor}, with an overview of the selected parameter ranges available in Table \ref{tab:MESA_param_table}.
For all MESA models, we assume the \citet{2009_Asplund} metal mixture.  We fixed the initial helium mass fraction as $Y_{\rm ini}=0.2795$ and varied the metallicity of the models as indicated in Table \ref{tab:MESA_param_table}. The initial hydrogen fraction was then adopted according to $X=1-Y_{\rm ini}-Z_{\rm ini}$. 
Further, the core hydrogen mass fraction $X_{\rm c}$
serves as an age indicator for stars on the MS, where a smaller $X_{\rm c}$ indicates a star nearer to the terminal-age MS (TAMS). We extract models near the zero-age MS (ZAMS; $X_{\rm c} \approx 0.675$), near the middle of the MS (mid-MS; $X_{\rm c} \approx 0.340$) and near the TAMS ($X_{\rm c} \approx 0.005$), for initial masses $M_{\rm ini} \in [1.3, 2.0, 3.0]$ M$_\sun$.

Convection is treated using the \citet{1968_Cox_Giuli} variant of mixing-length theory (MLT), where the efficiency of convection is parametrized by the mixing length parameter $\alpha_{\rm MLT}$. In their study of $g$ modes in $\gamma$~Dor stars, \citet{2016_Van_Reeth} take a fixed solar-calibrated value $(\alpha_{\rm MLT}=1.8)$, because there is considerable uncertainty on the value needed to model other stars \citep{2018_Viani}. More recently, \citet{Mombarg_2019_gamma_dor} pointed out that this parameter influences the asymptotic $g$-mode period spacing $\Pi_0$, the effective temperature $T_{\rm eff}$ and the surface gravity $\log g$. Therefore, we vary the $\alpha_{\rm MLT}$ parameter, slightly expanding our grid range compared to \citet{Mombarg_2019_gamma_dor}, because we consider an increased fundamental parameter range.

The way mixing processes due to rotational and magnetic instabilities are included in stellar evolution models such as MESA induces numerical discontinuities, which lead to $g$-mode behavior not observed in \textit{Kepler} data \citep{TruyaertKevin2016Tior}. 
We take into account that these mixing processes are not calibrated by approximating the envelope mixing with a constant value, denoted $D_{\rm mix}$, making the approach less model-dependent, and allowing us to constrain the diffusive mixing coefficients beyond the fully mixed convective cores \cite[e.g.][]{2016_Moravveji}. Convective core overshooting is parametrized using the exponential prescription of \citet{1996_Freytag} and \citet{2000_Herwig}, and is characterized by the overshoot parameter $f_{\rm ov}$, which is a measure of the distance over which convective eddies penetrate into the radiative zone. The values of $D_{\rm mix}$ and $f_{\rm ov}$ are not constrained from first principles and these parameters are therefore varied.  

The stellar atmosphere boundary condition we used for our MESA models (`\texttt{simple photosphere}') estimates the surface boundary conditions at optical depth $\tau = 2/3$.
We do not include stellar winds because they are not expected during the MS in this mass range.

\subsection{GYRE setup}\label{sec:3.2}

We compute dipolar gravity mode frequencies in the adiabatic approximation and take rotation into account using the TAR, with radial orders $n$ ranging from $-10$ to $-50$. This range is consistent with the radial order range of modes excited in $\gamma$ Dor stars according to non-adiabatic calculations in \citet{2005_Dupret_convection_pulsation_coupling_2}, \citet{2013_Bouabid} and \citet{2019_Ouazzani}. It also agrees with the observed radial orders of dipolar (prograde) $g$ modes in the large sample of 611 $\gamma$ Dor stars analyzed by \citet{2019_Gang_Li}. 
We vary the rotation rate in the form of the rotation ratio parameter $\mathcal{R}_{\rm rot} \in \left[0.01,0.25,0.50,0.75\right]$, defined as:
\begin{align}
    \mathcal{R}_{\rm rot} = \dfrac{\Omega}{\Omega_{\rm\, c,Roche}}~,
\end{align}
where $\Omega_{\rm\, c,Roche}$ is an approximation of the Roche critical angular rotation frequency of the model \cite[see e.g.][]{Maeder_book_rot_mag}, defined as:
\begin{align}
    \Omega_{\rm\, c,Roche} = \sqrt{\dfrac{8\ G M}{27\ R_{\rm p}^{\,3}}}~, \label{eq:crit_rot}
\end{align}
where $G$ is the gravitational constant and $R_{\rm p}$ is the polar radius, which is smaller than the equatorial radius $R_{\rm eq}$ if the star is rotating.
For our purposes, we assume $R_{\rm p}$ to be equal to the stellar radius of the MESA model, $R$.

Boundary conditions are required to close the (sub)set of pulsation equations solved by GYRE. The inner boundary condition is regularity-enforcing: it discards nonphysical behavior of the characteristic equations at the inner boundary \cite[see e.g.][]{GYRE_2013}. We varied the outer boundary condition, which affects the behavior of pulsations near the stellar surface, and the difference scheme used to solve the pulsation equations. These variations induce frequency shifts for our reference model {smaller than} $0.008$\,\textmu Hz. We therefore deem their influence to be negligible, as this is the frequency resolution of a 4-yr \textit{Kepler} data set. The default parameters used in this work are: outer boundary condition `\texttt{UNNO}' and difference scheme `\texttt{MAGNUS GL4}'. 

\subsection{Magnetic setup}

The confined magnetic field model we use in this work was semi-analytically derived by \citet{2010_Duez_Braithwaite_field} and \citet{Duez_field_2010}, and contains both poloidal and toroidal components. Numerical simulations by \citet{2004_Braithwaite_spruit} and \citet{2006_Braithwaite_Nordlund} obtain similar, stable, axisymmetric mixed poloidal-toroidal magnetic field configurations, hence confirming the stability of the \citet{2010_Duez_Braithwaite_field} field configuration. The magnetic field model represents an initial, confined configuration, avoiding destabilizing current sheets at the surface by enforcing the cancellation of both radial and latitudinal surface fields. In this way, magnetic helicity is conserved. We use the lowest-energy configuration of the \citet{2010_Duez_Braithwaite_field} field that adheres to these constraints. Such an initial axisymmetric configuration can over time evolve to an open, axisymmetric one, through long-term diffusive processes such as Ohmic diffusion. The \citet{2010_Duez_Braithwaite_field} magnetic field model strictly does not hold for convective regions. 
However, one cannot simply assume that the magnetic field strength inside the core is zero, because a convective core dynamo field is probably present within it and may couple to the fossil field at the core boundary \citep{2005_Brun,2009_Featherstone,Augustson_dynamos}. Therefore, we opted to compute the \citet{2010_Duez_Braithwaite_field} field within the whole stellar interior. 

We vary the near-core magnetic field strength by varying the magnetic field amplitude scaling factor $B_0$ (see P+19). To derive the range of this parameter, we base ourselves on fossil field strength estimates of HD 43317 \cite[][P+19]{Buysschaert_2018_magnetic_forward_model}, which range from $3 \cdot 10^{\,4}$ to $1.5 \cdot 10^{\,5}$ G. We therefore consider magnetic field amplitude scaling factors $B_0$ of $10^{\,4}$, $\,10^{\,5}$ and $10^{\,6}$ G. 
The corresponding near-core field strengths, approximated as the total magnetic field strength at the radius nearest to the convective core for which $N$ is positive, depend on the density profile of the stellar models, and therefore also depend on the evolutionary stage. 

The surface field strength ranges can be estimated from the radial profiles of magnetic energy density obtained from numerical simulations of the formation of stable magnetic field equilibria \cite[see figure~8 from][]{2008_Braithwaite}. 
\citet{2012_Briquet} performed a similar exercise, adopting a surface magnetic field strength that is approximately $30$ times weaker than the core field strength. We compute such a ratio, hereafter referred to as the surface scaling factor, by comparing the surface field according to \citet{2008_Braithwaite} to the near-core field strengths. Because we only consider fully axisymmetric configurations in this work, we consider these types of configurations that have relaxed over timescales \ga~$9$ Alfv\'en timescales $\tau_{\rm A} \equiv R \sqrt{M/2E}$, with $E$ the magnetic energy inside the star \cite[see figures 2 and 9 in][]{2008_Braithwaite}.
\citet{2008_Braithwaite} tapers the vector potential so that their magnetic field $B \sim \rho^{\,p}$, where $p = 2/3$ if the star is formed from a uniformly magnetized cloud and the same fraction of flux is lost from all fluid elements. The higher $p$, the lower the amount of magnetic flux lost throughout the surface, and only the configurations of \citet{2008_Braithwaite} for which $p=2/3$ or $p=1$ relax to roughly axisymmetric states. The $p=1$ model loses less magnetic flux through the surface, and therefore is more representative for the \citet{2010_Duez_Braithwaite_field} field model. However, no surface scaling factors can be computed for this configuration, and therefore no direct estimate can be made of the expected surface magnetic field strength using the near-core field strength. Instead, we compute such factors for the $p=2/3$ model of \citet{2008_Braithwaite}, providing us with an estimate of the upper limit of the dipole surface field strength ranges expected for a given stellar model.


\section{Parameter study}\label{sec:4}
We restrict ourselves to modes with the simplest geometry to understand how magnetic fields influence period spacing patterns: dipole zonal modes ($l=1,m=0$, where $l$ is the degree and $m$ is the azimuthal order) and sectoral dipole modes ($l = 1$,~ $\left|\,m\,\right| = 1$), hereafter referred to as retrograde ($m=-1$) and prograde ($m=1$) modes. 
We analyze the mode behavior based on the comparison of the rotationally and magneto-rotationally modified frequencies with the characteristic frequencies: the Brunt-V\"ais\"al\"a frequency $N\left(r\right)$, Lamb frequency $S_1\left(r\right)$, Coriolis frequency $2 \Omega/ 2\pi$ and the Alfv\'en frequency estimates described in Eqs. (\ref{eq:omega_a_prat}-\ref{eq:omega_a_first_ext}) and (\ref{eq:omega_a_second_ext}). This allows us to check the consistency of treating the magnetic field perturbatively (hereafter referred to as the perturbative criterion), see Eq. (\ref{eq:criterion}).

To assess magneto-rotational modification of period spacing patterns, we vary the input parameters of a background equilibrium MESA model and compare to a reference model.
We discuss the results for a reference model that has a mass of $3$ M$_\sun$ with solar metallicity ($Z_{\rm ini}=0.014$), core overshooting $f_{\rm ov} = 0.014$, envelope mixing $D_{\rm mix} =$  $1.0$ cm$^{\,2}$ s$^{\,-1}$, $\alpha_{\rm MLT} = 1.8$, with a \texttt{simple photosphere} appended to the interior structure. For the computation of the pulsation frequencies, we take a rotation ratio of $\mathcal{R}_{\rm rot} = 0.25$, and $B_0 = 10^{\,6}$ G (i.e. the strongest field in our range), unless otherwise specified. 

Any magnetic mode frequency shifts that are larger than or equal to 0.008~\textmu Hz, are resolved in 4-yr baseline photometric measurements obtained with the \textit{Kepler} telescope \citep{KEPLER_mission,Bowman_BOOK}. Additionally, the frequency resolution of a 1-yr baseline TESS continuous viewing zone (TESS-CVZ) time series is approximately $0.03$~\textmu Hz \citep{TESS_mission}, whereas the 2-yr baseline of long-duration observation phase time series obtained with the planned PLATO mission allow one to detect magnetic shifts of approximately $ 0.02$~\textmu Hz \citep{PLATO_mission}. \citet{2019_Gang_Li} derived $g$-mode period spacing patterns of $611$ $\gamma$~Dor stars from \textit{Kepler} time series, quoting typical frequency uncertainties of $0.01$\,\textmu Hz if both $g$ and Rossby modes are detected and $0.09$\,\textmu Hz if only $g$ modes were identified. We therefore quantify the detectability by converting magnetic mode period shifts to the frequency domain and compute the mode frequency differences $\Delta \omega_{\,\rm n_{i}}$ in the following way for a $g$ mode of radial order $n_{i}$:
\begin{equation}
    \Delta \omega_{\,\rm n_{i}} = \,\left|\,\omega_{\rm\, rot, n_{i}} - \omega_{\rm\, j, n_{i}}\,\right|~, \label{eq:diff_omega} 
\end{equation}
where $\omega_{\rm\, j, n_{i}}$ is the magneto-rotationally modified angular frequency of the specific $g$ mode that undergoes a magnetic shift when varying a parameter $j$ and where $\omega_{\rm\, rot, n_{i}}$ is the rotationally modified angular frequency of the same mode. The magnetic shifts can be computed in a perturbative manner if $\omega_{\rm\, rot, n_{i}} \ga \omega_{\rm A, n_{i}}$ is fulfilled, where $\omega_{\rm A, n_{i}}$ is the Alfv\'en frequency estimate from Eq.~(\ref{eq:omega_a_second_ext}). We define the perturbative mode frequency difference $\Delta \omega_{\,\rm per, n_{i}}$ as
\begin{equation} \label{eq:omega_allowed}
    \Delta \omega_{\,\rm per, n_{i}} = \left\{
  \begin{array}{lcr}
    \Delta \omega_{\,\rm n_{i}} & \text{if } & \omega_{\rm\, rot, n_{i}} \ga \omega_{\rm A, n_{i}}\\
    0 & \text{if } & \omega_{\rm\, rot, n_{i}} < \omega_{\rm A, n_{i}}
  \end{array}
\right. ~,
\end{equation}
and list the maximal perturbative mode frequency differences $\Delta \omega_{\,\rm per} / 2 \pi$ and maximal mode frequency differences $\Delta \omega / 2 \pi$ in Table \ref{tab:freq_deviation_table}, which are defined as:
\begin{subequations}
\par\noindent
\begin{gather}
     \Delta \omega_{\,\rm per} / 2 \pi\ = \max \left(\,\Delta \omega_{\rm\, per, n_{i}}/2\pi\, \forall\, n_{i} \in [-50,\ldots,-10]\,\right)~, \label{eq:max_omega_allowed}  \\
     \Delta \omega / 2 \pi\ = \max \left(\,\Delta \omega_{\,\rm n_{i}}/2\pi\, \forall\, n_{i} \in [-50,\ldots,-10]\,\right)~. \label{eq:max_omega} 
\end{gather}
\par\noindent
\end{subequations}

We verify whether the imposed magnetic fields can suppress rotational mixing or convective core overshoot and thermal convection according to stability criteria described in \citet{1999_Spruit} and \citet{2011_Zahn}, similar to what was done in \citet{2012_Briquet}. We compute the critical magnetic field strengths required to fulfill the criteria and discuss the outcomes in Appendix \ref{app:Bcrit}.

\subsection{Evolutionary phase: $X_{\rm c}$}
\begin{figure*}
\resizebox{0.49\hsize}{!}{\includegraphics{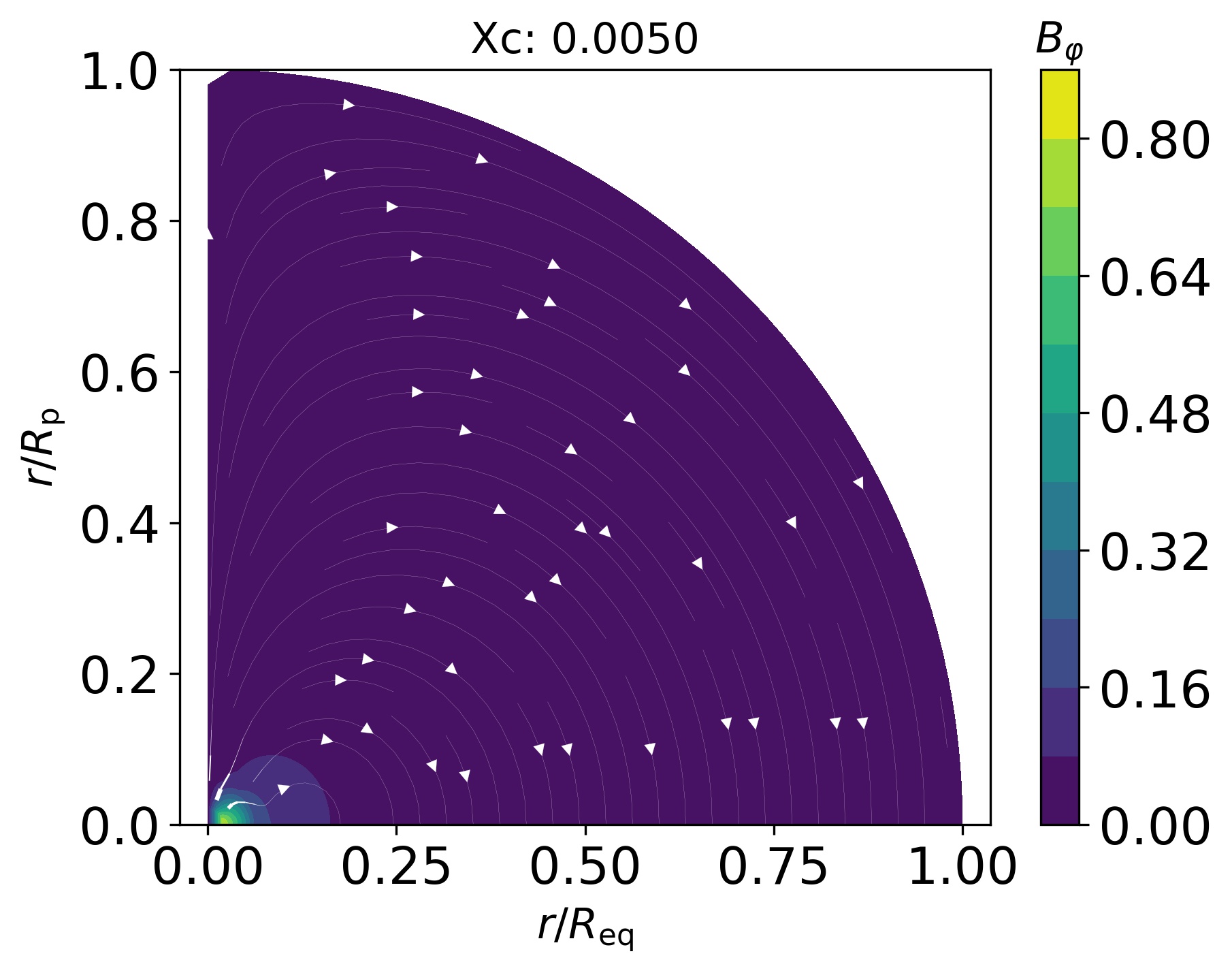}}
\resizebox{0.49\hsize}{!}{\includegraphics{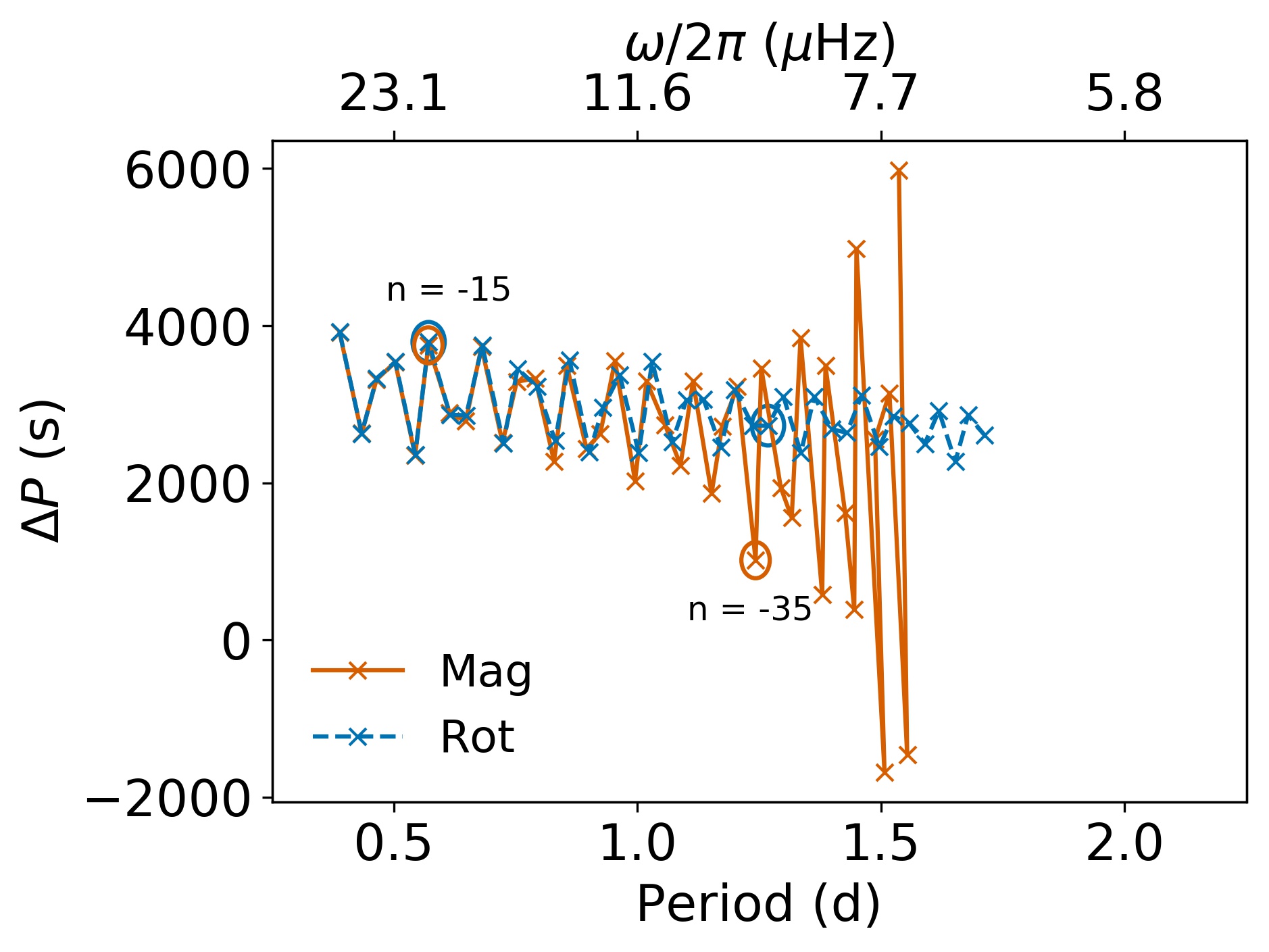}}\\
\resizebox{\hsize}{!}{\includegraphics{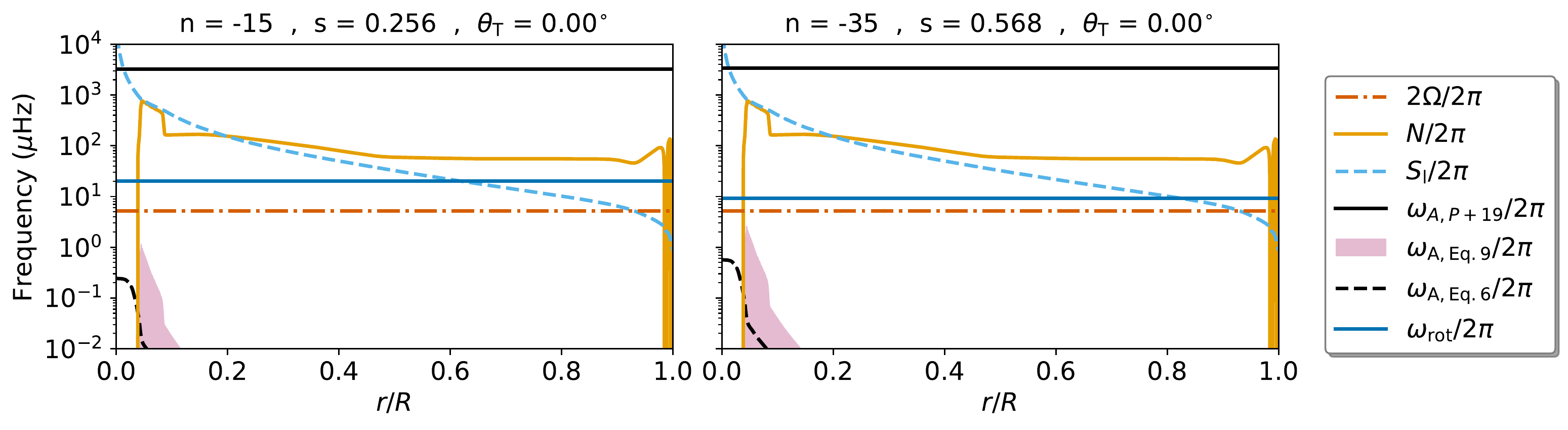}}
\caption{Normalized magnetic field model ($R_{\rm p}$ and $R_{\rm eq}$ denote the polar and equatorial radius, respectively; \textit{top left}) and period spacing pattern for dipole zonal modes (\textit{top right}) in the $3$-M$_\sun$ TAMS reference model discussed in the text. If $\omega_{\rm A}$ is computed using Eq. (\ref{eq:omega_a_second_ext}), $\omega_{\rm co} > \omega_{\rm A}$ is satisfied for every mode. The symbol $\omega$ denotes the corresponding angular pulsation frequency in the inertial frame. \textit{Bottom row:} Characteristic frequencies compared to the rotationally modified pulsation frequency $\omega_{\rm rot}/2\pi$ to assess the perturbative criterion: Coriolis frequency $2\Omega/2\pi$, Buoyancy frequency $N$, Lamb frequency $S_{\rm 1}$, and Alfv\'en frequency $\omega_{\rm A}/2\pi$ estimated using the P+19 approach (Eq. (\ref{eq:omega_a_prat})), our first extension (Eq. (\ref{eq:omega_a_first_ext})), and our second extension (Eq. (\ref{eq:omega_a_second_ext})).}
\label{fig:TAMS_evolution}
\end{figure*}

Throughout the evolution of a star on the MS, its internal structure and rotation change significantly.
The rotationally modified period spacing patterns, denoted on our figures as "Rot", therefore also change. To study how magneto-rotationally modified period spacing patterns, denoted on our figures as "Mag", change throughout MS evolution, we vary the $X_{\rm c}$ of our reference model.

\begin{figure*}
\resizebox{0.49\hsize}{!}{\includegraphics{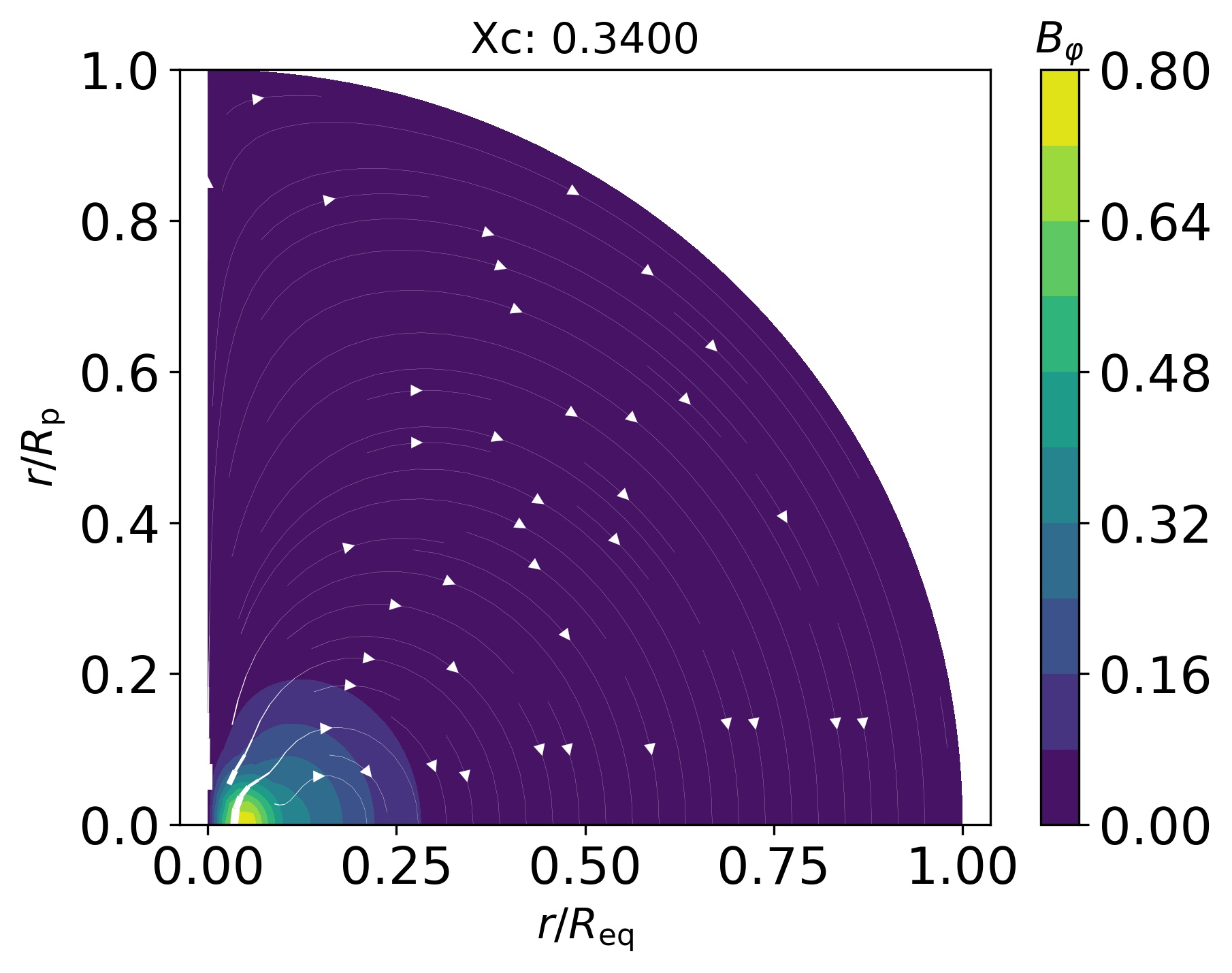}}
\resizebox{0.49\hsize}{!}{\includegraphics{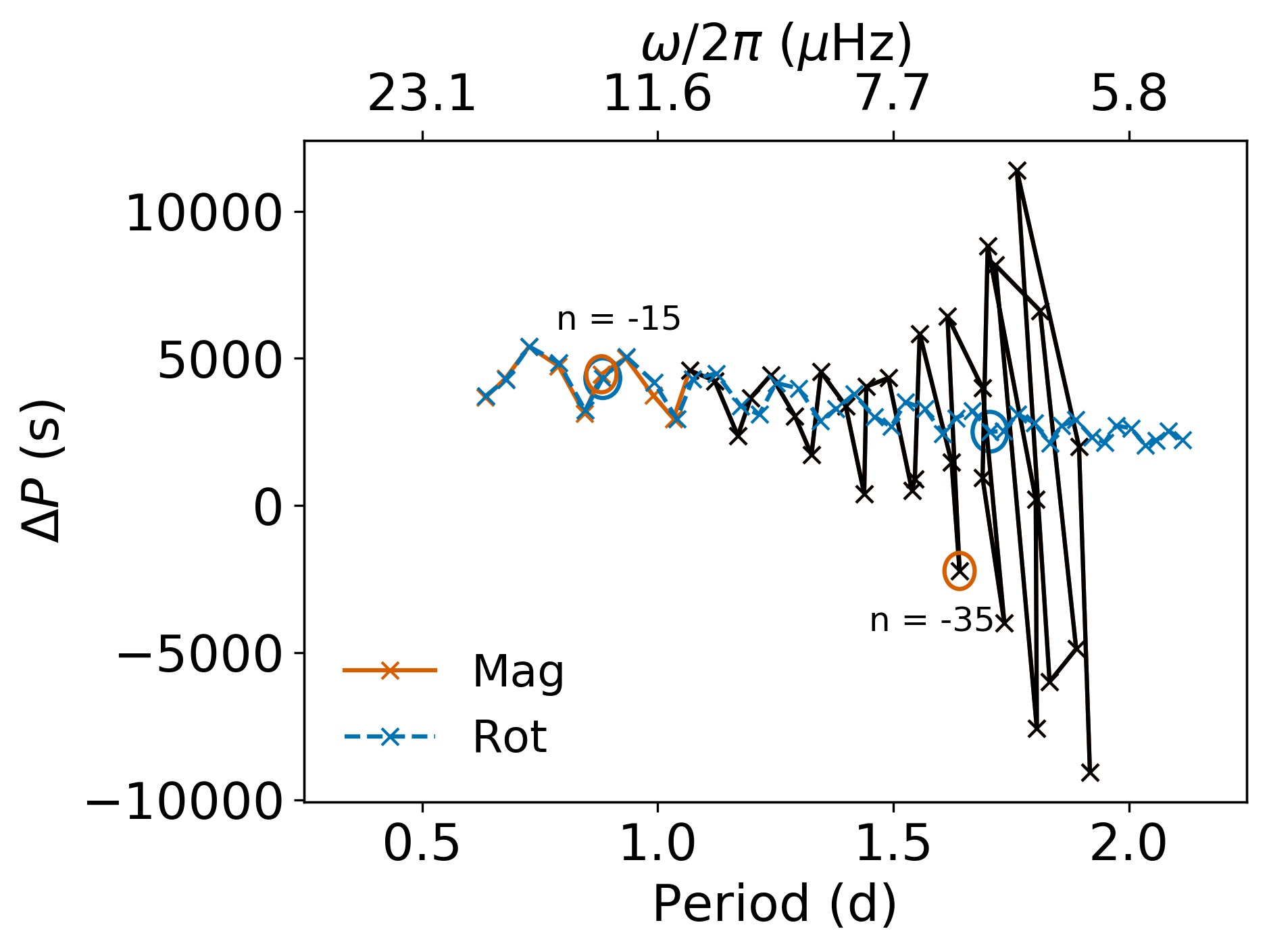}}\\
\resizebox{\hsize}{!}{\includegraphics{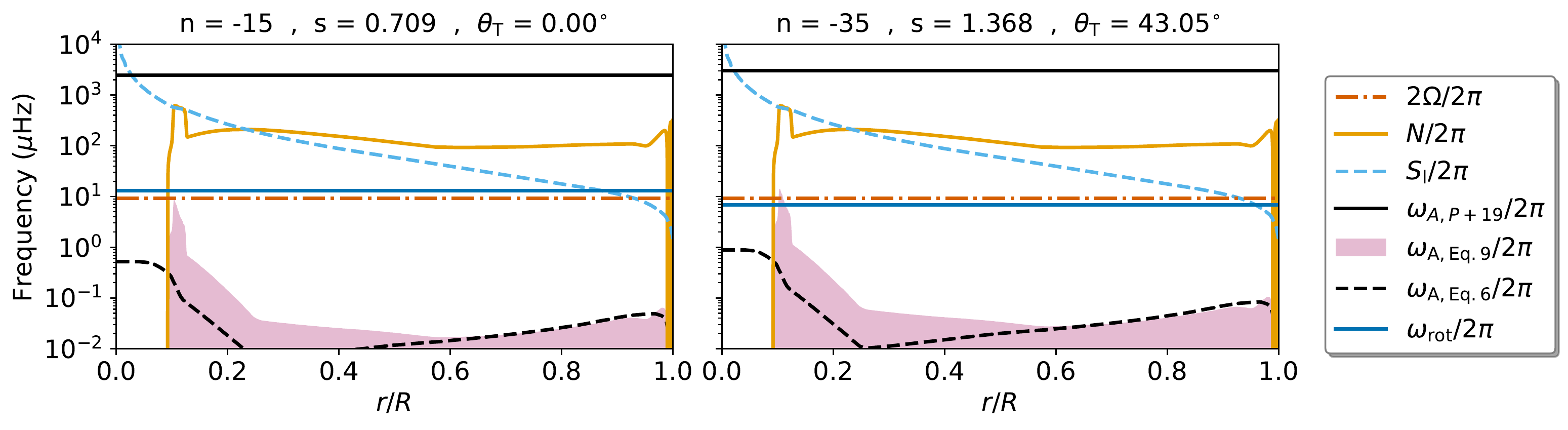}}
\caption{Same as Fig. \ref{fig:TAMS_evolution}, but for a mid-MS model. Black parts of the period spacing pattern indicate where $\omega_{\rm A} > \omega_{\rm co}$ if $\omega_{\rm A}$ is computed with Eq. (\ref{eq:omega_a_second_ext}).}
\label{fig:midMS_evolution}
\end{figure*}
 
\begin{figure*}
\resizebox{0.49\hsize}{!}{\includegraphics{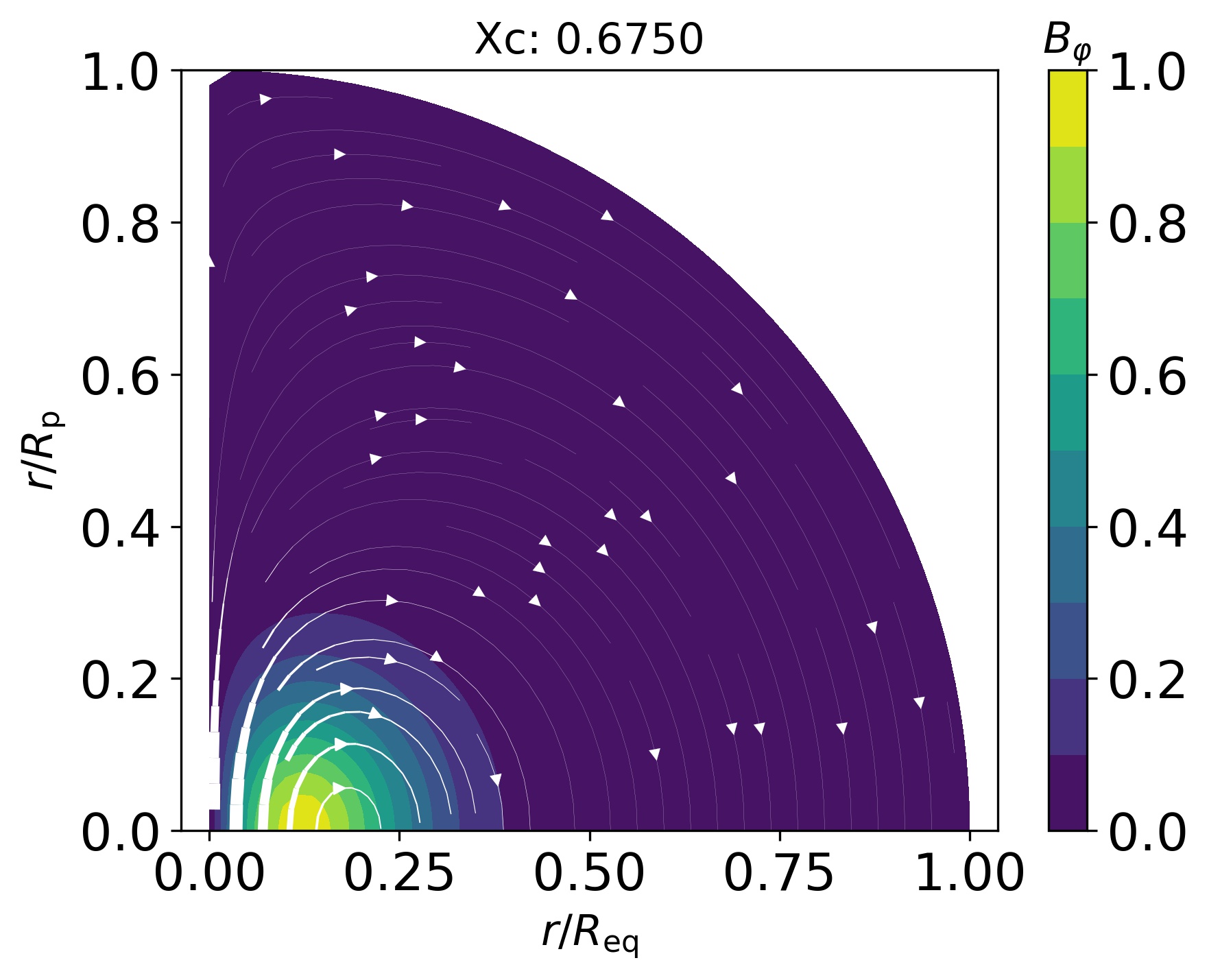}}ƒ
\resizebox{0.49\hsize}{!}{\includegraphics{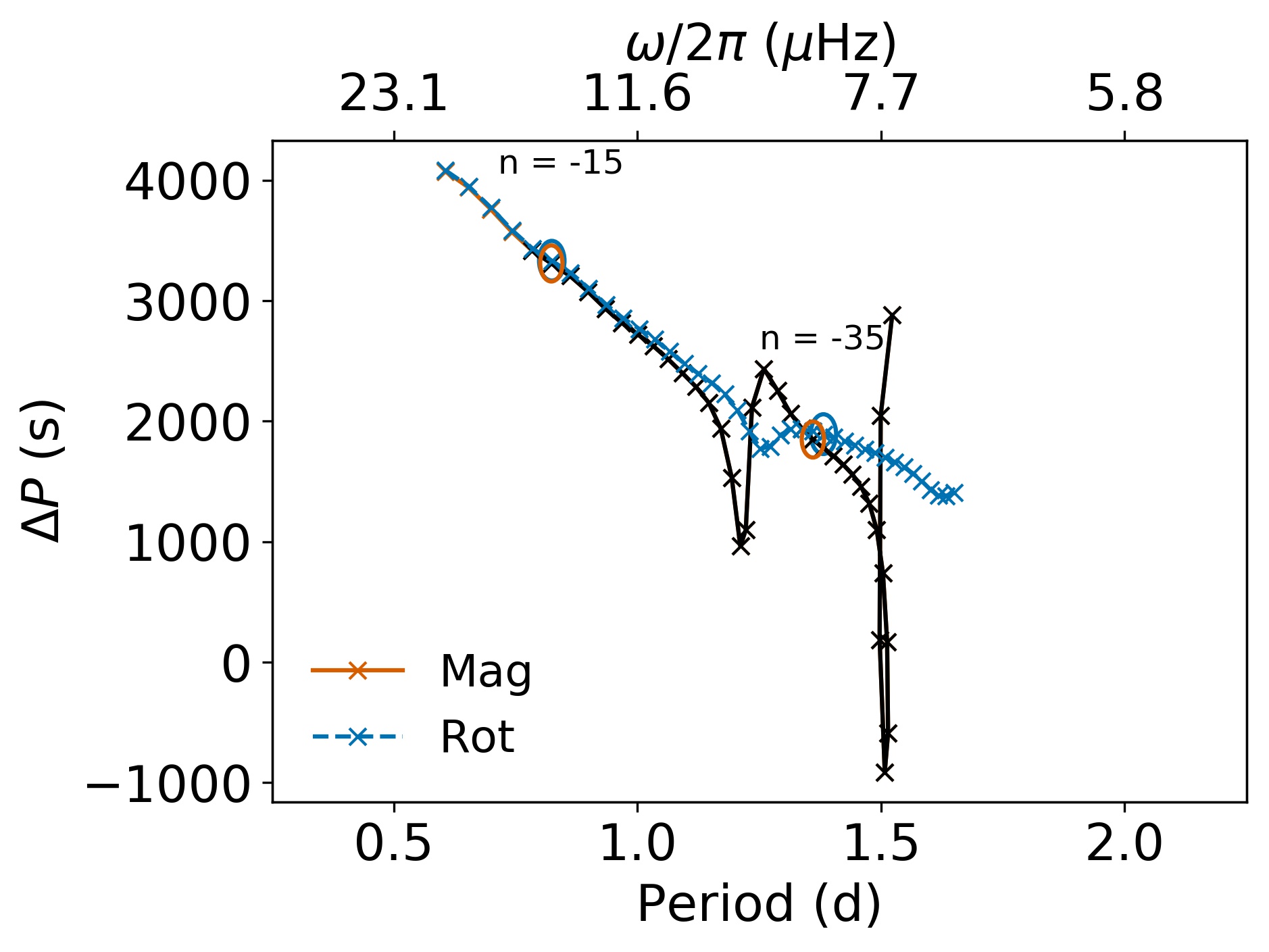}}\\
\resizebox{\hsize}{!}{\includegraphics{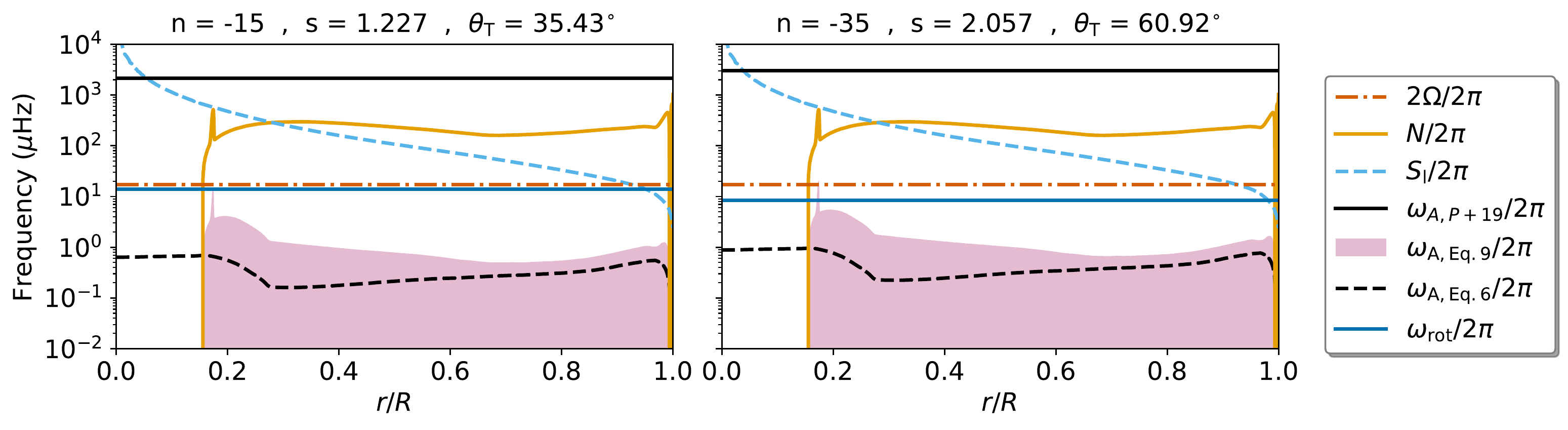}}
\caption{Same as Fig. \ref{fig:TAMS_evolution}, but for a ZAMS model. Black parts of the period spacing pattern indicate where $\omega_{\rm A} > \omega_{\rm co}$ if $\omega_{\rm A}$ is computed with Eq. (\ref{eq:omega_a_second_ext}).}
\label{fig:ZAMS_evolution}
\end{figure*}

Near the TAMS the perturbative criterion holds for every mode when $\omega_{\rm A}$ is calculated with Eq. (\ref{eq:omega_a_first_ext}), which ignores the influence of rotation and the Brunt-V\"ais\"al\"a frequency on the radial wave number. Although this estimate is crude, we notice a similar trend when the Alfv\'en frequency is estimated using Eq. (\ref{eq:omega_a_second_ext}).
We consider both estimates when analyzing mode behavior, but also show the less appropriate Eq.~(\ref{eq:omega_a_prat}) $\omega_{\rm A}$ estimate in Figs. \ref{fig:TAMS_evolution}-\ref{fig:ZAMS_evolution} to point out how significantly P+19 overestimated the Alfv\'en frequency. In general, mode frequencies computed for mid-MS and ZAMS models require a non-perturbative description of rotational influence, indicated by the spin factors being larger than unity and corresponding large values of $\theta_{\rm T}$ for the selected modes in Figs.~\ref{fig:midMS_evolution}-\ref{fig:ZAMS_evolution}. For multiple selected modes in Figs.~\ref{fig:TAMS_evolution}~and~\ref{fig:midMS_evolution}, $\theta_{\rm T}$ is equal to $0\degree$ because they are not equatorially trapped. 

The magneto-rotationally and rotationally modified mode frequencies overlap if both were to be plotted in the pulsation mode cavity diagrams in Figs. \ref{fig:TAMS_evolution}-\ref{fig:ZAMS_evolution}, which also show the characteristic frequency profiles in the stellar interior. For clarity, we only show the rotationally modified mode frequency. In practice, we implement Eq. (\ref{eq:criterion}) by verifying whether ${\omega_{\rm A}<\omega_{\rm co}}$.
If any of the estimated Alfv\'en frequencies (i.e. from Eqs. (\ref{eq:omega_a_prat}-\ref{eq:omega_a_first_ext}) and (\ref{eq:omega_a_second_ext})) are larger than $\omega_{\rm co}$, the magnetic shift is no longer perturbative. 

We find significant differences between period spacing patterns of magneto-rotationally and rotationally modified gravity modes. Higher radial order modes are confined more to the center \cite[see e.g. figure 10 in][]{2016_Moravveji} and some modes are trapped in the near-core region due to chemical gradients left behind by a receding convective core \citep{2008_Miglio}. The magnetic frequency shifts increase with increasing radial order, with these near-core trapped modes being affected the most, as the magnetic field is strongest in their trapping region. A characteristic magnetic sawtooth-like feature therefore develops for higher radial order modes. We are confronted with the limitations of our perturbative approach where period spacings become negative. For mid-MS models and near-ZAMS models, loops in the magneto-rotationally modified period spacing patterns become apparent. 
This is reflected in the maximal perturbative frequency deviations. For the near-TAMS model, maximal perturbative frequency deviations are obtained for the higher radial order modes (e.g. $n=-50$). For the mid-MS and near-ZAMS models, however, the maximal perturbative frequency deviations listed in Table \ref{tab:freq_deviation_table} are much smaller, because only low-radial order modes (e.g. $n=-10$) undergo shifts that can be treated as a perturbation, as can be observed in the pulsation mode cavity diagrams in Figs. \ref{fig:midMS_evolution}-\ref{fig:ZAMS_evolution}.

A clear difference in normalized magnetic field models derived from stellar models at different evolution stages (near-TAMS, mid-MS, near-ZAMS) can be noticed in Figs. \ref{fig:TAMS_evolution}-\ref{fig:ZAMS_evolution}, where the toroidal field strength is indicated with color, and poloidal field lines are indicated by the streamlines, whose thickness is proportional to the poloidal field strength. Near-TAMS fields are (much) more confined to the near-core regions than their near-ZAMS counterparts, and hence only exert significant influence in these small regions. This influence can be characterized by the ratio $\omega_{\rm A}/\omega_{\rm co}$, increased in regions of significant influence. For a given $\omega_{\rm co}$, this ratio is proportional to $B\left(r,\theta\right)N\left(r\right) / \sqrt{\rho\left(r\right)}$. All of these profiles peak near the core, and $B\left(r,\theta\right)$ is proportional to $\rho\left(r\right)$, indicating that the validity of the inequality represented in Eq.~(\ref{eq:criterion}) is typically decided in the near-core region.
The Brunt-V\"ais\"al\"a frequency $N\left(r\right)$ is most important in checking the consistency of magnetic frequency shifts in near-ZAMS stars: it is sharply peaked because only a small region near the core has significant chemical gradients. This implies that the perturbative criterion is less easily fulfilled, as confirmed by the mode cavity diagrams shown on the panels of the bottom row of Fig. \ref{fig:ZAMS_evolution}.

We derive surface scaling factors of approximately $20$ to $30$ for the $p=2/3$ model of \citet{2008_Braithwaite} after relaxation times of approximately $25$ and $14$ $\tau_{\rm A}$. The estimated ranges for the surface magnetic field strengths that correspond to our derived near-core field strengths, computed with these surface scaling factors, are available in Table \ref{tab:surface_fields_reference}. If these values are representative for stellar magnetic surface fields, only the $B_{\rm 0} = 10^{\,5}$ G model seems to be realistic, according to the dipole surface magnetic field strengths derived for the early B-type stars, Ap and Bp stars, and magnetic O-type stars \cite[see figure~6 from][]{2019_Schultz}. 
Because of the uncertainty in deriving the surface magnetic field strength for the $p=1$ model of \citet{2008_Braithwaite} we also consider values of $B_{\rm 0}$ higher than the limit imposed by the \citet{2019_Schultz} observations. According to our simulations, these large field strengths do not exceed the critical field strength limits imposed by \citet{1999_Spruit} and \citet{2011_Zahn}. Hence, thermal convection and rotational mixing are not suppressed by the imposed \citet{2010_Duez_Braithwaite_field} magnetic fields.

\begin{table}[t!]\centering
\caption{Derived surface magnetic field strengths $B_{\rm surf}$ according to the surface scaling factors obtained from the $p=2/3$ model of \citet{2008_Braithwaite} and estimated near-core magnetic field strengths $B_{\rm n-c}$ for the \citet{2010_Duez_Braithwaite_field} field at different life phases of the reference model.}
\label{tab:surface_fields_reference}
\begin{tabular}{@{} Sl Sl Sl Sl @{}}
\hline \hline 
 Life phase & $B_{\rm surf}$ (kG) & $B_{\rm n-c}$ (G) & $B_{\rm 0}$ (G)\\ \hline
& $0.390-0.619$ & $1.10 \cdot 10^{\,4}$ & $10^{\,4}$ \\
 near-ZAMS & $3.90-6.19$ & $1.10 \cdot 10^{\,5}$ & $10^{\,5}$\\
  & $39.0-61.9$ & $1.10 \cdot 10^{\,6}$  & $10^{\,6}$\\ \hline
& $0.324-0.513$ & $0.913 \cdot 10^{\,4}$ & $10^{\,4}$ \\
 mid-MS & $3.24-5.13$ & $0.913 \cdot 10^{\,5}$ & $10^{\,5}$\\
  & $32.4-51.3$ & $0.913 \cdot 10^{\,6}$  & $10^{\,6}$\\ \hline
  & $0.176-0.278$ & $0.495 \cdot 10^{\,4}$ & $10^{\,4}$ \\
 near-TAMS & $1.76-2.78$ & $0.495 \cdot 10^{\,5}$ & $10^{\,5}$\\
  & $17.6-27.8$ & $0.495 \cdot 10^{\,6}$  & $10^{\,6}$\\ \hline
\end{tabular}
\end{table}

The magnetic shifts in near-TAMS stars can be treated perturbatively within our approximations and significant deviations between rotationally and magneto-rotationally modified period spacing patterns are observed. We therefore conclude that gravity-mode-pulsating stars nearer to the TAMS are excellent probes of perturbative internal magnetic fields. We therefore focus the following discussions on the implications of varying the MESA parameters for near-TAMS models.

\subsection{Field strength: $B_0$, and rotation ratio: $\mathcal{R}_{\rm rot}$}

\begin{figure*}
\includegraphics[width=6cm,height=4.5cm]{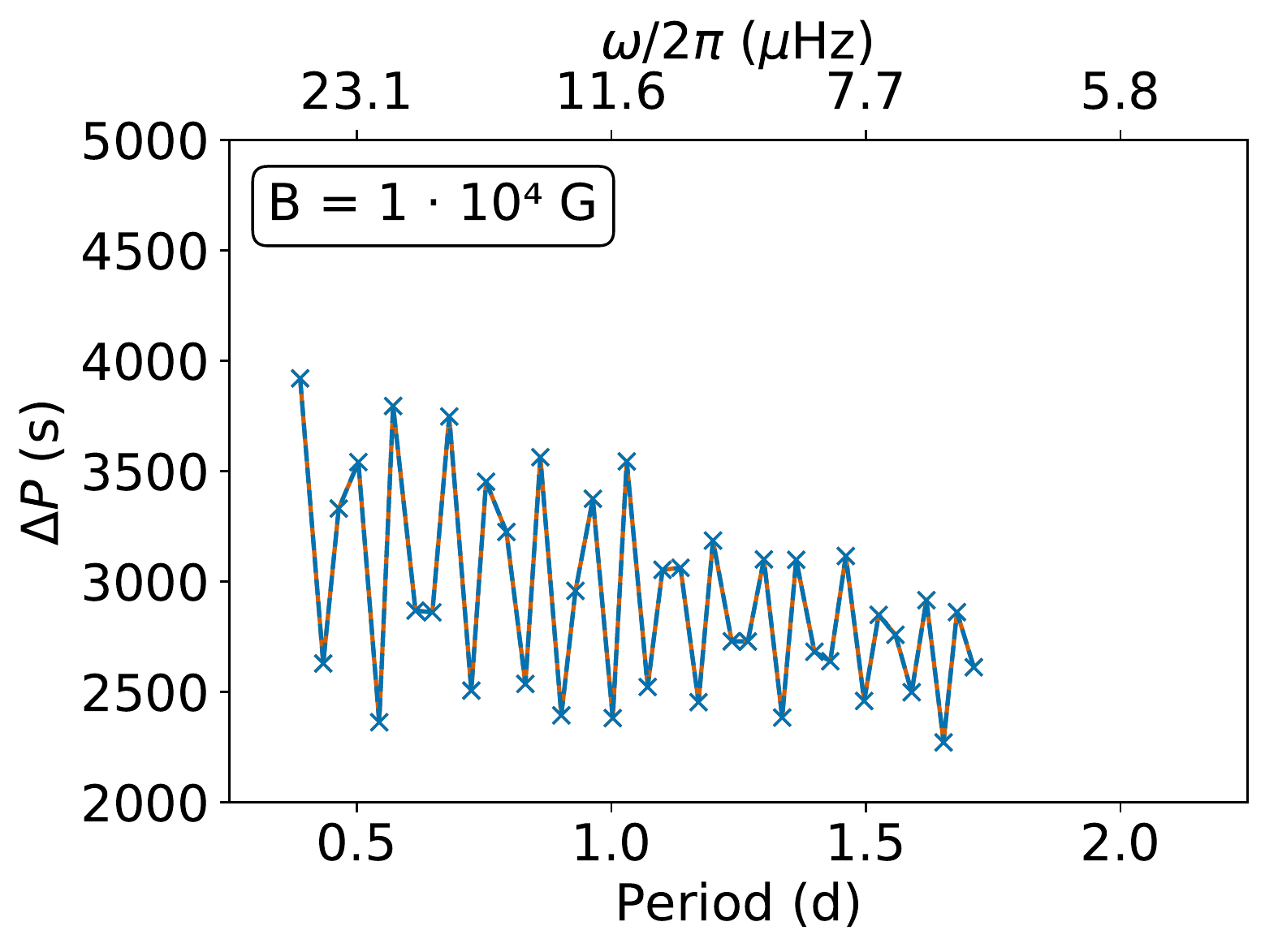}\includegraphics[width=6cm,height=4.5cm]{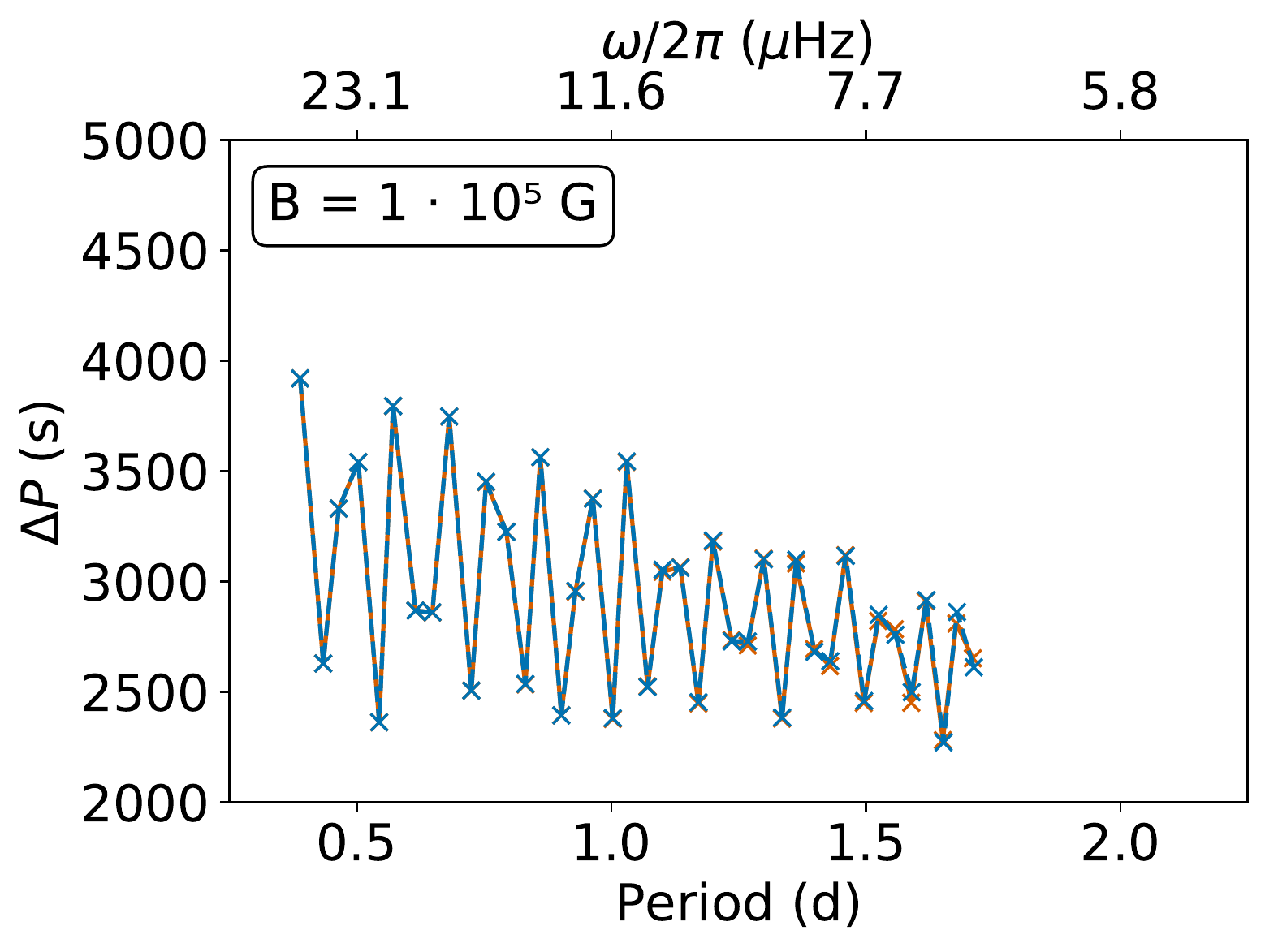}\includegraphics[width=6cm,height=4.5cm]{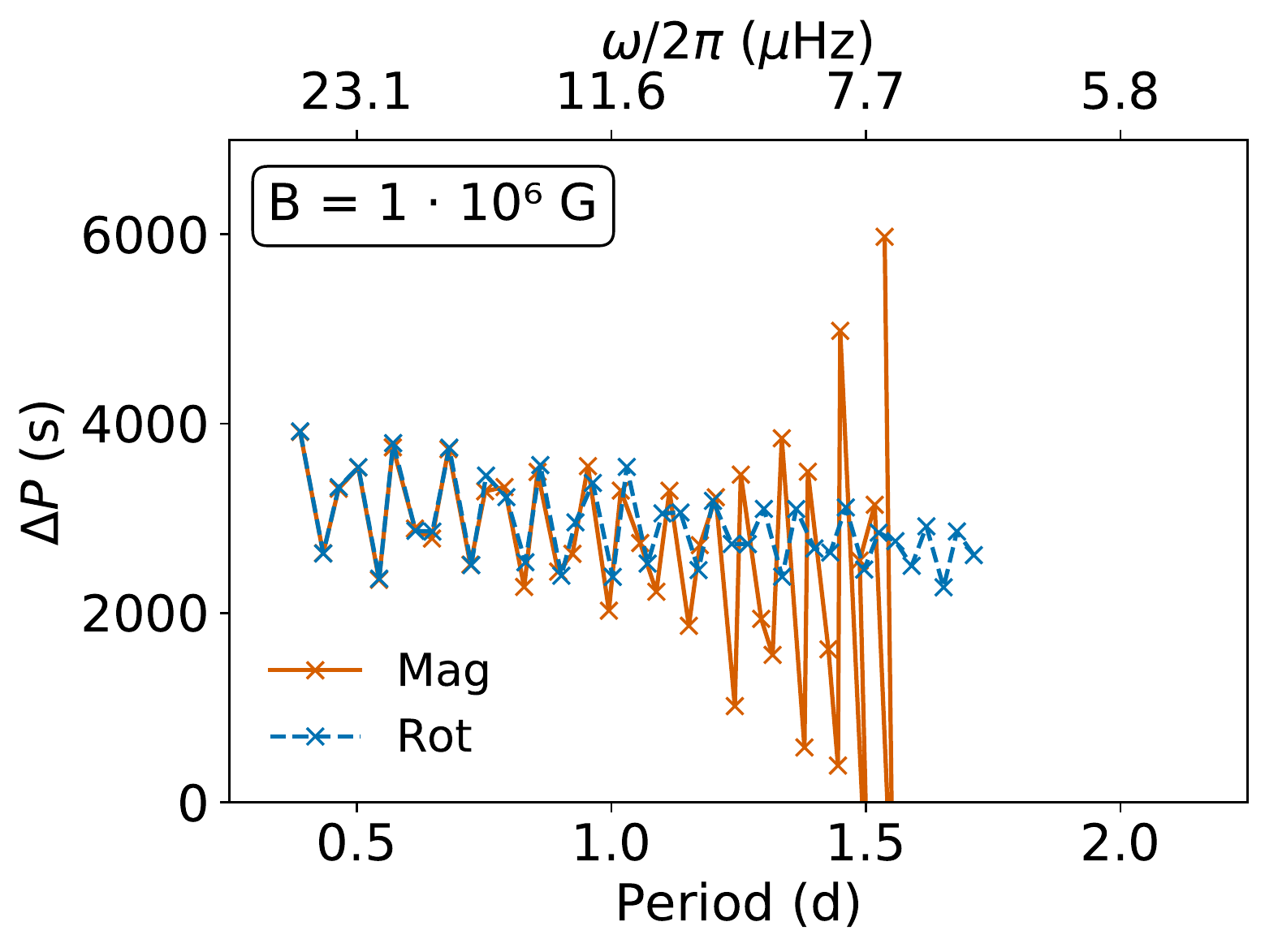}
\caption{ Period spacing patterns of zonal dipole modes of the $3$-M$_\sun$ TAMS reference model, varying $B_0$. The color scheme is the same as in the top right panel of Fig. \ref{fig:midMS_evolution}.}
\label{fig:B0_differences}
\end{figure*}

We discuss how the parameter $B_{\rm 0}$ and the rotation ratio affect the magnetic shifts in the same section, because these parameters were varied in P+19, although they considered a smaller parameter range.
Figure \ref{fig:B0_differences} shows the effect of variation in $B_0$ on zonal modes for the reference model. 

Both sectoral and zonal mode magnetic shifts are heavily affected by $B_0$. 
Whereas the deviations due to a $10^{\,6}$ G magnetic field near the core would be readily detectable at even the largest frequency uncertainties determined by \citet{2019_Gang_Li} (they are smaller than $\Delta\omega_{\rm per}/2\pi$), the frequency shifts due to near-core fields of strengths of approximately $10^{\,4}$ and $10^{\,5}$ G are at least an order of magnitude smaller than the 4-yr \textit{Kepler} frequency resolution. The derived surface magnetic field strengths according to the $p=2/3$ simulations of \citet{2008_Braithwaite} are in the observed range of \citet{2019_Schultz} only if $B_{\rm 0}$ in our models is $10^{\,5}$ G. However, if the \citet{2010_Duez_Braithwaite_field} magnetic field model is more similar to the magnetic field configuration obtained from the $p=1$ simulations of \citet{2008_Braithwaite}, no observable surface magnetic field strength is expected.
The derived surface magnetic field strengths, the magnetic $g$-mode frequency shifts, and the difficulties in detecting the amount of $g$ modes necessary to derive period spacing patterns of these magnetic pulsating stars, in which these magnetic features can be observed, explain why such internal magnetic fields have not yet been detected.

Stellar rotation makes the detection of magnetic fields more complicated: magnetic shifts of zonal modes are largest if $\mathcal{R}_{\rm rot}=0.25 - 0.50$, with the magnetic shift being slightly larger in the latter case, contrary to what might be expected from the conclusions of P+19. Only for $\mathcal{R}_{\rm rot}=0.75$ do we find the typical magneto-rotational behavior observed in P+19: magnetic shifts become less strong for increasing rotation rate. 
However, the validity of the TAR assumption in this regime for these types of modes is questionable, and most zonal modes are detected in the rotational regime $\mathcal{R}_{\rm rot}=0.01 - 0.50$ \cite[e.g.][]{Aerts2019_ARAA}. 
Retrograde modes are typically detected in stars rotating at $\mathcal{R}_{\rm rot}=0.01 - 0.25$, whereas prograde modes are observable at all considered rotation ratios \cite[e.g.][]{Aerts2019_ARAA}, yet are most easily identified when $\mathcal{R}_{\rm rot}=0.25 - 0.75$. We therefore consider $\mathcal{R}_{\rm rot} = 0.25 - 0.75$ to be the relevant rotation ratio domain for detecting prograde modes. Within these $\mathcal{R}_{\rm rot}$ domains, retrograde mode shifts increase and prograde mode shifts decrease with increasing $\mathcal{R}_{\rm rot}$.  Information on the magnetic shifts over the full rotation ratio domain considered in the grid can be found in Table \ref{tab:rot_var_freq_table}, and Figs. \ref{fig:field_rotation_evolution_prograde} - \ref{fig:field_rotation_evolution_retrograde} represent the magneto-rotationally modified period spacing patterns when varying $\mathcal{R}_{\rm rot}$ within its observationally constrained domain.

As explained by P+19, the Brunt-V\"ais\"al\"a frequency usually is much larger than the Coriolis frequency, so the lower frequency bound for gravito-inertial waves can be approximated by $\omega_- \approx 2\ \Omega\ \cos\theta$ \cite[e.g.][]{Prat_2016}. The larger the difference between the pulsation mode frequency and the estimated Alfv\'en frequency, the less sensitive pulsation modes are to the (slower) magnetic fluctuations, and the smaller the magnetic signatures in the period spacing pattern.
Estimated Alfv\'en frequencies for $B_0 = 10^{\,4} - 10^{\,5}$ G are much lower than the lower frequency bound $\omega_-$\,, so that magnetic shifts are small. Increasing the rotation rate further increases $\omega_-$. As a result, the magnetic signatures in the period spacing pattern decrease with increasing rotation rate. 

This led P+19 to conclude that one should search for magnetic signatures in period spacing patterns of slow rotators. Their conclusion was based on the best-fitting stellar model of HD~43317, which has a $X_{\rm c} = 0.54$ and an initial mass $M_{\rm ini} = 5.8$~M$_\sun$ \citep{Buysschaert_2018_magnetic_forward_model}. Hence, this model is situated firmly outside our grid. According to our results, the conclusion of P+19 still holds for the lower-mass end of intermediate-mass stars. Moreover, with increasing rotation rate, the perturbative criterion, already satisfied for most zonal modes in near-TAMS models at low $\mathcal{R}_{\rm rot}$, is also found to be more readily satisfied in models not near the TAMS. The perturbative criterion is satisfied for all sectoral modes in all models considered here.
We thus extend the conclusion of P+19, and infer the strengths of (strong) internal magnetic fields of rapidly-rotating intermediate-mass stars using the P+19 formalism in a wider parameter domain than considered by P+19.

If these strong fossil fields exist in the set of MS stars for which a perturbative magnetic influence is expected, they would readily be detectable from \textit{Kepler} and TESS-CVZ photometric light curves, and are expected to be detectable from PLATO photometric light curves because their induced frequency shifts are greater than the frequency resolutions. Even for the $\gamma$ Dor stars with the largest frequency uncertainties, such as the $0.09$\,\textmu Hz as obtained by \citet{2019_Gang_Li} for 4-yr \textit{Kepler} photometry, they remain detectable. So far, however, those deviations have not been detected.

\begin{table}[t!]\centering
\caption{Maximal perturbative frequency deviations $\Delta \omega_{\,\rm per}/2 \pi$ and maximal frequency deviations $\Delta \omega/2 \pi$ for frequencies of dipole modes with radial orders $n \in [-50,\ldots,-10]$, computed with Eqs. (\ref{eq:max_omega_allowed}) and (\ref{eq:max_omega}), for parameter variations of the reference model.}
\label{tab:freq_deviation_table}
\begin{tabular}{@{}Sl Sl Sl Sl @{}}
\hline \hline 
Parameter & Value & $\Delta \omega_{\,\rm per} / 2 \pi$ (\textmu Hz) & $\Delta \omega / 2 \pi$ (\textmu Hz)\\ \hline
& $0.005$ & $0.77$ & $0.77$ \\
$X_{\rm c}$ &  $0.340$ & $0.08$\, \tablefootmark{a}  & $0.98$\, \tablefootmark{a} \\
& $0.675$ & $0.01$\, \tablefootmark{a} & $0.64$\, \tablefootmark{a} \\ \hline
& $1 \cdot 10^{\,4}$ & $7.7\cdot10^{\,-5}$ & $7.7\cdot10^{\,-5}$ \\
$B_0$ (G) & $1 \cdot 10^{\,5}$ & $7.7\cdot10^{\,-3}$ & $7.7\cdot10^{\,-3}$\\
& $1 \cdot 10^{\,6}$ & $0.77$ & $0.77$\\ \hline
& $0.25$ &$0.89$ & $0.89$ \\
$\mathcal{R}_{\rm rot}$& $0.50$ & $0.63$\, \tablefootmark{a} & $0.63$\, \tablefootmark{a}\\
& $0.75$ & $0.49$\, \tablefootmark{a}  & $0.49$\, \tablefootmark{a} \\ \hline
& $-1$ & $1.80$ & $1.80$ \\
$m$& $0$ & $0.77$ & $0.77$ \\
& $1$ & $0.89$ & $0.89$\\ \hline
&  $0.004$ & $0.52$ & $0.84$  \\
$f_{\rm ov}$ &  $0.014$ & $0.77$ & $0.77$ \\
&  $0.024$ & $0.68$ & $0.68$ \\ \hline
&  $0.1$ & $0.81$ & $0.81$ \\
$D_{\rm mix}$ (cm$^{\,2}$ s$^{\,-1}$) &  $1.0$ & $0.77$  & $0.77$ \\
&  $10.0$ & $0.70$ & $0.70$ \\ \hline
&  $1.5$ & $0.73$ & $0.73$ \\
$\alpha_{\rm MLT}$&  $1.8$ & $0.77$ & $0.77$ \\
&  $2.0$ & $0.75$ & $0.75$ \\ \hline
&  $0.010$ & $0.83$ & $0.83$ \\
$Z_{\rm ini}$ & $0.014$ &$0.77$ & $0.77$   \\
&  $0.018$ & $0.70$ &  $0.70$\\ \hline
&  $1.3$ & $0.29$ & $0.29$ \\
$M_{\rm ini}$ (M$_\sun$) &  $2.0$ & $0.43$ & $0.43$  \\
&  $3.0$ & $0.77$ & $0.77$  \\ \hline
\end{tabular}
\tablefoot{All quantities, except for $B_{0}$, $M_{\rm ini}$ and $D_{\rm mix}$, are unitless. The influence is assessed for zonal modes ($m=0$), except when varying $m$ ($m\in [-1,0,1]$) or $\mathcal{R}_{\rm rot}$ ($m=1$).\\
\tablefoottext{a}{Not all computed spin parameters were $< 1$.}}
\end{table}

\subsection{ Non-radial mode geometry: $m$}

\begin{figure*}
\includegraphics[width=6cm,height=4.5cm]{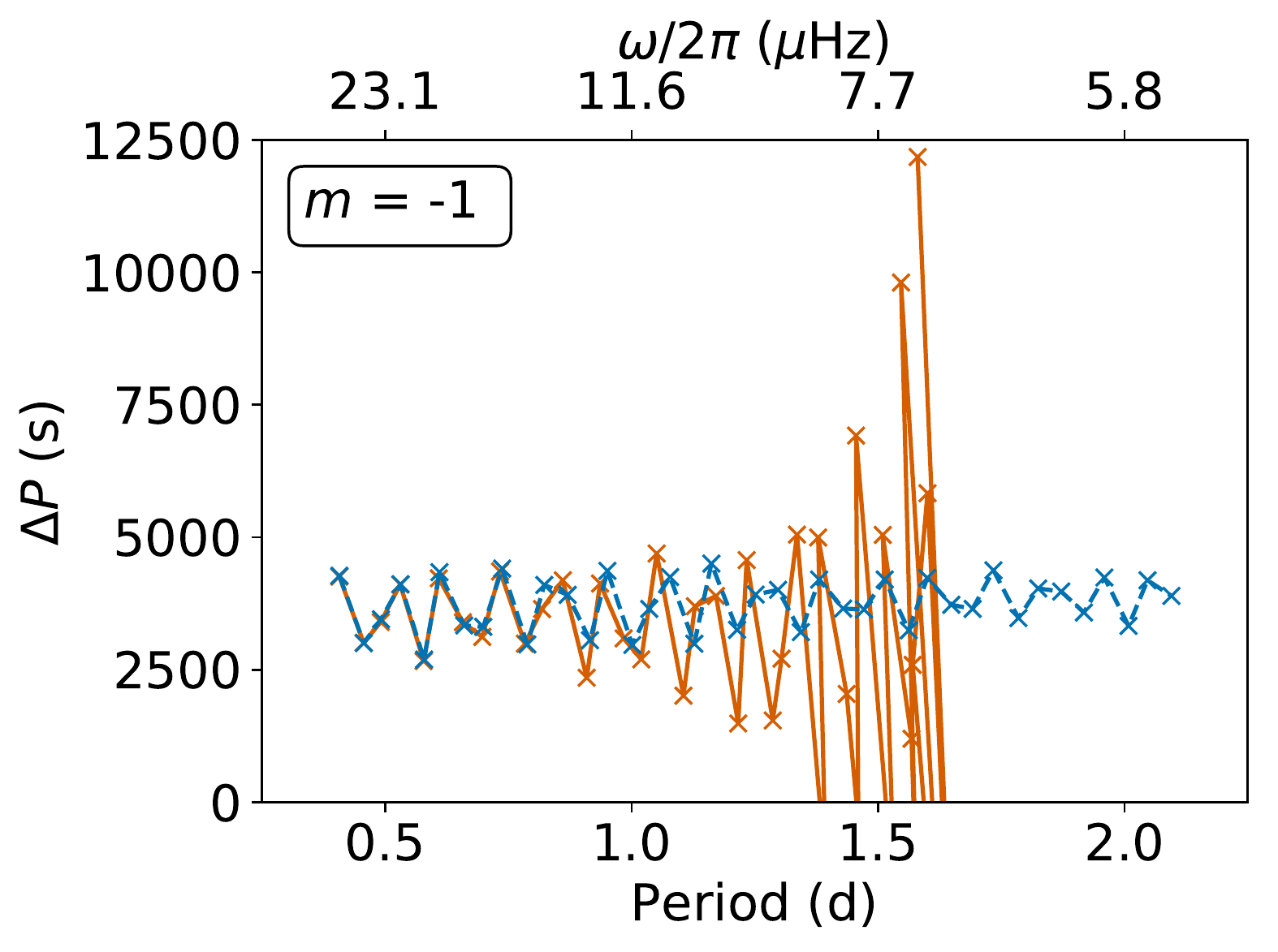}\includegraphics[width=6cm,height=4.5cm]{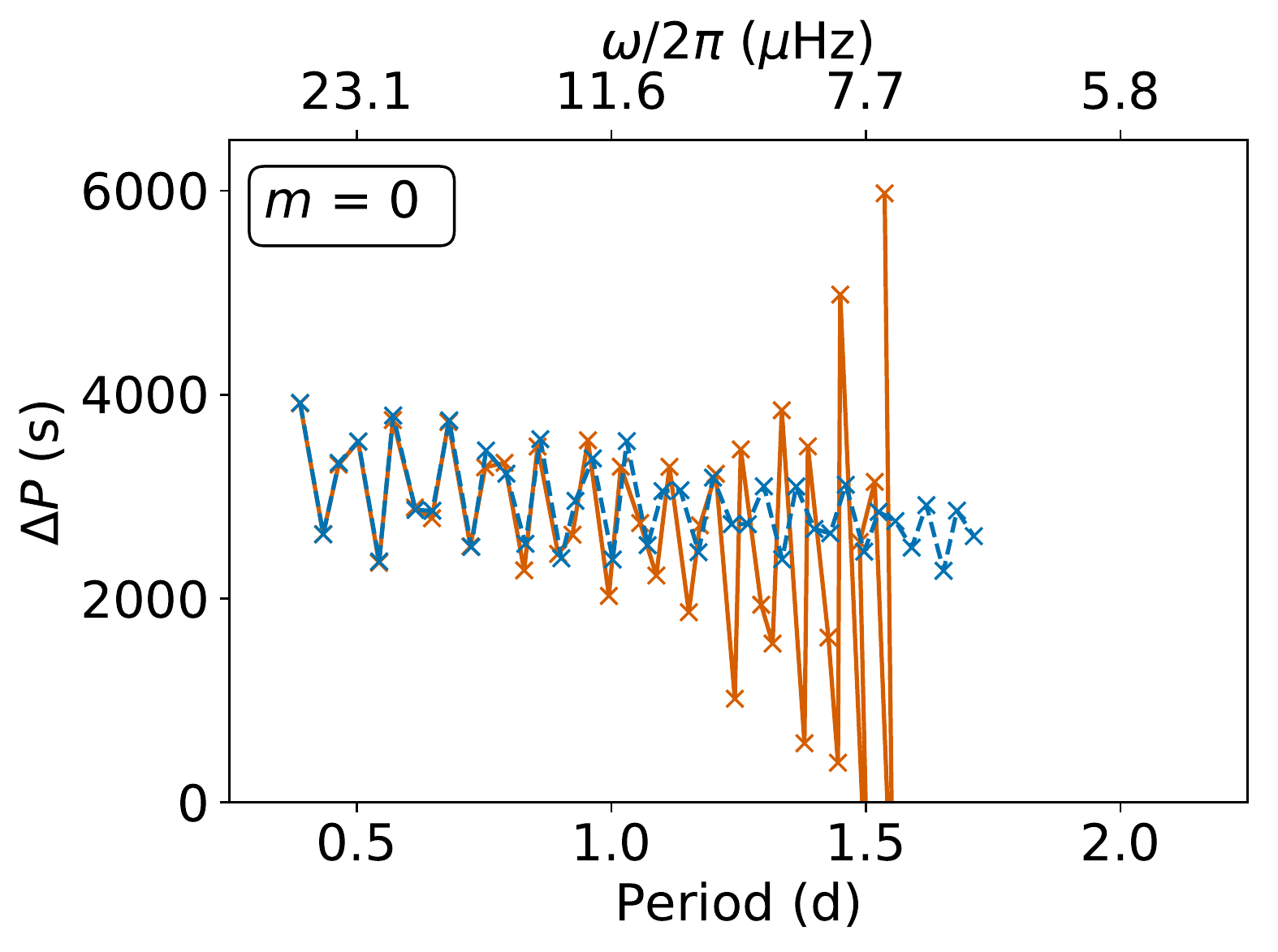}\includegraphics[width=6cm,height=4.5cm]{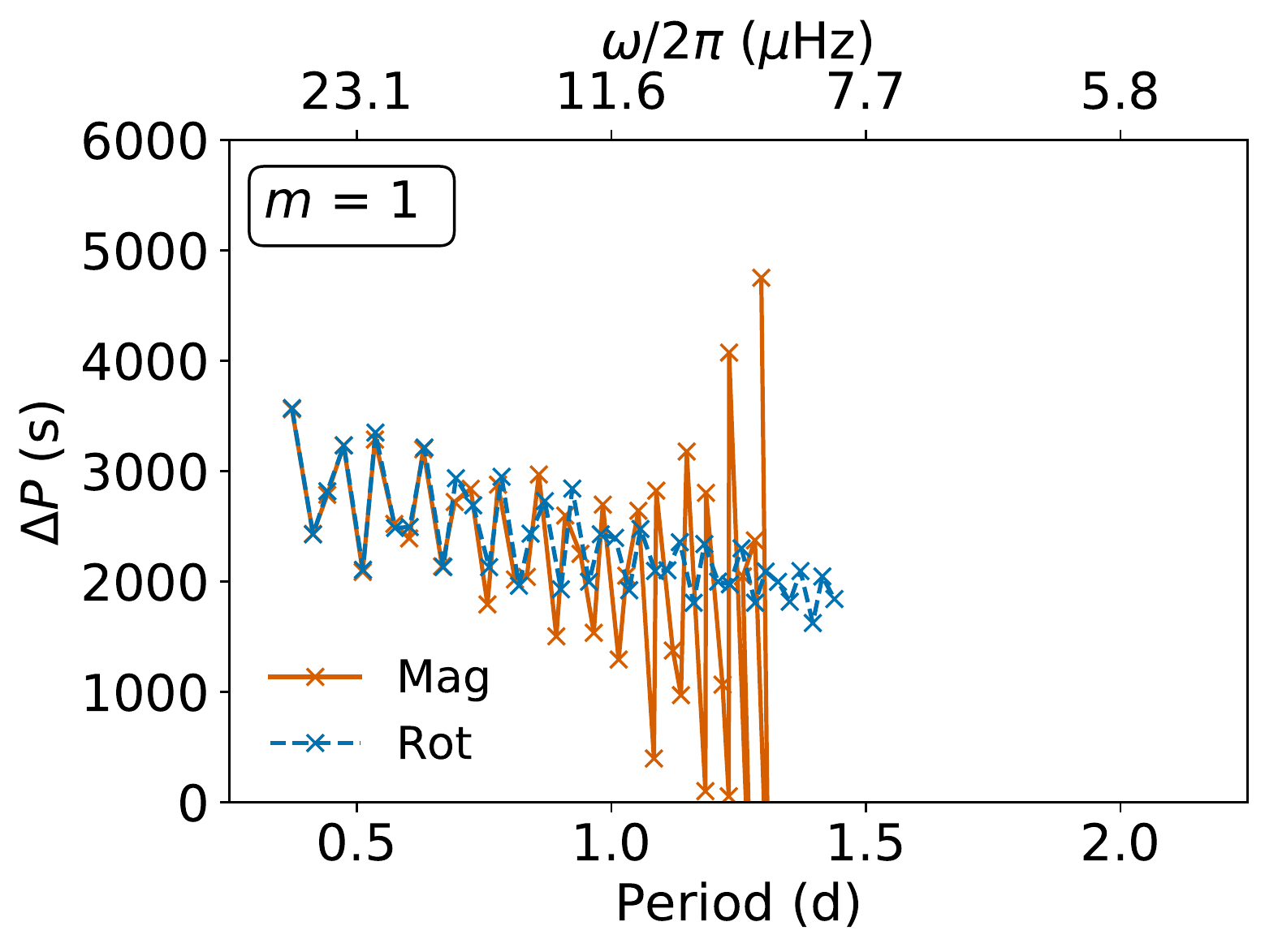}
\caption{ Same as Fig. \ref{fig:B0_differences}, but varying the non-radial mode geometry.}
\label{fig:mode_id_differences}
\end{figure*}

Period spacing patterns of modes of differing non-radial geometry are affected by rotation in a dissimilar way \cite[e.g.][]{2013_Bouabid,2015_Van_Reeth_BIS,2015_Van_Reeth,2017_papics}.
In this section, we investigate how mode geometry affects the magneto-rotationally modified period spacing patterns of our reference model, which is shown in Fig. \ref{fig:mode_id_differences}. 

In the zonal and prograde magneto-rotationally modified period spacing patterns, the sawtooth-like feature appears for modes of radial order $n \la -40$. For retrograde $g$ modes, this feature already appears for radial orders $n \la -35$. Trapped higher-radial-order modes display negative period spacings. The magnetic shift value is significantly changed for differing non-radial geometry, as can be observed from the $\Delta P$ values on the y-axes of the period spacing patterns shown in Fig. \ref{fig:mode_id_differences}.
From the maximal perturbative frequency deviations listed in Table \ref{tab:freq_deviation_table}, we learn that sectoral modes undergo larger shifts for near-TAMS models at the rotation ratio $\mathcal{R}_{\rm rot} = 0.25$. In fact, of all the dipolar non-radial modes considered, the retrograde modes undergo the largest shifts for $\mathcal{R}_{\rm rot} = 0.01 - 0.25$, whereas zonal mode shifts are largest for $\mathcal{R}_{\rm rot} = 0.50 - 0.75$.

\subsection{Mixing: $f_{\rm ov}$, $D_{\rm mix}$ and $\alpha_{\rm MLT}$}\label{sec:4.1.5}
We assess the influence of $D_{\rm mix}$, $\alpha_{\rm MLT}$ and $f_{\rm ov}$ by varying one parameter and fixing the others in the reference model. Extra mixing changes the evolution of a star on a long-term scale, because it affects local composition and the energy transport. This impacts the local density, thus affecting the magnetic field model \cite[through equation (10) of][]{Duez_field_2010} and the pulsation modes. All three parameters ($f_{\rm ov}$, $D_{\rm mix}$, $\alpha_{\rm MLT}$) affect mixing inside the star to a certain degree, and are therefore discussed together.

\begin{figure*}
\includegraphics[width=8.5cm]{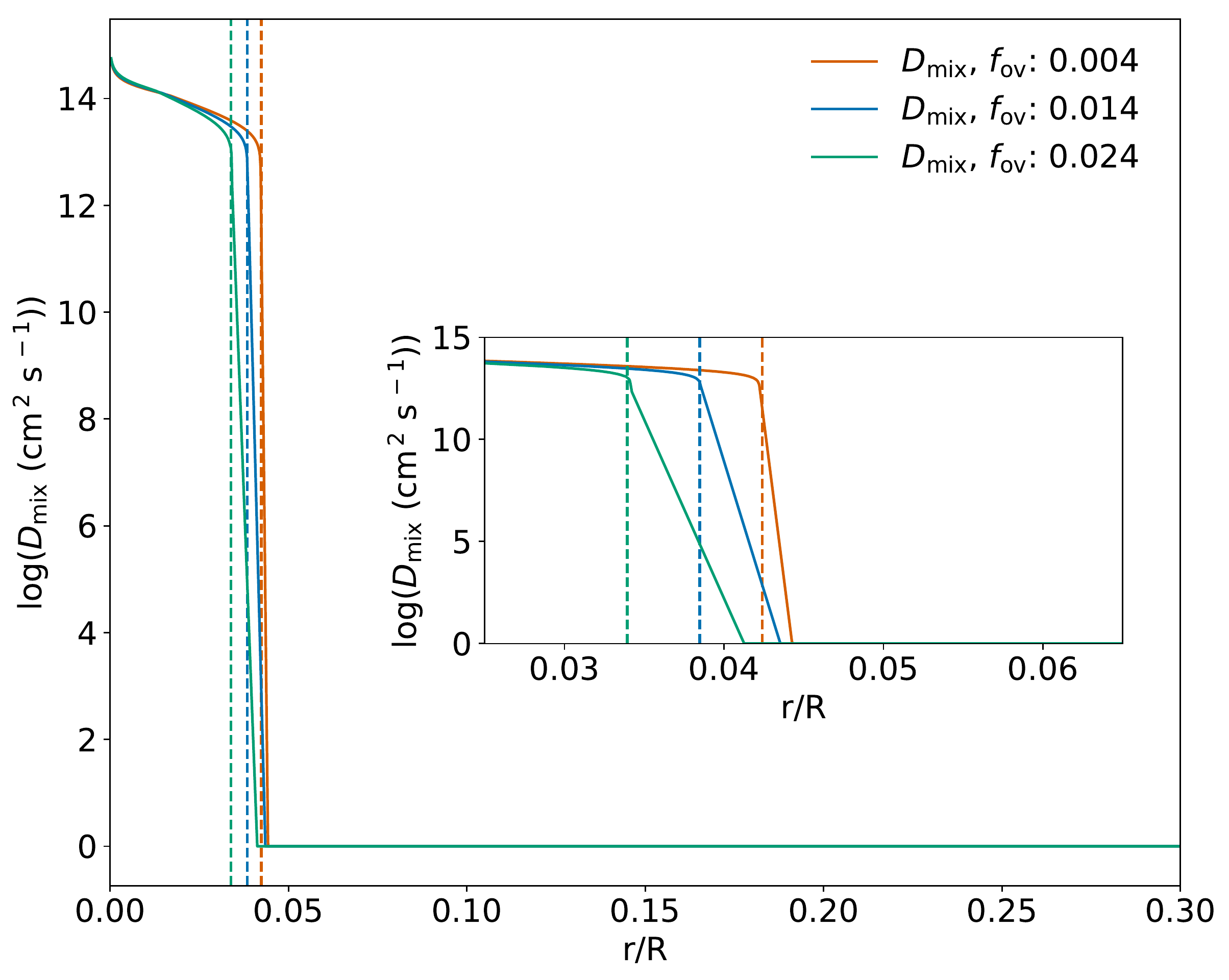}\hfill\includegraphics[width=8.5cm]{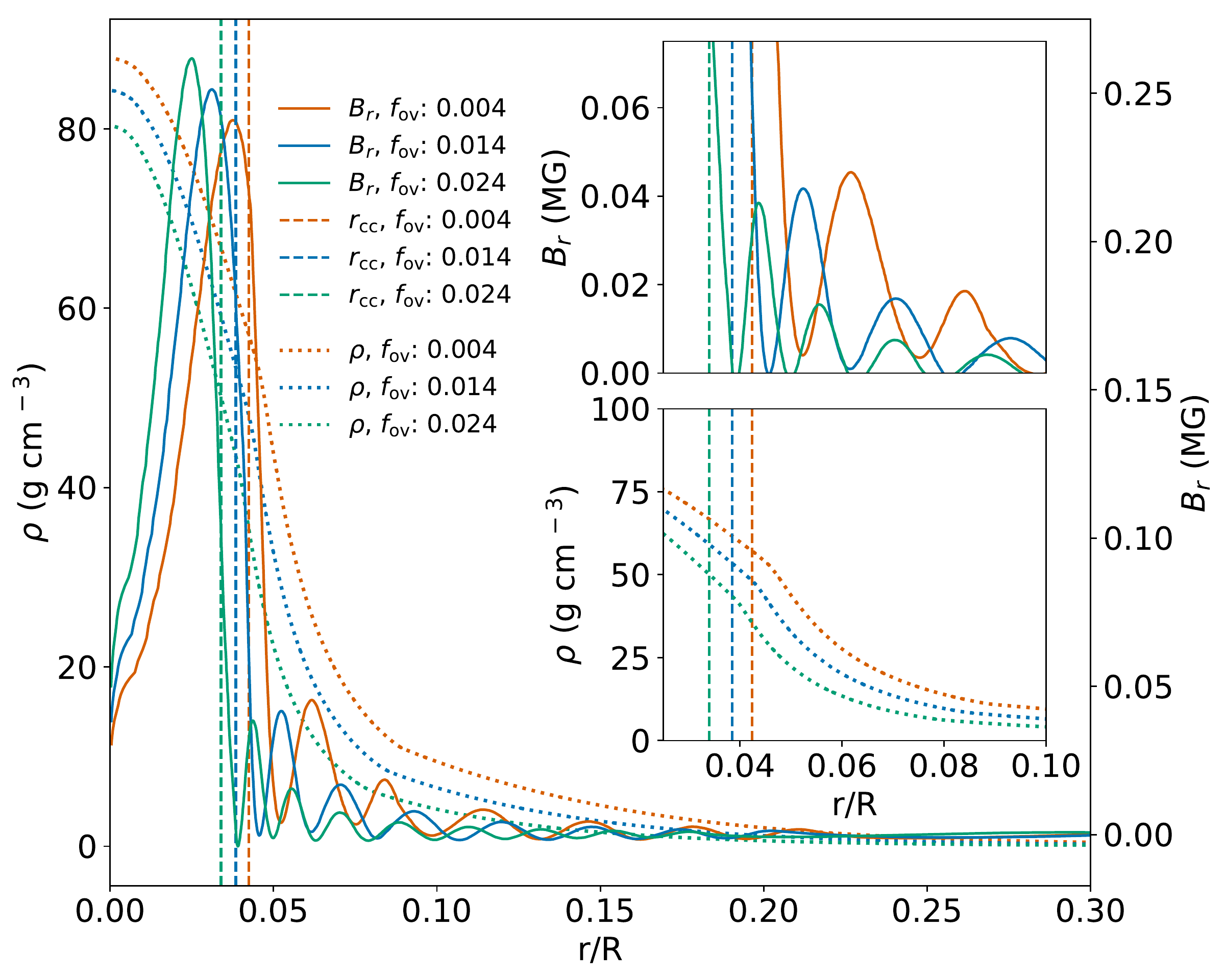}\\
\includegraphics[width=6cm,height=4.5cm]{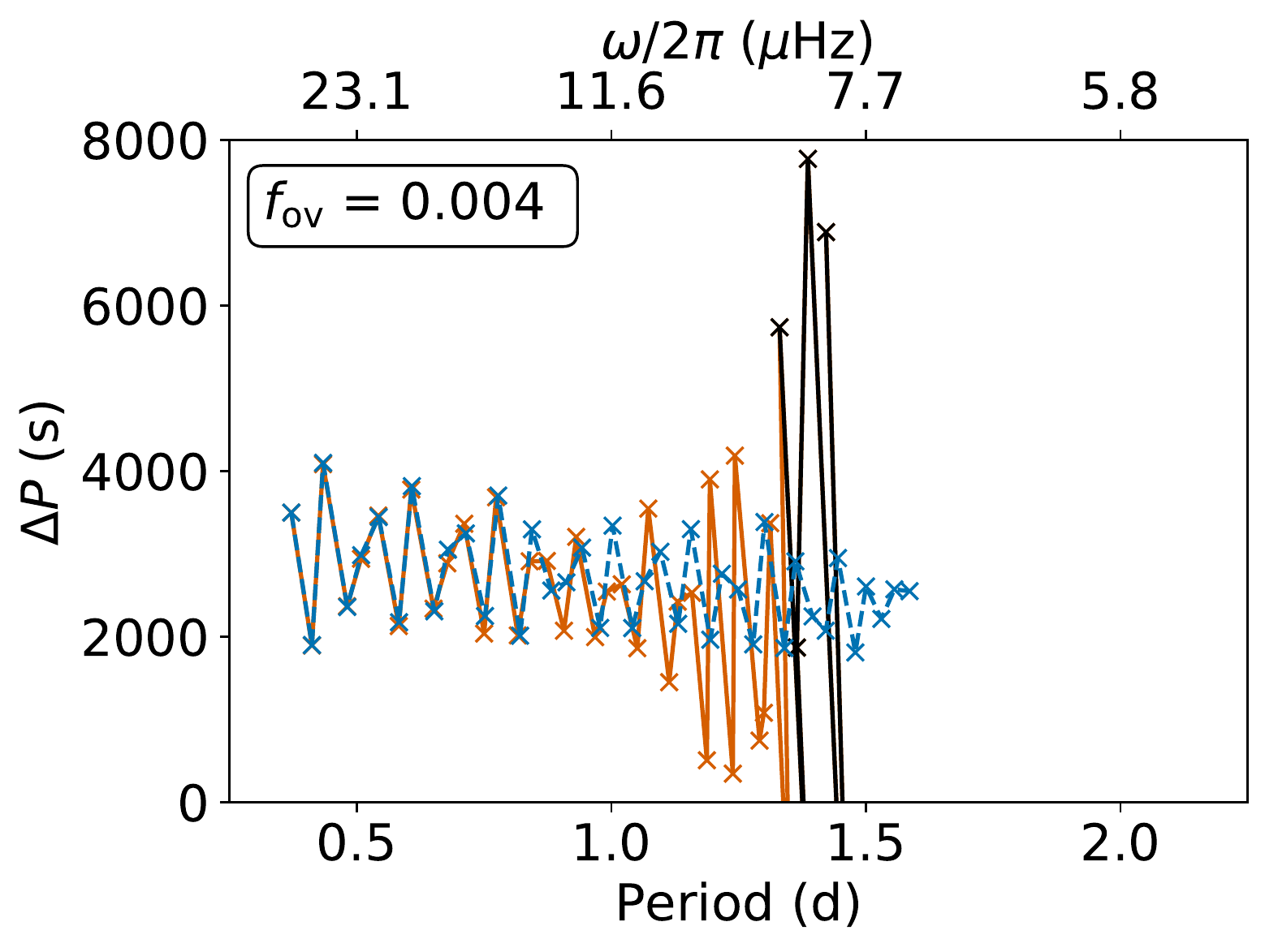}\includegraphics[width=6cm,height=4.5cm]{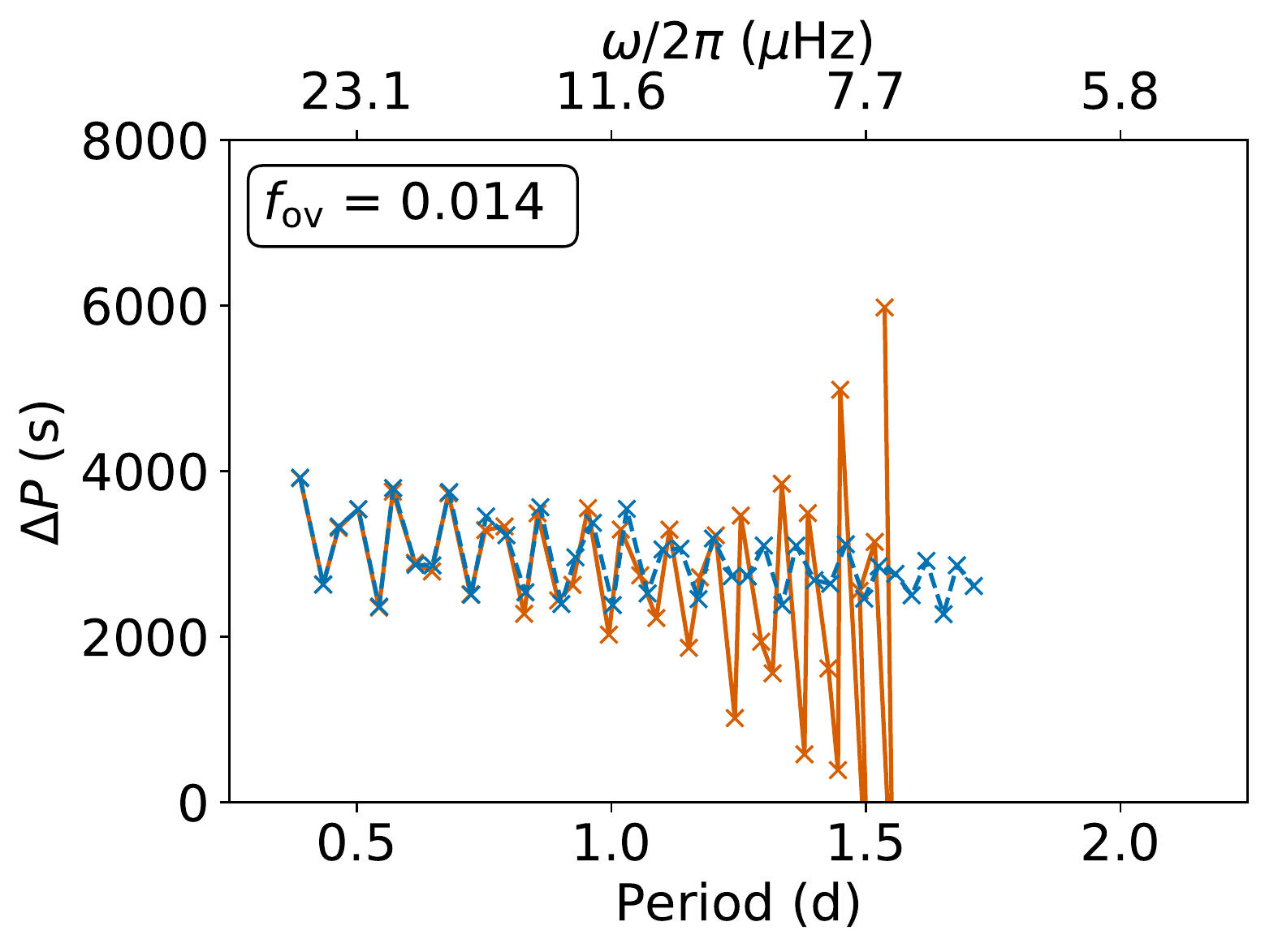}\includegraphics[width=6cm,height=4.5cm]{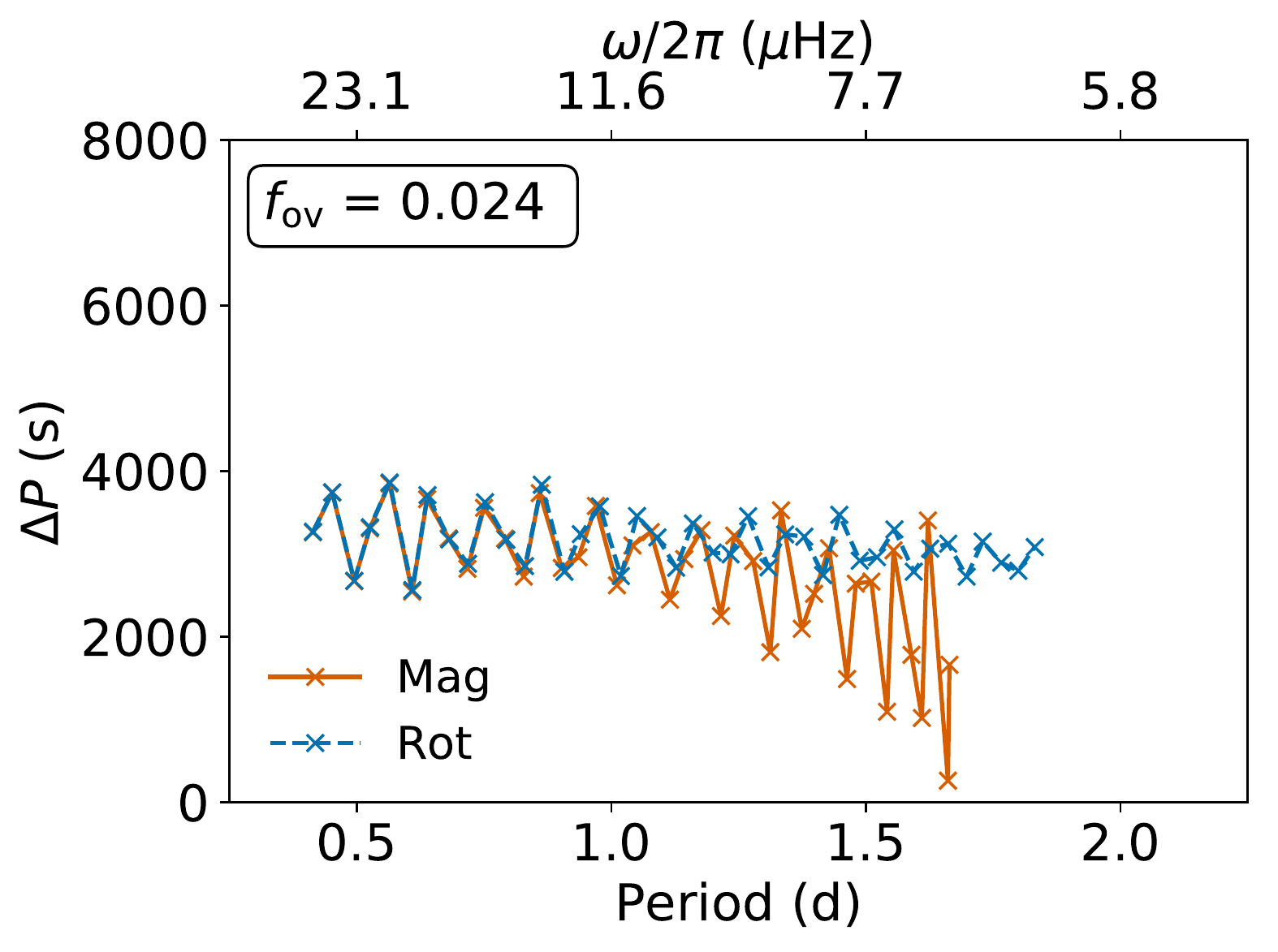}
\caption{\textit{Top left:} Mixing coefficient profiles $D_{\rm mix}$ (full lines) as a function of normalized radius when varying the overshoot parameter $f_{\rm ov}$ in the $3$-M$_\sun$ reference model. Dashed vertical lines indicate convective core boundaries $r_{\rm cc}$, obtained from the Schwarzschild criterion \cite[e.g.][]{SSE_KippenHahn_book}. \textit{Top right:} density $\rho$ (dotted lines) and radial magnetic field component $B_r$ (at $\theta = 5^{\circ}$; full line) profile for the same stellar model. Dashed vertical lines indicate the same as on the left. \textit{Bottom row:} same as Fig. \ref{fig:B0_differences}, but varying $f_{\rm ov}$.}
\label{fig:fov_influence}
\end{figure*}

Constraining convective core overshooting involves strong model degeneracies \cite[e.g.][]{Pedersen_2018_overshooting,Aerts_2018_forward_modelling}, because overshooting entrains hydrogen into the convective core, so the MS lifetime is significantly affected by the overshoot parameter $f_{\rm ov}$. A change in MS lifetime necessarily means that, for a given $X_{\rm c}$, the radius, $T_{\rm eff}$ and the density profile are affected.
The panels in the top row of Fig. \ref{fig:fov_influence} show mixing, density and radial magnetic field component profiles extracted at an arbitrarily chosen colatitude $\theta = 5 \degr$ for near-TAMS models.
The convective core size, obtained from the Schwarzschild criterion \cite[e.g.][]{SSE_KippenHahn_book}, is also shown in Fig. \ref{fig:fov_influence}. Given the changes in radial magnetic field component in the stellar radiative zone, it is reasonable to anticipate strong changes in magnetic shifts when comparing the models while varying $f_{\rm ov}$. Mixing in the core overshoot zone also dissipates chemical gradients that were left behind by the receding convective core during main sequence evolution, affecting the pulsation modes strongly.

The period spacing patterns shown on the panels in the bottom row of Fig. \ref{fig:fov_influence} display a trend: the smaller the amount of overshooting the larger the magnetic frequency deviations near the TAMS. This is reflected by the values of $\Delta\omega/2\pi$ in Table \ref{tab:freq_deviation_table}, which mainly assess the magnetic shifts of the trapped modes. The highest radial order modes even undergo non-perturbative magnetic shifts in the lowest $f_{\rm ov}$ model. This trend of increasing magnetic shift with decreasing $f_{\rm ov}$ is expected, because magnetic shifts of trapped modes are larger than those of untrapped modes. Increasing $f_{\rm ov}$ results in a larger overshoot region (because of increasing pressure scale height and the increased value of $f_{\rm ov}$), changes the convective core mass and size, smooths chemical gradients in the near-core region, and decreases mode trapping. The mixing level varies in the overshoot region, and is denoted by $D_{\rm ov} (r)$. This parameter, evaluated near the overshoot region boundary, together with the mixing within the radiative zone, $D_{\rm mix}$, are most important for mode trapping. The former is affected by all previously mentioned parameters that are influenced by $f_{\rm ov}$, whereas the latter is constant in our models. Magnetic shifts are also affected by the change in magnetic field structure induced by the dissimilar near-core density profile. The effect of increasing $f_{\rm ov}$ is two-fold: although the magnetic shifts of the modes that are trapped in models with lower $f_{\rm ov}$ are smaller in models with higher $f_{\rm ov}$, we gain an additional diagnostic in the form of a slope deviation from the rotationally modified period spacing pattern that is easily recognizable and increases with radial order (see the right-most panel of the bottom row of Fig. \ref{fig:fov_influence}). We show the analog of Fig. \ref{fig:fov_influence} for the $2$-M$_\sun$ reference model in Fig. \ref{fig:fov_influence_M2}, indicating that the effect of the magnetic field is similar. The magnetic slope deviations and sawtooth-like features are yet to be discovered in photometric time series but it seems promising that core overshooting could be better constrained in the presence of strong near-core magnetic fields. 

The dominant contributing term to the magnetic frequency shift is the one depicted in Eq. 17 of P+19, because the toroidal magnetic field strength in our models is smaller than the poloidal magnetic field strength. It is similar to the main contribution considered in \citet{2005_Hasan}. Physically, it describes the effect of a Lorentz force in the direction of wave propagation that is larger than the average local value, which accelerates the gravito-inertial wave in that same direction, increasing its frequency. The magnitude of this term depends on the absolute values of radial derivatives of the radial magnetic field strength and the horizontal wave displacement $\xi_{\rm h}$. The absolute values of derivatives of $\xi_{\rm h}$ increase with radial order, which results in larger, positive frequency shifts for high-radial-order modes. Hence, the perturbative mode frequency differences ($\Delta\omega_{\rm per, ni}/2\pi$) increase with radial order, leading to a steeper, downward slope in the period spacing diagram.

\begin{figure}
\includegraphics[width=88mm,height=6.25cm]{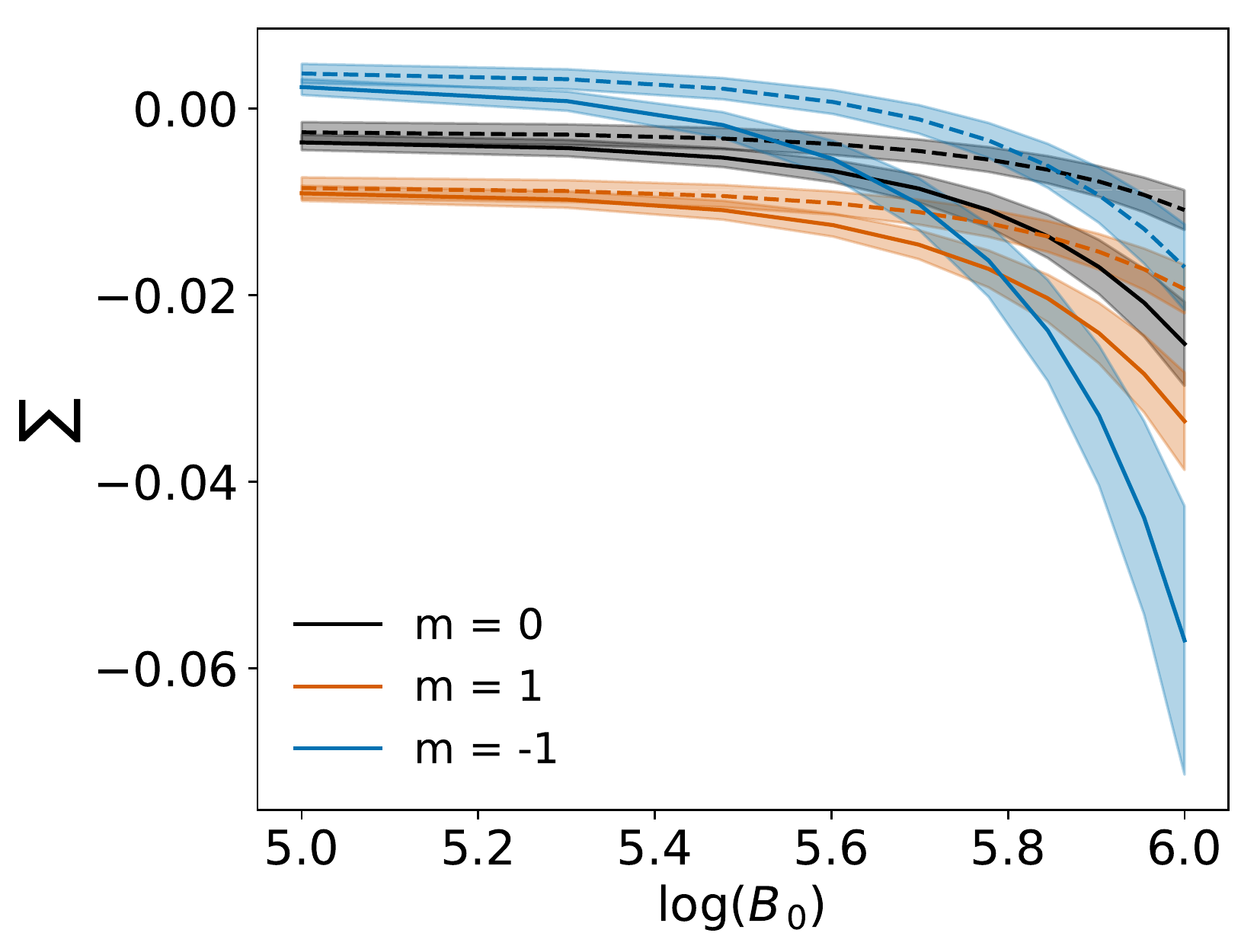}
\caption{Inferred magneto-rotationally modified period spacing pattern slopes $\sum$ as a function of $B_0$ (in G) for different non-radial mode geometry of dipole modes propagating inside the $f_{\rm ov}~=~0.024$ model. Dashed lines and solid lines indicate the slope estimated around the $n=-30$ and $n=-40$ mode, respectively. The shaded bands indicate $95\%$ confidence intervals.}
\label{fig:slopes_fov}
\end{figure}

By computing the moving averages of the magneto-rotationally modified mode frequencies we can infer the general trend of the period spacing pattern.
Therefore, similar to \citet{Ouazzani_2017},
we infer the ordinary least-squares slope $\bm{\sum}$ of the magneto-rotationally modified period spacing pattern of the $f_{\rm ov}=0.024$ stellar model for the $n=-30$ and $n=-40$ modes for values of $B_{\rm 0}$ ranging from $10^{\,5}$ to $10^{\,6}$ G. The slopes are inferred from the moving averages for radial orders $[n_{\rm s}-10,\ldots,n_{\rm s},\ldots,n_{\rm s}+10]$, where $n_{\rm s}$ are $30$ and $40$, respectively. The inferred slopes are found to be predominantly negative, as shown in Fig. \ref{fig:slopes_fov}, due to the large magnetic shifts induced by the magnetic field, and increase with radial order. Magnetic slope deviations become apparent for a value of $B_{\rm 0}$ of $7 \cdot 10^{\,5}$ G for zonal modes, $8\cdot 10^{\,5}$ G for prograde modes, and $4\cdot 10^{\,5}$ G for retrograde modes, with the latter being affected the most.
Slopes can be inferred with higher precision for models with larger $f_{\rm ov}$ values. Inferred slopes are smaller for lower-mass models, as can be seen in Fig. \ref{fig:slopes_fov_M2}. The values of $B_{\rm 0}$ correspond to near-core magnetic field strengths of approximately $0.269\cdot 10^{\,6}$, $0.307\cdot 10^{\,6}$, and $0.154\cdot 10^{\,6}$ G, with derived surface magnetic field strength ranges of $9.54-15.1$, $10.9-17.3$, and $5.46-8.66$ kG. These are larger than the typical surface magnetic field strengths observed by \citet{2019_Schultz}.

The effect of varying $D_{\rm mix}$ is similar to that of varying $f_{\rm ov}$: higher levels of envelope mixing $D_{\rm mix}$ lead to less pronounced sawtooth-like features and a magnetic slope deviation (as can be observed on the panels in the bottom row of Fig. \ref{fig:fov_influence}). Changes in the density profile and radial magnetic field component profiles are minimal. 
In this case, the changes in period spacings occur mainly because of changes in mode trapping due to increased or decreased mixing near the core overshoot boundaries. Varying $D_{\rm mix}$ is less effective at inducing changes in the period spacing patterns of $3$-M$_\sun$ models, as is quantified by the maximal frequency deviations listed in Table \ref{tab:freq_deviation_table}. For lower-mass models, as shown in Fig. \ref{fig:fov_influence_M2} for the $2$-M$_\sun$ reference model, $D_{\rm mix}$ is more effective than $f_{\rm ov}$ at inducing (clear) magnetic slope deviations in period spacing patterns.

The $\alpha_{\rm MLT}$ parameter changes the size of the convection zones, affecting the $g$-mode propagation in the stellar interior. In our case it primarily affects the pressure scale height at the core boundary, and the size and mass of the convective core. Hence, it changes the size of the overshoot region and slope of the $D_{\rm ov}(r)$ profile. Moreover, because we assume convection zones to be completely mixed, this also affects the density profile, and thus the magnetic field model. 
The magneto-rotationally modified period spacing patterns change when varying $\alpha_{\rm MLT}$, but no features additional to the sawtooth-like feature and slope change catch the eye. Only for the $\alpha_{\rm MLT} = 1.5$ model do we notice a clear slope change at higher radial orders, compared to the rotationally modified period spacing pattern, which is similar to that observed when varying $D_{\rm mix}$ and $f_{\rm ov}$. This is either related to a change in near-core mixing level or to a changing radial magnetic field component.

\begin{figure*}
\includegraphics[width=8.5cm]{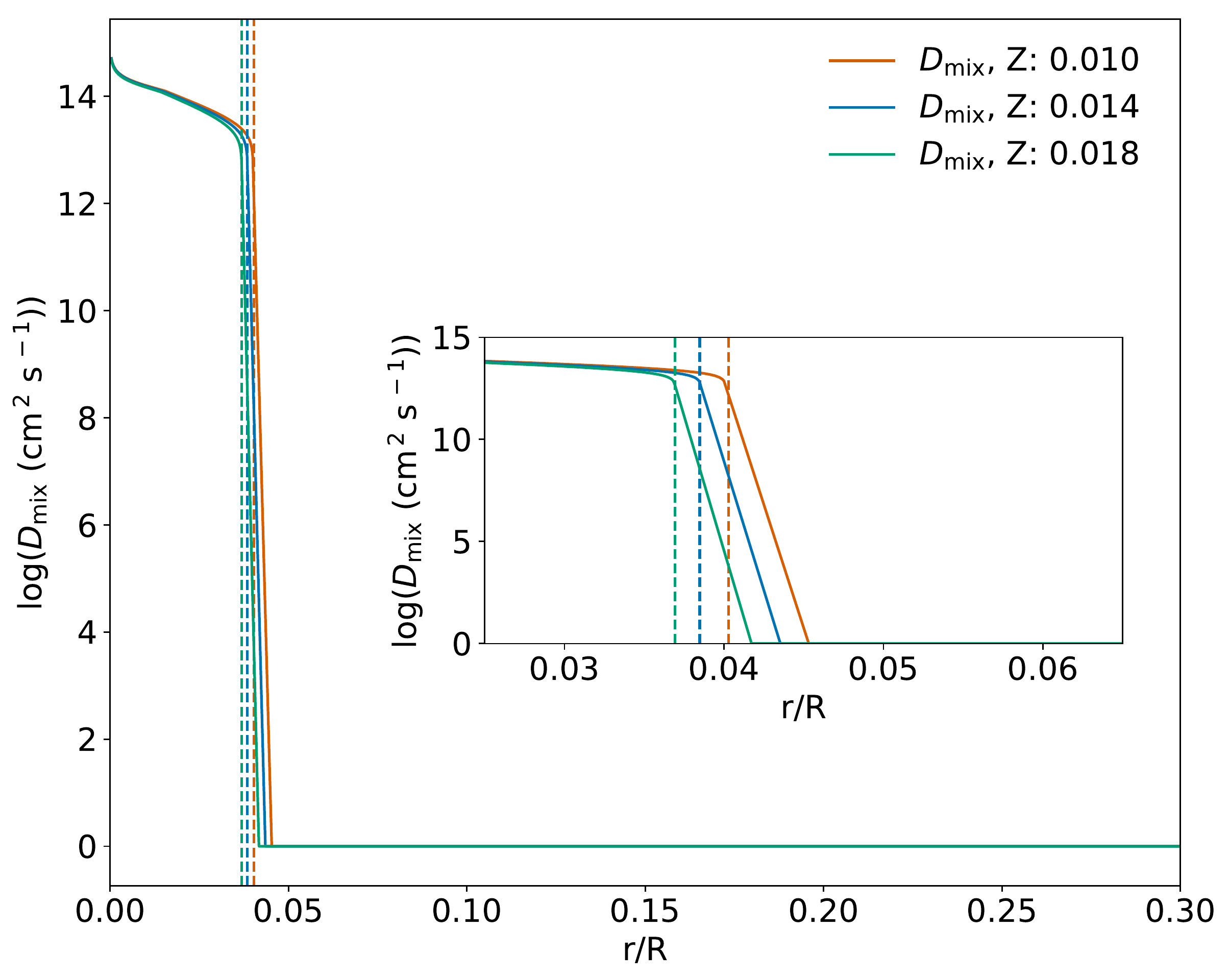}\hfill\includegraphics[width=8.5cm]{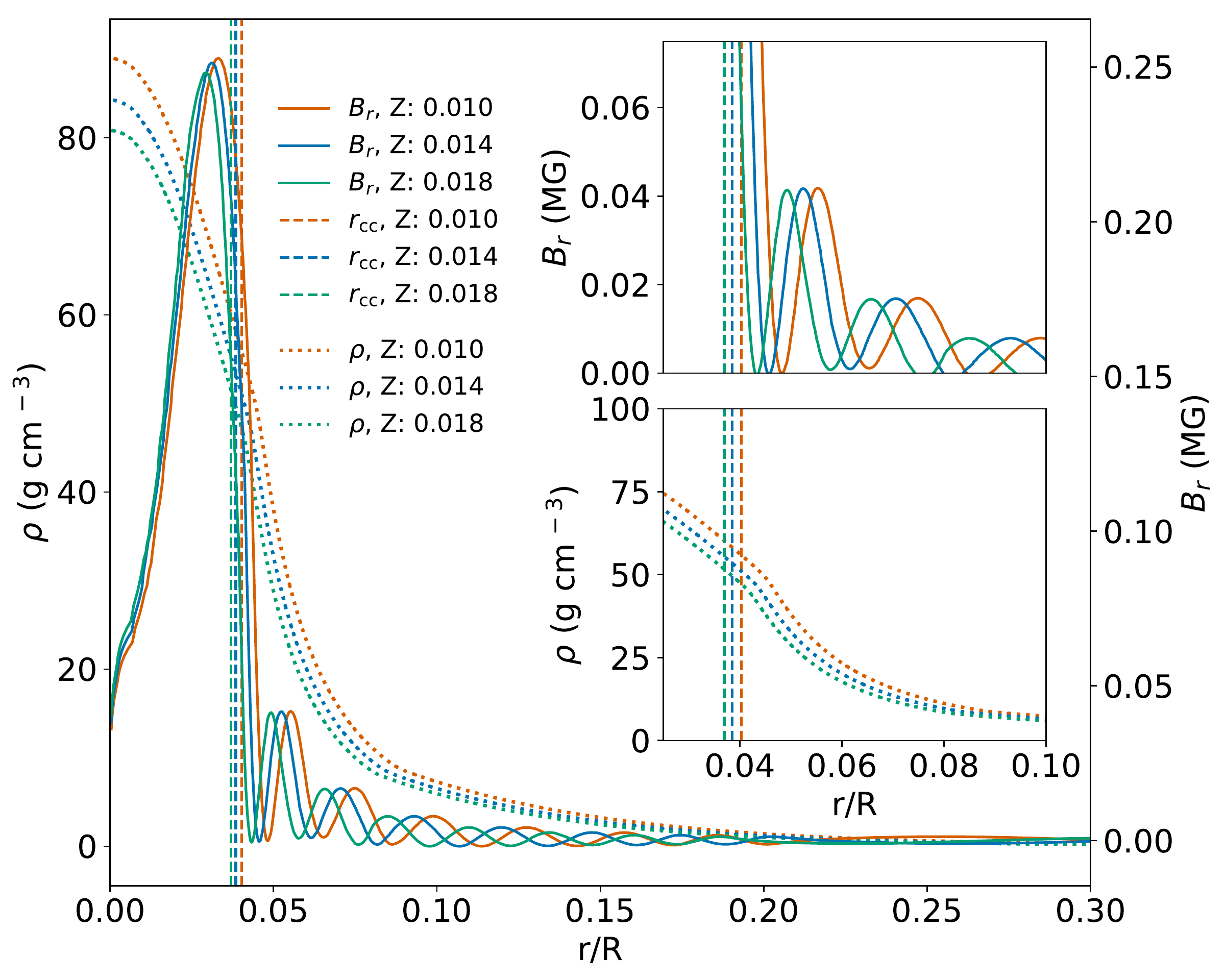}\\
\includegraphics[width=6cm,height=4.5cm]{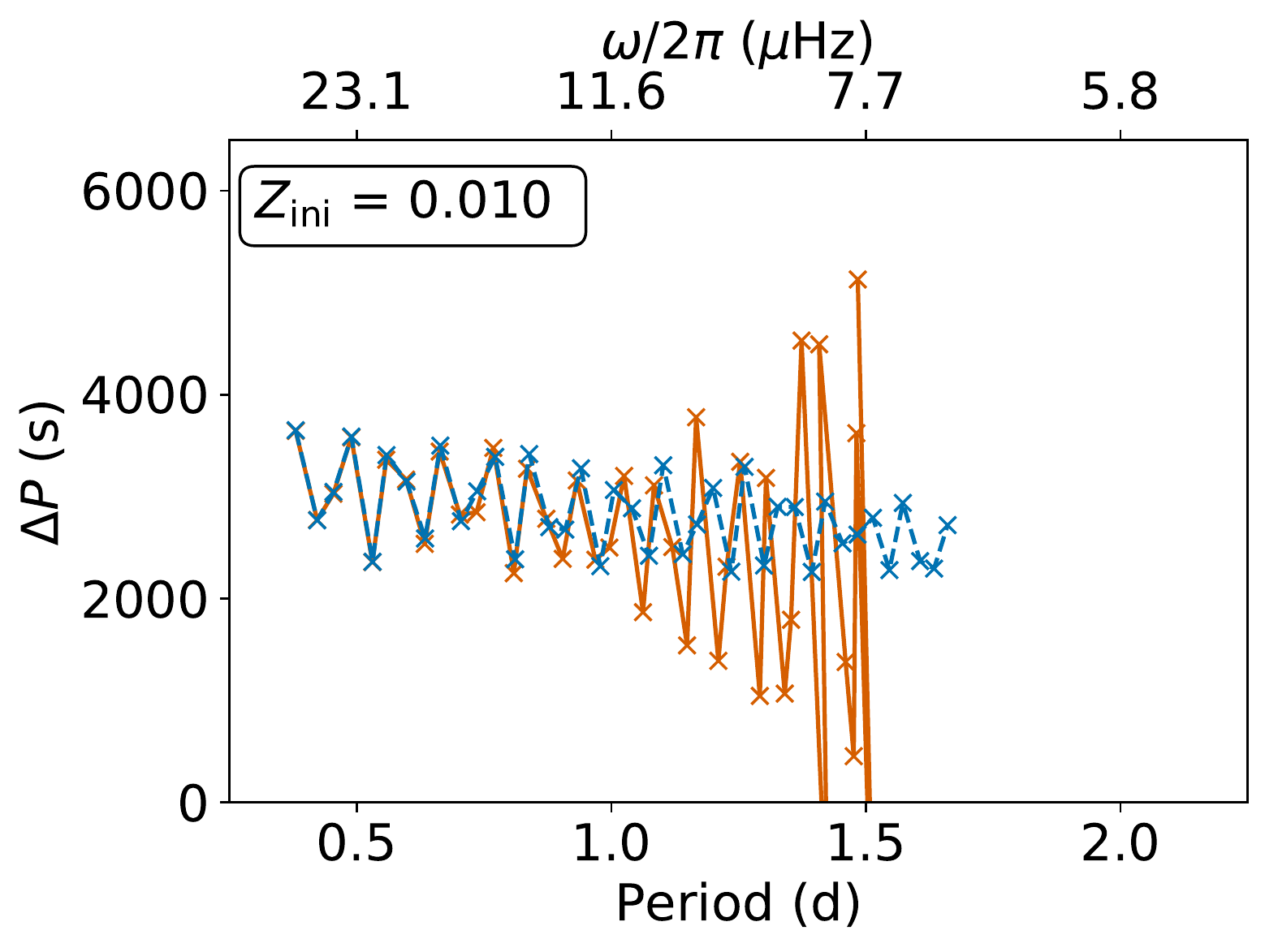}\includegraphics[width=6cm,height=4.5cm]{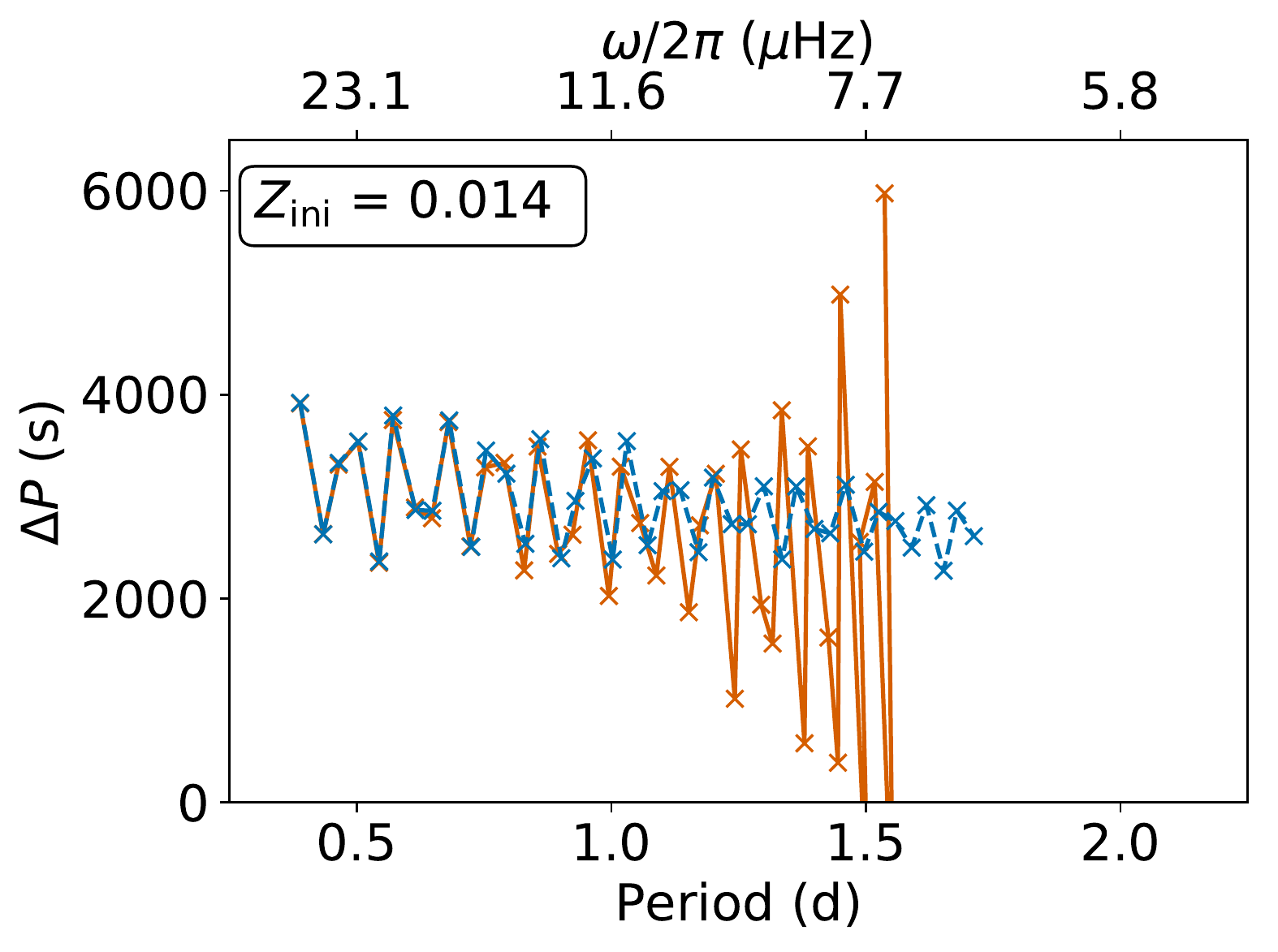}\includegraphics[width=6cm,height=4.5cm]{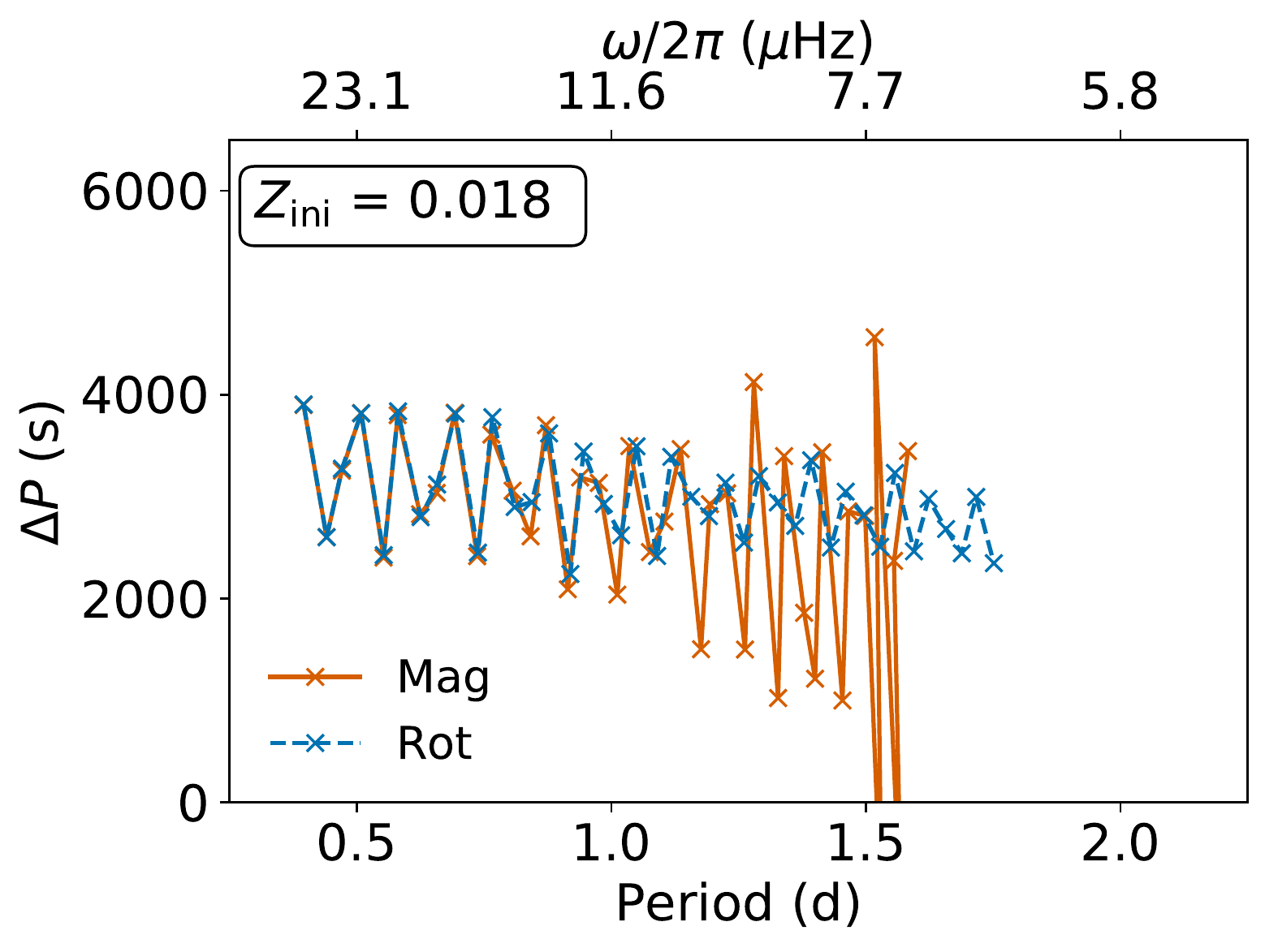}
\caption{Same as Fig. \ref{fig:fov_influence}, but varying $Z_{\rm ini}$ (indicated with $Z$ in the figures on top).}
\label{fig:Z_influence}
\end{figure*}

If we decrease $\alpha_{\rm MLT}$ and keep all other parameters constant, the pressure scale height at the core boundary increases. This leads to a less steep decay of the mixing level in the near-core overshooting region, as can be seen in the top left panel of Fig. \ref{fig:AMLT_influence}, changing the eigenmode cavities. A changing radial magnetic field component can be attributed to a differing density profile. A similar result was obtained for lower-mass models, as can be observed in Fig. \ref{fig:AMLT_influence_M2}.
The radial magnetic field varies in a stronger way in the near-core region for the $\alpha_{\rm MLT} = 1.5$ model than for the other models in our comparison. These dissimilar gradients of the radial magnetic field in the near-core regions are likely responsible for the changes in the period spacing patterns. Maximal frequency deviations from the variation in $\alpha_{\rm MLT}$ are however small compared to the frequency deviations induced by the variation in the other stellar fundamental parameters (see Table \ref{tab:freq_deviation_table}).

\subsection{Initial metallicity: $Z_{\rm ini}$} \label{sec:4.1.6}

Significant deviations in rotationally modified period spacing patterns occur for models of varying $Z_{\rm ini}$, because the mode cavity of the $g$ modes are affected by the differing core mass \cite[see e.g. figure~7 in][]{2015_Moravveji}. Here, we investigate how magneto-rotationally modified period spacing patterns are affected by the initial metallicity $Z_{\rm ini}$.

When we look at the magnetically modified period spacing patterns we observe no clear slope deviation for the reference model (see the panels in the bottom row of Fig. \ref{fig:Z_influence}). Maximal frequency deviations are larger for lower $Z_{\rm ini}$ models, as shown in Table \ref{tab:freq_deviation_table}, and are all one order of magnitude greater than the 4-yr \textit{Kepler}, 1-yr TESS, and the 2-yr PLATO frequency resolutions. 
The largest frequency uncertainty quoted by \citet{2019_Gang_Li} for \textit{Kepler} light curves is approximately $7$ times smaller than the magnetic frequency shift for the $B_{\rm 0}~=~10^6$ G reference model.
Magnetic field models slightly differ in their toroidal components but this does not induce changes in the magneto-rotationally modified period spacing patterns similar to the one(s) noticed when varying the fundamental mixing parameters for the reference model. This slope deviation is present in $2$-M$_\sun$ and $1.3$-M$_\sun$ models for higher radial orders, and can be partially linked to the mixing efficiency (see Sect. \ref{sec:4.6}). As an example, we show the period spacing pattern for the $2$-M$_\sun$ reference model in Appendix~\ref{sec:app_B}.

Most of the changes observed in magneto-rotationally modified period spacing patterns due to variation in $Z_{\rm ini}$ are in the high radial order regime. The change in magnetic shifts induced by varying metallicity does not seem to greatly affect the shape of the period spacing pattern for low radial order modes, but it does affect the magnetic shift values.

\subsection{Initial stellar mass: $M_{\rm ini}$}\label{sec:4.6}
We keep all parameters but $M_{\rm ini}$ constant to assess the influence of stellar mass. As noted in Table \ref{tab:freq_deviation_table}, the maximal frequency deviations steadily decrease with decreasing mass. A stellar model with a different initial mass $M_{\rm ini}$ attains an internal structure dissimilar to that of the reference model at the same $X_{\rm c}$. Hence, varying mass changes the density profile and therefore the magnetic field model, which in turn changes the magnetic shifts. Some similarities between the magneto-rotationally modified patterns can however be noticed: most near-TAMS magnetic shifts pass the consistency checks and most high-radial-order modes are affected, such that the slope deviates from the rotationally modified period spacing pattern. Some distinct changes in magneto-rotationally modified period spacing patterns are noticed as well: the sawtooth-like feature of the $3$-M$_\sun$ model is more pronounced than that of the $2$-M$_\sun$ model, whereas no such pattern is observed for the $1.3$-M$_\sun$ model (see Fig. \ref{fig:mini_influence}).

\begin{figure*}
\includegraphics[width=\textwidth]{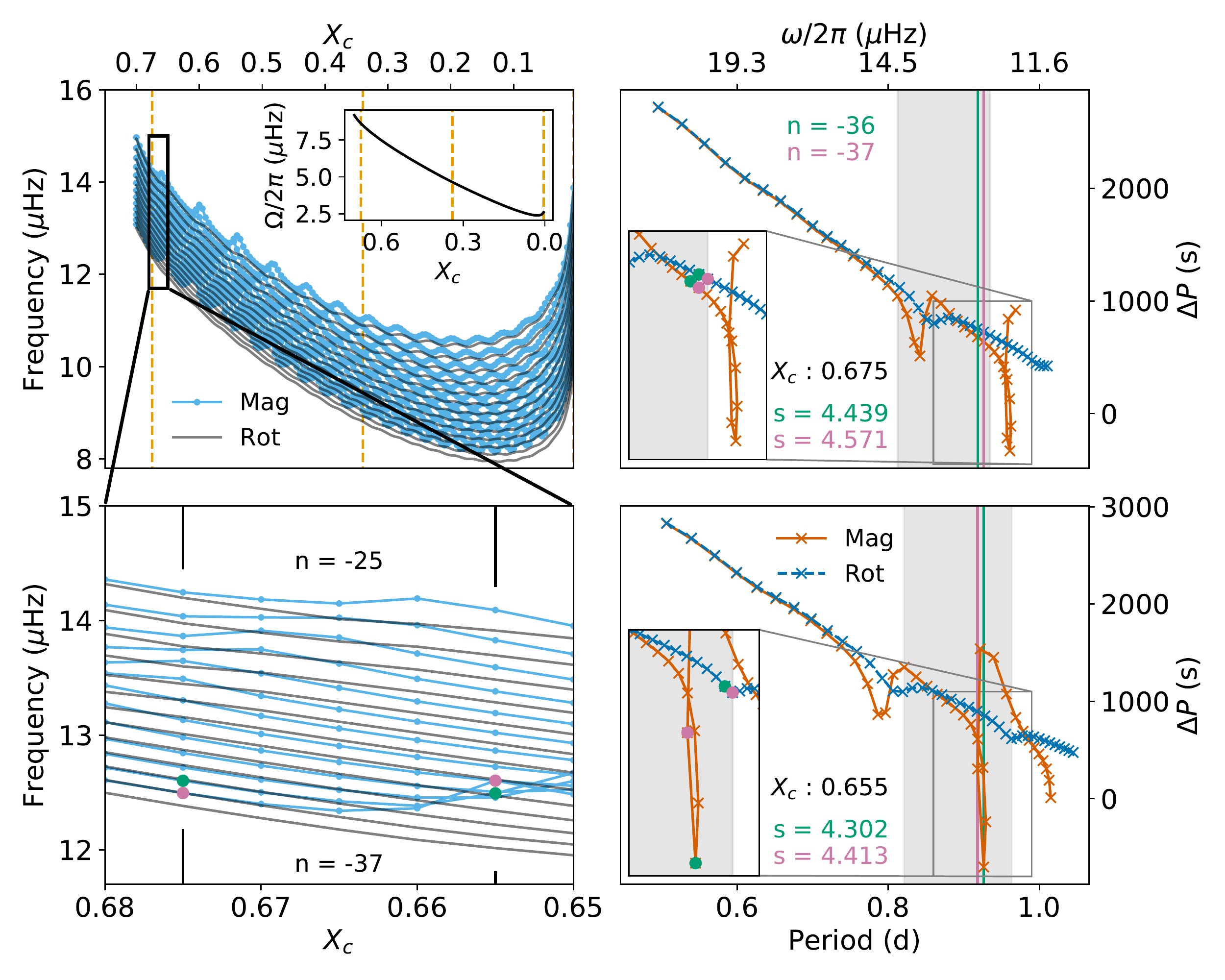}
\caption{Prograde dipole mode frequencies for the $3$-M$_\sun$ reference model, with the $X_{\rm c}$ values of our model grid (Sect. \ref{sec:4}) indicated with orange dashed vertical lines, and the rotation profile ($\mathcal{R}_{\rm rot} = 0.25$) indicated in the inset on the \textit{top left}. Radial order boundaries are indicated in the near-ZAMS zoom (\textit{bottom left}), as well as the $X_{\rm c}$ values for which period spacing patterns are extracted (\textit{right hand column}). That panel also shows an observed frequency crossing. Specific modes and spin parameters are indicated in color, and the color scheme for the period spacing patterns is the same as in the top right panel of Fig. \ref{fig:midMS_evolution}.}
\label{fig:bump_mag_rot}
\end{figure*}

The lower the mass of the stellar model, the longer the model evolves on the MS. 
Because of this longer time spent on the MS and the smaller amount of mass contained within the envelope, mixing processes more efficiently smooth chemical gradients in the near-core region during evolution in between the evolutionary models sampled at a specific $X_{\rm c}$.
Moreover, the $1.3$-M$_\sun$ models in our grid have a growing convective core throughout a significant part of the MS evolution \cite[see figure~2 in][]{Mombarg_2019_gamma_dor}, and hence do not produce chemical gradients in the near-core region through convective core recession. Modes are therefore less efficiently trapped in lower-mass models. 

Both the occurrence of a slope deviation for higher radial orders and the less pronounced sawtooth-like features in the period spacing patterns of lower-mass stars indicate that decreasing the initial mass affects magnetic shifts in a similar way to increasing the amount of mixing. 
The $1.3$-M$_\sun$ stellar structure model does however contain a convective envelope, which is not included in the magnetic field model of \citet{2010_Duez_Braithwaite_field}. As was indicated already by \citet{1967_Kraft}, stars with masses $M \la 1.3$ M$_\sun$ can harbor envelope dynamos that cause stellar spin-down by affecting the stellar wind (so-called magnetic braking). In this work we do not consider any envelope dynamo or stellar wind to be present (and hence $M = M_{\rm ini}$). The validity of our conclusions drawn based on this magnetic field model for these lower-mass stars should thus be confirmed in future work.

\subsection{Mode interaction and mode bumping}\label{subsect:mode_bump}

The phenomenon of mode bumping can be observed in the context of gradually evolving stellar background models, following the change in mode frequencies throughout the stellar evolution \cite[e.g.][]{1977_aizenman_smeyers,1979_roth_weigert,1981_JCD,1992_Gautschy,2010_BOOK_Smeyers_VanHoolst}. This is reminiscent of the phenomenon of avoided crossings that takes place between modes of two coupled oscillators that can be decoupled (mathematically), as was done in \citet{1977_aizenman_smeyers}, to study these interacting modes. They did not take the influence of rotation into account.
Mode bumping has mostly been studied for massive stars on the MS \cite[e.g.][]{1977_aizenman_smeyers,1979_roth_weigert} or for more evolved stars such as subgiants and red giants \cite[e.g.][]{2011_Bedding_modebump_redgiant,2012_Benomar_bumping_subgiant,2012_Mosser_modebump_redgiant,2013_Benomar_modebump_subgiant}, involving bumping events between $g$ and $p$ dipole modes. With each bumping event, interacting modes gain an additional node \cite[e.g.][]{2010_BOOK_Smeyers_VanHoolst}. 

Our simulations show that mode bumping between magneto-rotationally modified prograde dipole $g$~modes, a phenomenon we define as magnetic mode bumping and which we observe in the form of frequency crossings, can be substantial. Preferentially, degenerate perturbation theory should be employed to describe these situations where frequency shifts become comparable to mode spacing. However, this involves treating the unperturbed eigenfunctions as a superposition of the near-degenerate modes \cite[as mentioned in e.g.][]{2020_Loi}. Its use thus requires more involved calculations than the ones represented in the P+19 formalism.

The magnetic bumping events are indicated in Fig. \ref{fig:bump_mag_rot}, where the evolution of $g$-mode frequencies in the $3$-M$_\sun$ reference model is shown for radial orders $n \in [-37,\ldots,-25]$ and $B_0 = 10^{\,6}$~G, keeping a constant $\mathcal{R}_{\rm rot} = 0.25$ (i.e. $\Omega/2\pi \sim 8.6$ \textmu Hz at near-ZAMS: $X_{\rm c} \approx 0.675$) throughout MS evolution.
Magnetic mode bumping is different from the classical bumping events considered by \citet{1977_aizenman_smeyers}, because it is an interaction between two magneto-rotationally modified $g$~modes instead of interacting $p$~and~$g$~modes. It only occurs if a certain (critical) near-core field strength is attained. 
This already occurs close to the near-ZAMS phase for high-radial-order modes in our reference model, rendering correct mode identification hard, even for the limited subset of modes displayed in Fig. \ref{fig:bump_mag_rot}. 

The typical behavior of the magneto-rotationally modified patterns discussed in this section can clearly be distinguished: trapped mode frequencies are strongly modified, with some shifted frequencies even generating characteristic loops in the period spacing patterns. In such loops, higher-radial-order modes would have lower frequencies than their lower-order companions (i.e. their frequencies have crossed). Most of these loops in zonal period spacing patterns are not perturbative according to our consistency check (see the upper right panels in Figs. \ref{fig:midMS_evolution} and \ref{fig:ZAMS_evolution}). However, for sectoral mode period spacing patterns, most magneto-rotationally modified period spacings are considered to be perturbative according to our criterion, including those that form loops. The period spacing patterns displayed in the panels of the right hand column of Fig. \ref{fig:bump_mag_rot} are a good example: all magnetic frequency shifts are perturbative.  

The generation of these loops can be explained by looking at how the frequencies of two example modes ($n=-36$ and $n=-37$), which we refer to as the $\alpha$ modes, change during near-ZAMS evolution: trapping of $g$~modes changes throughout MS evolution, as a result of the chemical gradient left behind by the receding convective core. 
The magneto-rotationally modified frequencies approach each other more closely when evolving from $X_{\rm c} = 0.675$ to $X_{\rm c} = 0.660$, crossing each other at $X_{\rm c} = 0.655$, because a chemical gradient-related dip passes by that traps the modes in the near-core region. 
Quantitatively, the mode frequency differences (i.e. $\left|\,\omega_{\,n=-36} - \omega_{\,n=-37}\,\right|/2\pi$) decrease from $\sim 0.11$ \textmu Hz ($X_{\rm c} = 0.675$) to $\sim 0.02$ \textmu Hz ($X_{\rm c} = 0.660$), rendering them indistinguishable with respect to the frequency uncertainties quoted by \citet{2019_Gang_Li}. 
We observe 18 additional frequency crossings between these $\alpha$ modes throughout the MS of our $3$-M$_\odot$ reference model. 

We thus find that magnetic mode bumping, which we observe in the form of frequency crossings, is important throughout the entire MS evolution in the case of strong internal magnetic fields in intermediate-mass stars.
It certainly is important for our reference model, which is representative of a low-mass SPB pulsator with a $10^{\,6}$ G near-core magnetic field. In general, however, the degree to which $g$ modes undergo magnetic mode bumping depends on the fundamental parameters of the stellar model and its rotational and magnetic field evolution. 

\section{Conclusions and prospects}\label{sec:5}

Internal stellar magnetic fields are poorly characterized, because there is no direct way to observe them. 
These fields can contribute significantly to the transport of angular momentum inside stars and might be key to resolving the problem of transport in current models \citep{Aerts2019_ARAA}. Moreover, they may make significant contributions to (local) chemical and general energy transport in stellar interiors. 

We investigated how the \citet{Duez_field_2010} and \citet{2010_Duez_Braithwaite_field} mixed poloidal-toroidal dipolar axisymmetric large-scale internal magnetic field model modifies the frequency of dipole gravity-mode pulsations in rotating, magnetic stars, relying on the TAR. The formalism we use was developed by P+19, who treat the effect of the magnetic field as a perturbation. Here, we performed a parameter study of intermediate-mass MS stars with masses ranging from 1.3 to 3.0 M$_\sun$, linking modifications of magnetic features in period spacing patterns to fundamental stellar parameters and improving the consistency check of P+19. Our fundamental parameter grid is representative of both $\gamma$~Dor and low-mass SPB stars and can thus be used to trace axisymmetric magnetic field influences in these pulsators. 

The simulations in this work show that $g$-mode period spacing patterns are an excellent tool for detecting and characterizing strong near-core magnetic fields ($B_{\rm 0} > 10^{\,5}$~G) in stars near the TAMS. 
We find observables that are distinct from those attributed to the chemical gradient left behind by the receding convective core \citep{2008_Miglio} and rotation \citep{2016_Van_Reeth,Van_Reeth_2018_differential_rotation}.
The higher-order modes are more confined to the near-core region \citep{2015_Moravveji}, where the stellar magnetic field is strongest. Hence, they are influenced by a strong magnetic field over a larger part of their mode cavity, compared to modes that are less confined to this region, and therefore undergo larger magnetic shifts. Pulsation modes that are trapped by chemical gradients in the stellar near-core region, undergo even larger magnetic frequency shifts.
An increase in the strength of mixing, typically due to a decrease in slope of the $D_{\rm ov}(r)$ profile in the near-core region in a stellar model, washes out such gradients, so that only the slope of higher-radial-order modes is affected. We find that the value of $D_{\rm mix}$ near the core overshooting region boundary is especially important in this process. 
P+19 found that an increase in near-core rotation rate decreases the magnetic shifts values for their 5.8-M$_\sun$ stellar model.
We draw a similar conclusion for our lower-mass models, making it essential to include rotational influence when computing magnetic shifts. 
Magnetic shifts are smaller than the 4-yr \textit{Kepler} frequency resolution for near-core field strengths up to $10^{\,5}$~G, but stronger fields induce detectable shifts in period spacing patterns from 4-yr \textit{Kepler}, 1-yr TESS and 2-yr PLATO light curves \citep{KEPLER_mission,TESS_mission,PLATO_mission}.
The features are similar for stars with different metallicity, although individual magnetic frequency shift values change.

We find that many of the magnetically modified zonal frequencies in near-ZAMS ($X_{\rm c} \approx 0.675$) and mid-MS ($X_{\rm c} \approx 0.340$) models undergo shifts that fall outside of the allowed range if the near-core field strength is $10^{\,6}$ G.
Faster-rotating near-ZAMS and mid-MS pulsators have more modes that undergo perturbative magnetic shifts throughout the entire MS, irrespective of the value of $B_{\rm 0}$ .

This work allows detailed forward modeling of magnetic, pulsating MS stars. The $\gamma$~Dor stars modeled by \citet{Mombarg_2019_gamma_dor} are predominantly near the ZAMS, and those found by \citet{2019_Gang_Li} cover the entire main sequence. Most of these stars exhibit prograde dipole $g$-mode patterns, for which our simulations show that most magnetic shifts are perturbative according to Eq. (\ref{eq:omega_allowed}). They are therefore prime candidates for future magneto-asteroseismic forward modeling of $\gamma$~Dor pulsators.

The \citet{2010_Duez_Braithwaite_field} and \citet{Duez_field_2010} initial magnetic field configuration considered in this work only depends on the stellar density profile, and strongly changes throughout the MS evolution, becoming more confined to the near-core region for more evolved stars.
Once such an initial magnetic field configuration is formed, it evolves on longer timescales through Ohmic diffusion, moves outward and gradually changes its axisymmetric, confined configuration into a axisymmetric, open one \citep{2010_Duez_Braithwaite_field}. Moreover, many observed large-scale surface magnetic fields are oblique: they are inclined with respect to the stellar rotation axis \citep{1970_Landstreet,1990_Moss,2012_Walder,2016_MIMES_WADE}.
\citet{2020_Prat} recently extended the P+19 formalism to non-axisymmetric oblique dipolar mixed magnetic fields. Establishing the influence on $g$-mode period spacing patterns of the non-axisymmetric configurations encountered in the evolution of such fields constitutes the next step in linking internal magnetic fields to spectropolarimetrically observable large-scale surface magnetic fields.

\begin{acknowledgements} The research leading to these results received funding from
the European Research Council (ERC) under the European Union’s Horizon
2020 research and innovation program (grant agreements No. 647383: SPIRE
with PI S. M. and No. 670519: MAMSIE with PI C. A.) We thank the anonymous referee for valuable input, and the MESA and GYRE
code development team members for all their efforts put into the
public stellar evolution and pulsation codes. J. V. B. would like to thank Jim Fuller, Tim Van Hoolst and Tom Van Doorsselaere for their useful comments on his master's thesis text that made this manuscript better. T. V. R. gratefully acknowledges support from the Research Foundation Flanders (FWO) through grant 12ZB620N. \end{acknowledgements}

\bibliographystyle{bibtex/aa.bst} 
\bibliography{bibtex/ourbiblio.bib} 

\begin{appendix}

\section{MESA and GYRE inlists}\label{sec:C1}
Example MESA and GYRE inlists used for this work are available from the MESA inlists section of the MESA Marketplace: 
\href{http://cococubed.asu.edu/mesa_market/inlists.html}{\texttt{http://cococubed.asu.edu/mesa\_market/inlists.html}}

A large jump in period spacing for the $\alpha_{\rm MLT} = 1.5$ model (Sect. \ref{sec:4.1.5}) was apparent if the standard MESA input file was used, which could be attributed to numerical inaccuracy in obtaining the $n=-13$ mode. 
We remedied the problem by decreasing the `\texttt{varcontrol\_target}' parameter value to \texttt{2d-5} in the MESA inlist (the default parameter value in our setup is \texttt{5d-5}).

\section{Magnetic suppression of near-core mixing} \label{app:Bcrit}

Strong magnetic fields can suppress thermal convection, hence, we verify whether the convective instability is suppressed by the locally imposed magnetic field \citep{2011_Zahn} at different evolutionary phases of the $3$-M$_\sun$ reference model. If this is the case, convective processes will be suppressed locally, affecting for example convective core overshooting. We therefore compute the critical field strength at and above which convection is suppressed, according to the convective stability condition of a stratified fluid in the presence of a magnetic field derived by \citet{1961_Chandrasekhar}, and find that only near the surface the total magnetic field strength of the \citet{2010_Duez_Braithwaite_field} field, $B = \sqrt{B_r^2 + B_\theta^2 + B_\varphi^2}$, is larger than this critical value. Thus, only the field near the surface convection zone is likely suppressed.

The amount of overshooting measured by asteroseismology is generally smaller for magnetic stars \citep{2012_Briquet,Buysschaert_2018_magnetic_forward_model}. Core overshooting is expected to be stronger if the star is rotating more rapidly, as was shown by the simulations of \citet{2004_Browning} and \citet{2012_Neiner_overshoot}. Similar to \citet{2012_Briquet}, we evaluate two criteria that tell us approximately for which field strength core overshooting is suppressed. The first criterion is based on the AM transport equation that takes into account a magnetic field \citep{2005_Mathis}, from which a critical field strength $B_{\rm crit,\,Z}$ for suppression of rotational mixing (i.e. $B>B_{\rm crit,\,Z}$) is defined \citep{2011_Zahn}. The second criterion computes the initial field strength $B_{\rm crit,\,S}$ above which the magnetic field remains non-axisymmetric and rotation becomes uniform due to magnetic torques that suppress differential rotation by the process of phase mixing \cite[see e.g.][]{1999_Spruit}. Similar to what is obtained for suppression of convective mixing at different evolutionary phases of the $3$-M$_\sun$ reference model, only (rotational) mixing near the surface is suppressed at these evolutionary phases.

How suppression of mixing near the surface influences $g$ modes is beyond the scope of the current study. Moreover, we must take into account the fact that the \citet{2010_Duez_Braithwaite_field} field only represents an initially confined configuration, which evolves over time. This field evolution further changes the regions in the stellar interior for which magnetic suppression of mixing is expected.

\section{Additional Material} \label{sec:app_B}

In this appendix we present additional plots and tables on some $2$-M$_\sun$ and $3$-M$_\sun$ models in the model grid that were not shown in the main text, but which have been mentioned in discussions. They further support our choice of mainly discussing $3$-M$_\sun$ models in the main text, because the effect of the magnetic field is similar for models of different masses, even though its magnitude depends on the stellar structure. We do not show period spacing patterns of $1.3$-M$_\sun$ models, except for one period spacing pattern in Fig. \ref{fig:mini_influence}, because of the uncertainties in the derived mode periods, related to the presence of the convective envelope.

\begin{table}[t!]\centering
\caption{Maximal perturbative frequency deviations $\Delta \omega_{\,\rm per}/2 \pi$ and maximal frequency deviations $\Delta \omega/2 \pi$ for frequencies of dipole modes with radial orders $n \in [-50,\ldots,-10]$, computed with Eqs. (\ref{eq:max_omega_allowed}) and (\ref{eq:max_omega}) at the TAMS, when varying $\mathcal{R}_{\rm rot}$ of the reference model, for different $m$.}
\label{tab:rot_var_freq_table}
\begin{tabular}{@{} Sl Sl Sl Sl @{}}
\hline \hline 
 $m$ & $\mathcal{R}_{\rm rot}$ & $\Delta \omega_{\,\rm per} / 2 \pi$ (\textmu Hz) & $\Delta \omega / 2 \pi$ (\textmu Hz)\\ \hline
 $1$& $0.01$ & $1.37$ & $1.37$ \\
$1$& $0.25$ & $0.89$ & $0.89$ \\
 $1$& $0.50$ & $0.63$\, \tablefootmark{a} & $0.63$\, \tablefootmark{a}\\
 $1$ & $0.75$ & $0.49$\, \tablefootmark{a}  & $0.49$\, \tablefootmark{a} \\ \hline
  $0$& $0.01$ & $0.70$ & $0.70$ \\
$0$& $0.25$ & $0.77$ & $0.77$ \\
 $0$& $0.50$ & $0.78$\, \tablefootmark{a} & $0.78$\, \tablefootmark{a}\\
 $0$ & $0.75$ & $0.65$\, \tablefootmark{a}  & $0.65$\, \tablefootmark{a} \\ \hline
  $-1$& $0.01$ & $1.42$ & $1.42$ \\
$-1$& $0.25$ & $1.80$ & $1.80$ \\
 $-1$& $0.50$ & $1.65$ & $1.65$\\
 $-1$ & $0.75$ & $1.22$\, \tablefootmark{a} & $1.22$\, \tablefootmark{a} \\ \hline
\end{tabular}
\tablefoot{
\tablefoottext{a}{Not all computed spin parameters were $< 1$.}}
\end{table}

\begin{figure}
\includegraphics[width=88mm]{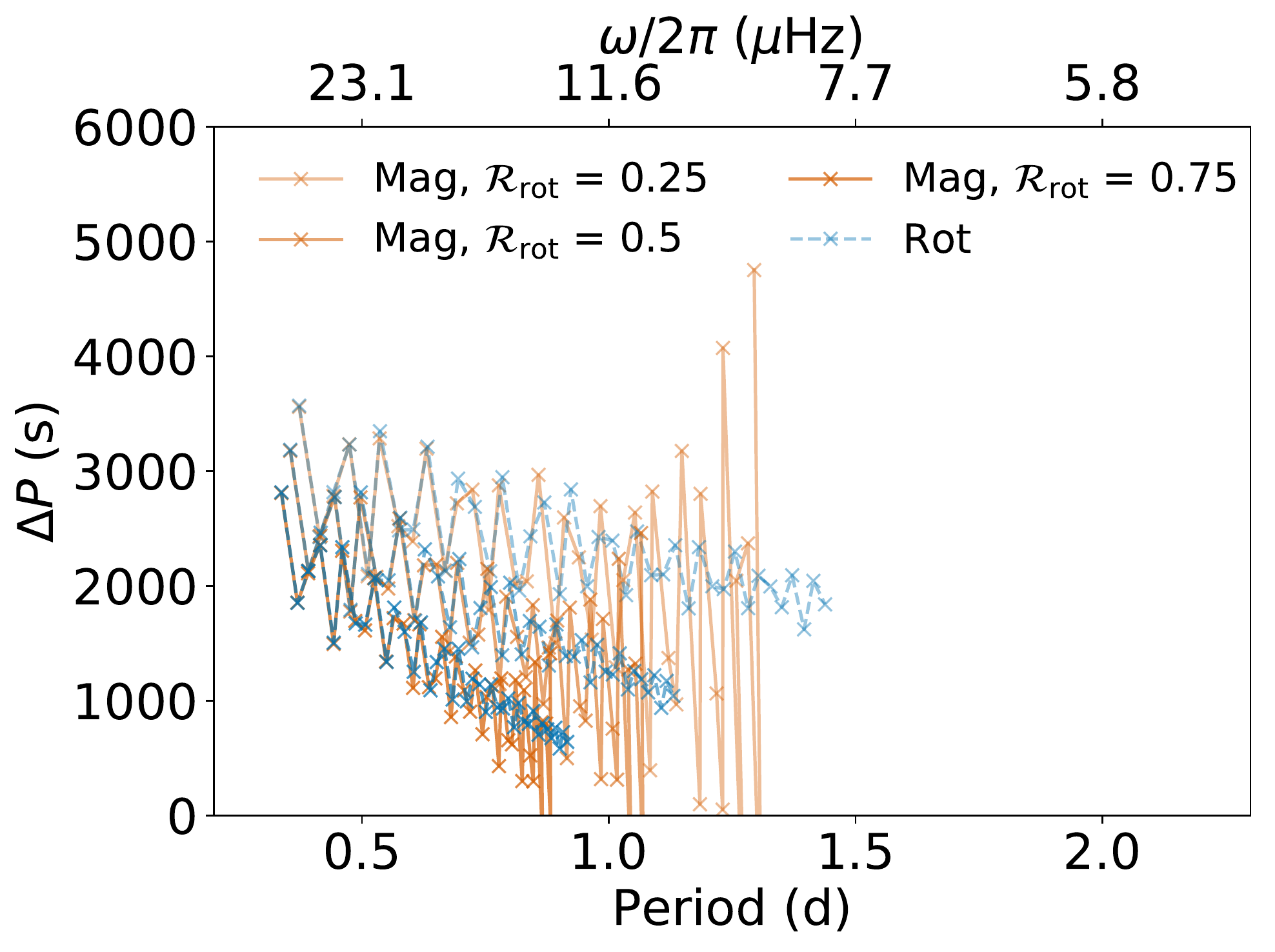}
\caption{Period spacing patterns of prograde dipole modes of the $3$-M$_\sun$ near-TAMS reference model, varying the rotation ratio $\mathcal{R}_{\rm rot}$. The color scheme is the same as in the right hand panel of the top row of Fig. \ref{fig:midMS_evolution}.}
\label{fig:field_rotation_evolution_prograde}
\end{figure}

\begin{figure*}
\includegraphics[width=8.5cm]{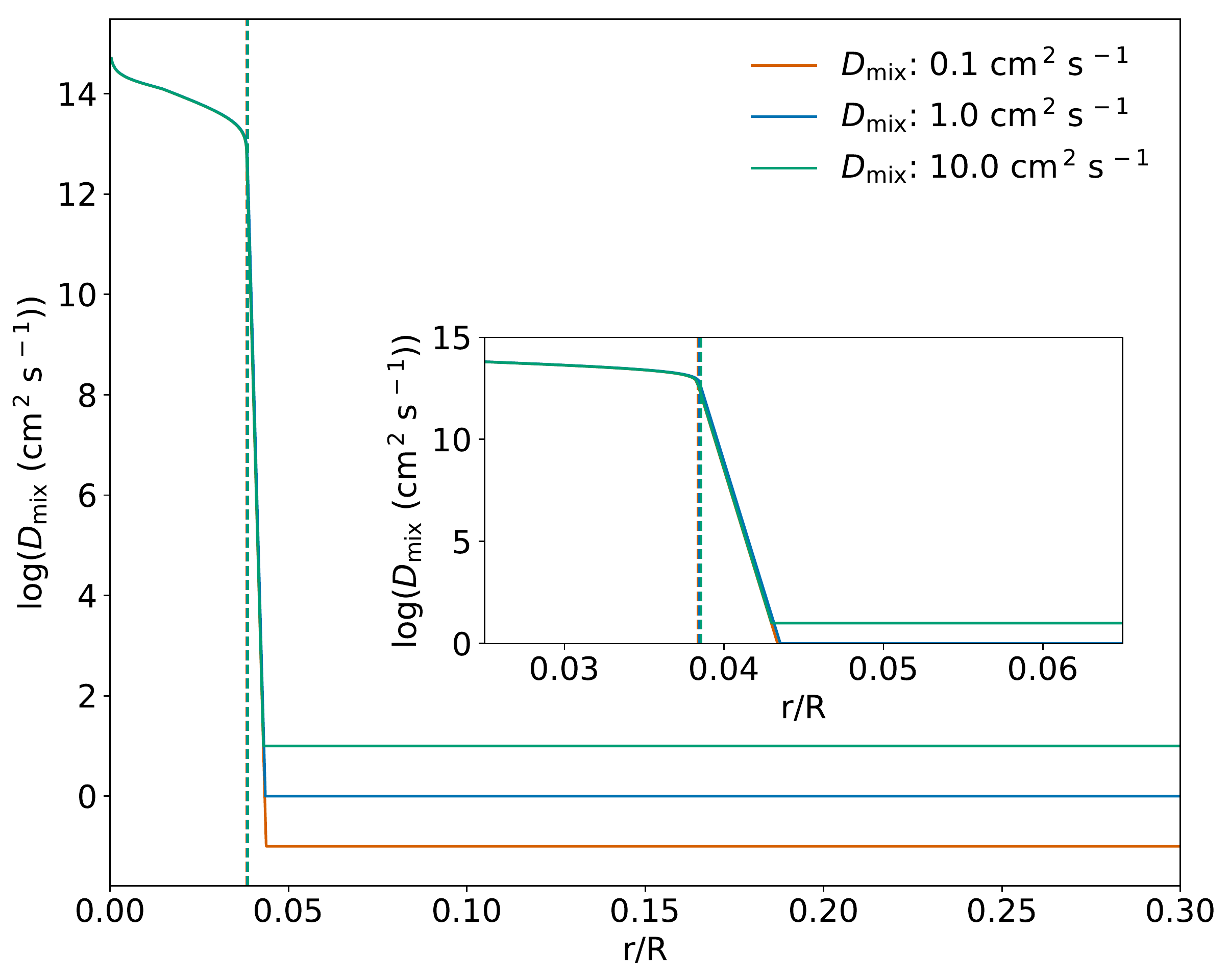}\hfill\includegraphics[width=8.5cm]{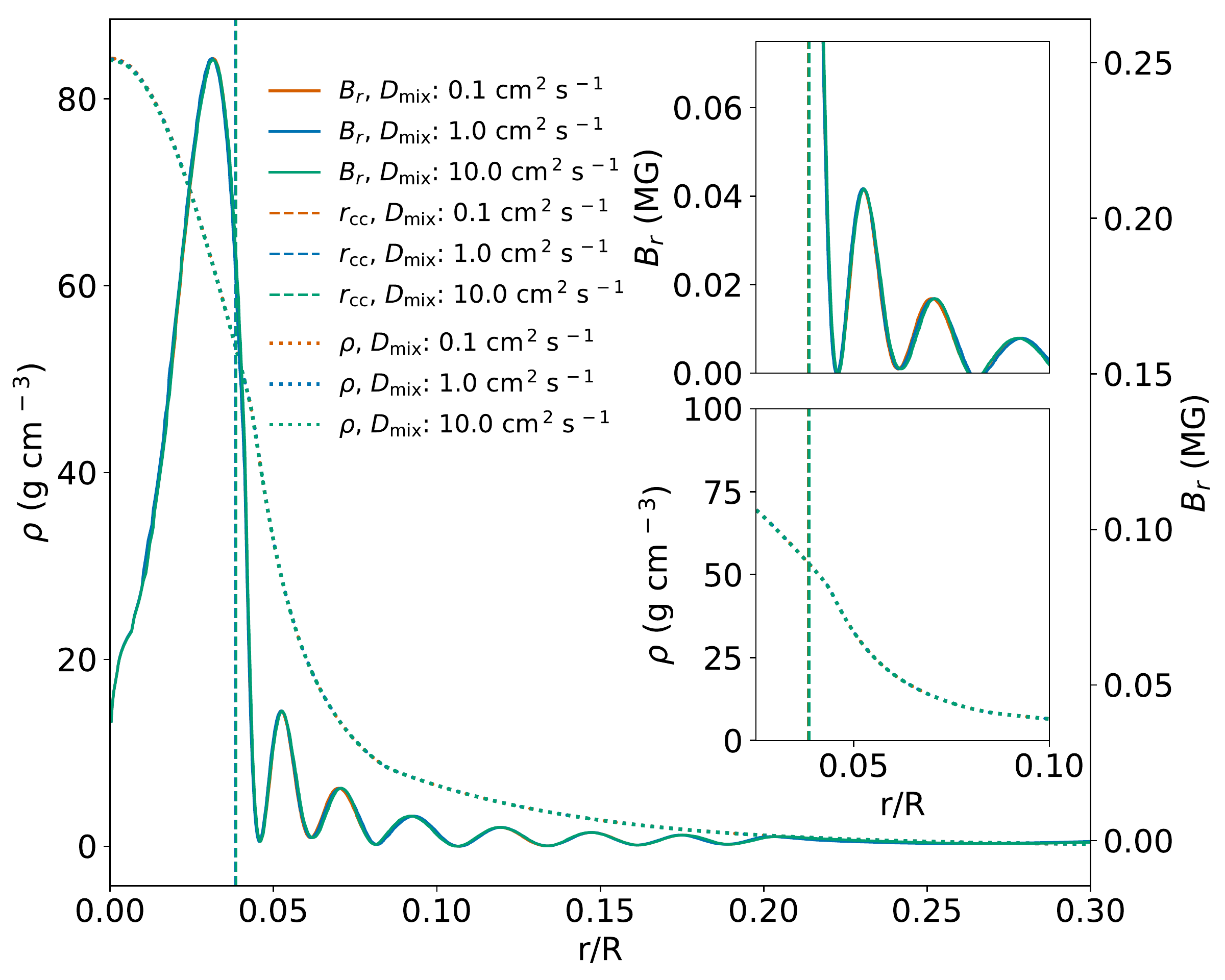}\\
\includegraphics[width=6cm,height=4.5cm]{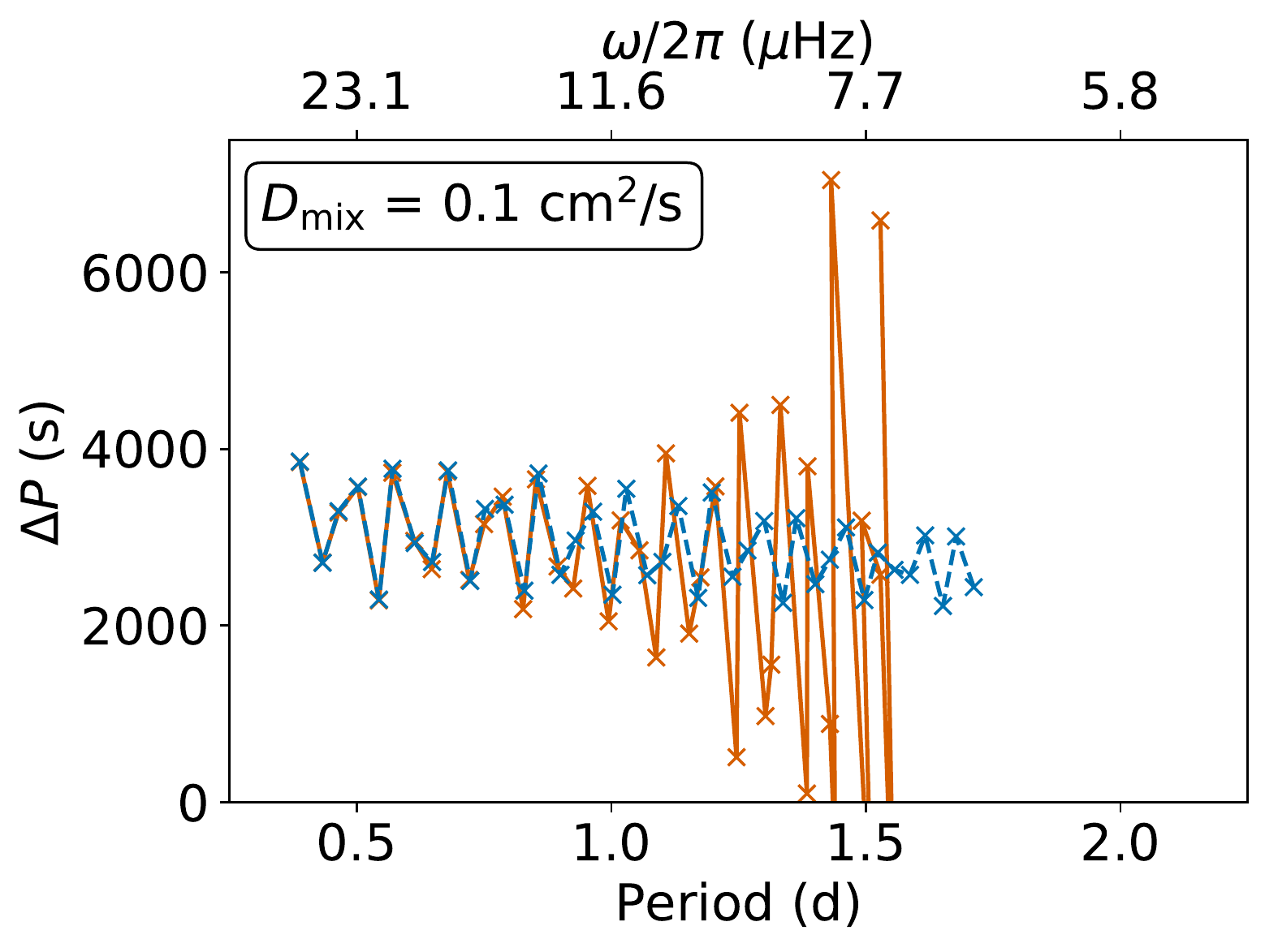}\includegraphics[width=6cm,height=4.5cm]{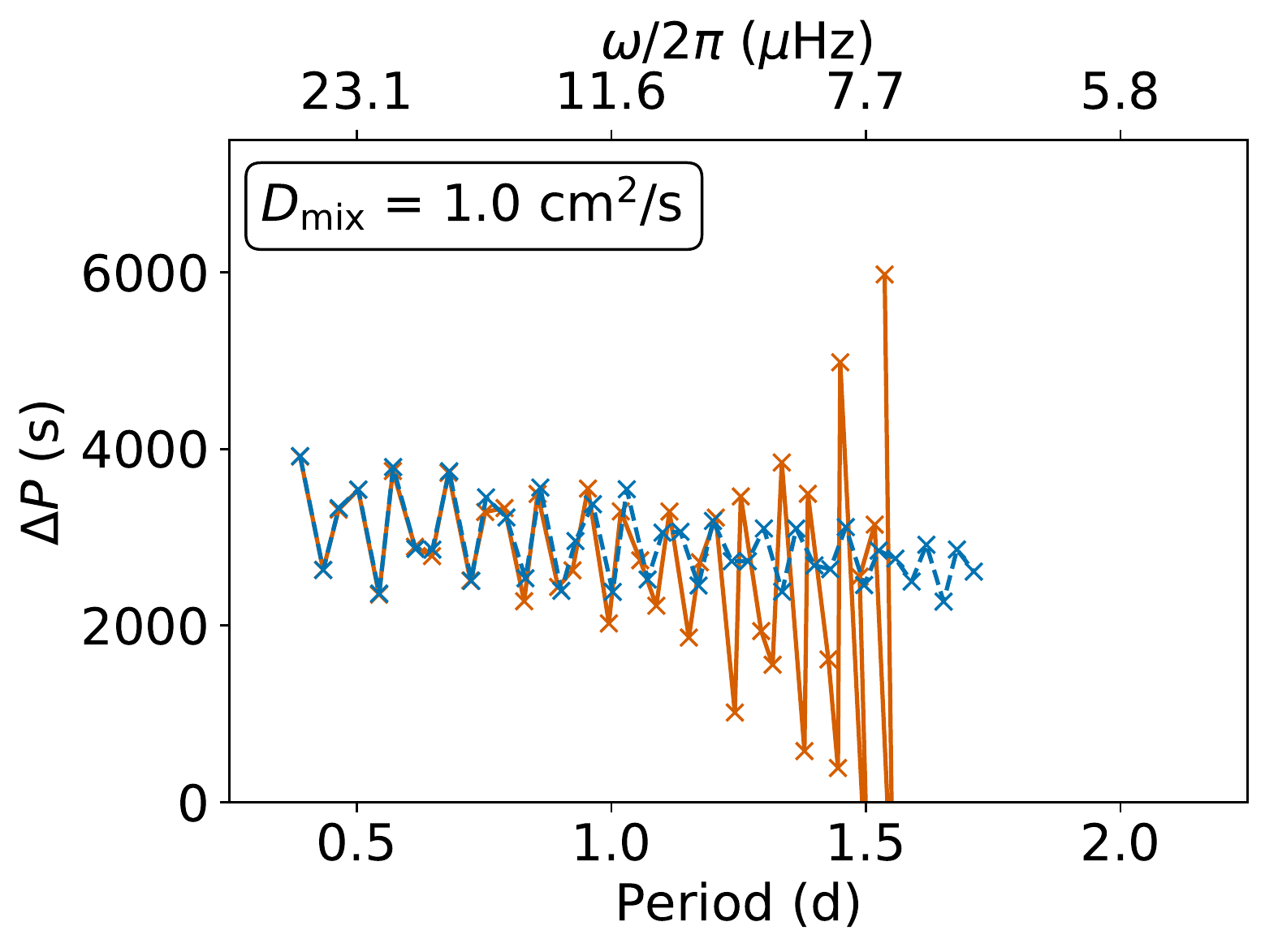}\includegraphics[width=6cm,height=4.5cm]{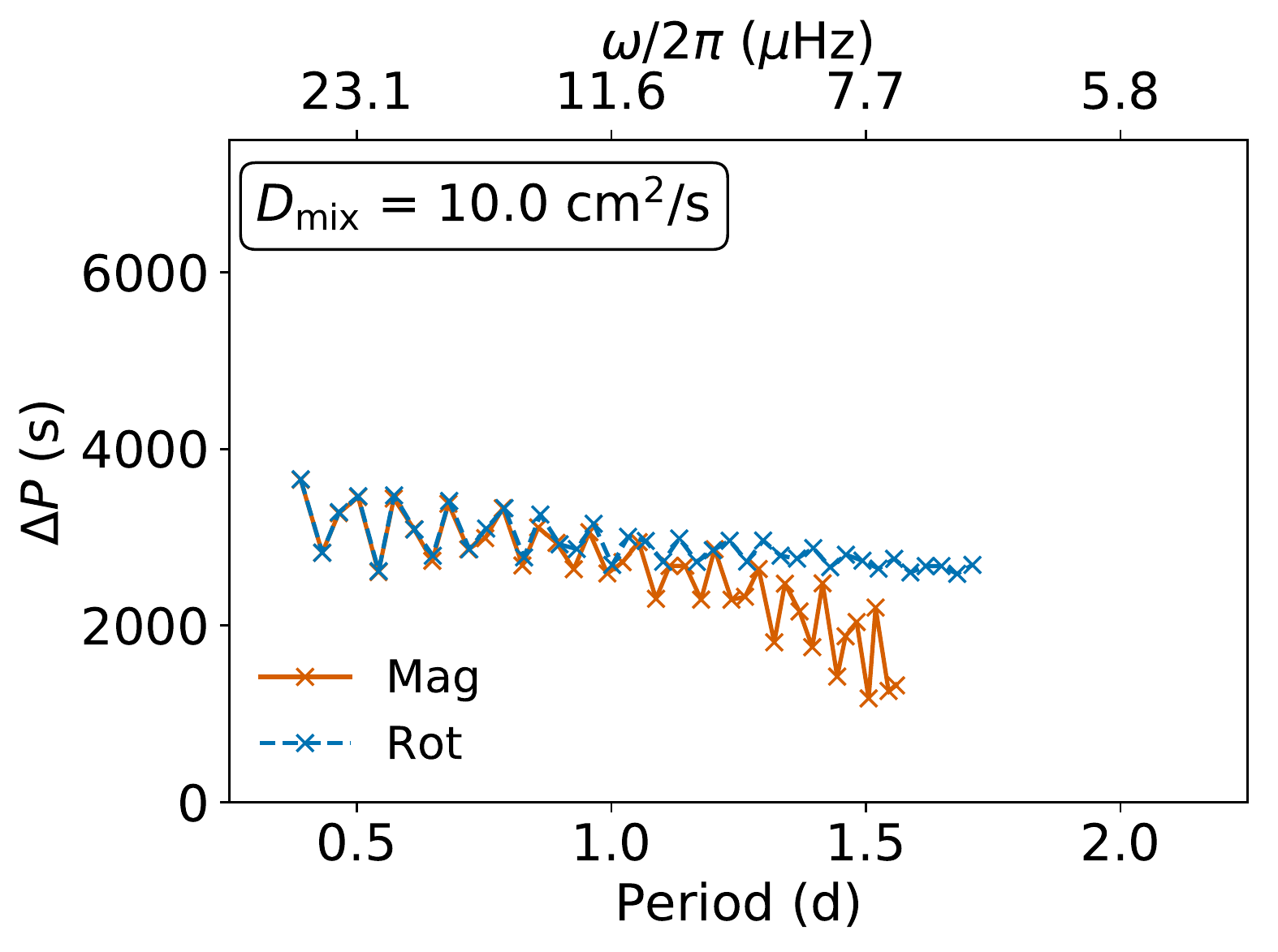}
\caption{Same as Fig. \ref{fig:fov_influence}, but varying $D_{\rm mix}$.}
\label{fig:Dmix_influence}
\end{figure*}

\begin{figure}
\includegraphics[width=88mm]{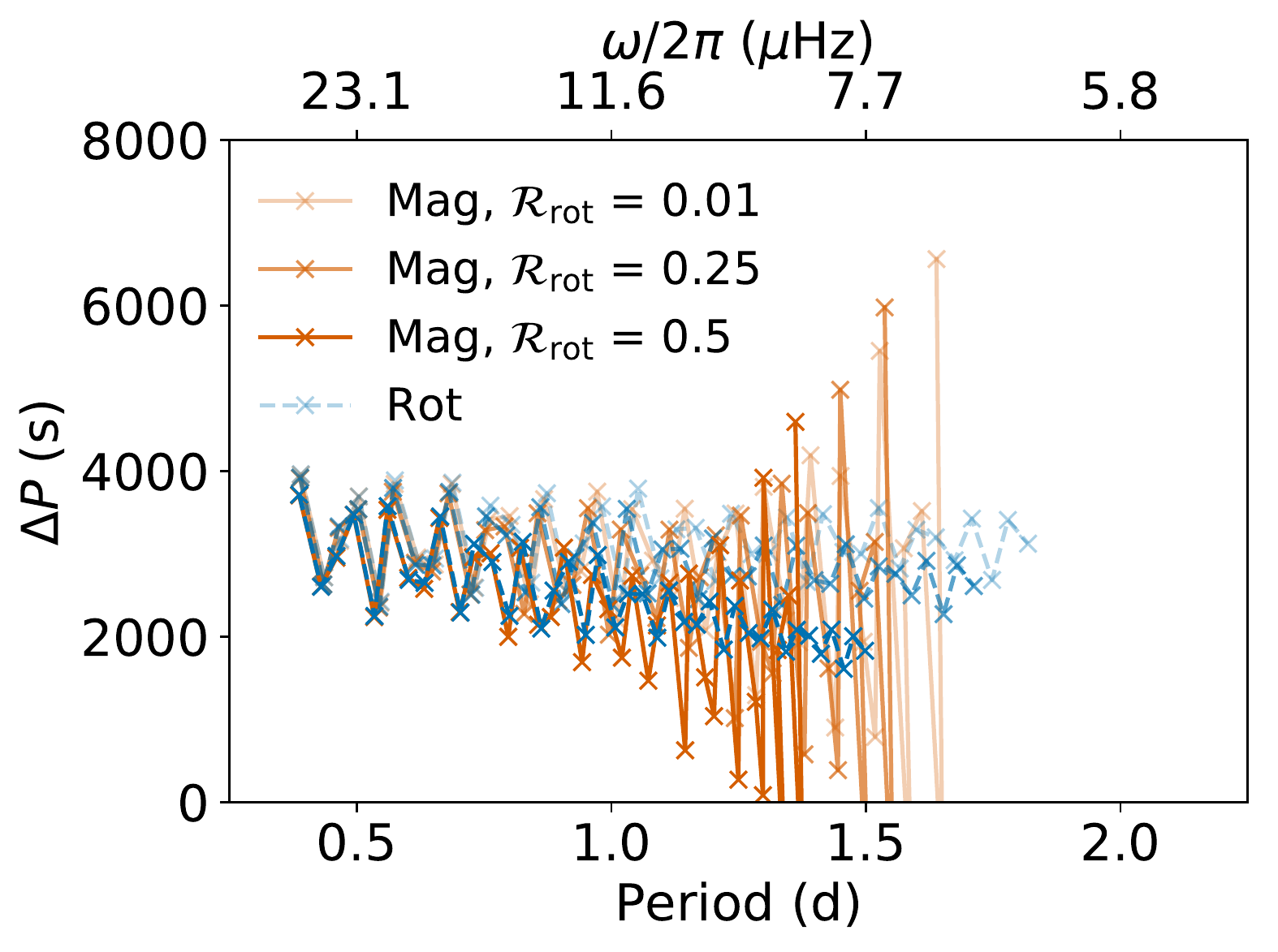}
\caption{Same as Fig. \ref{fig:field_rotation_evolution_prograde}, but for zonal modes.}
\label{fig:field_rotation_evolution_zonal}
\end{figure}

\begin{figure}
\includegraphics[width=88mm]{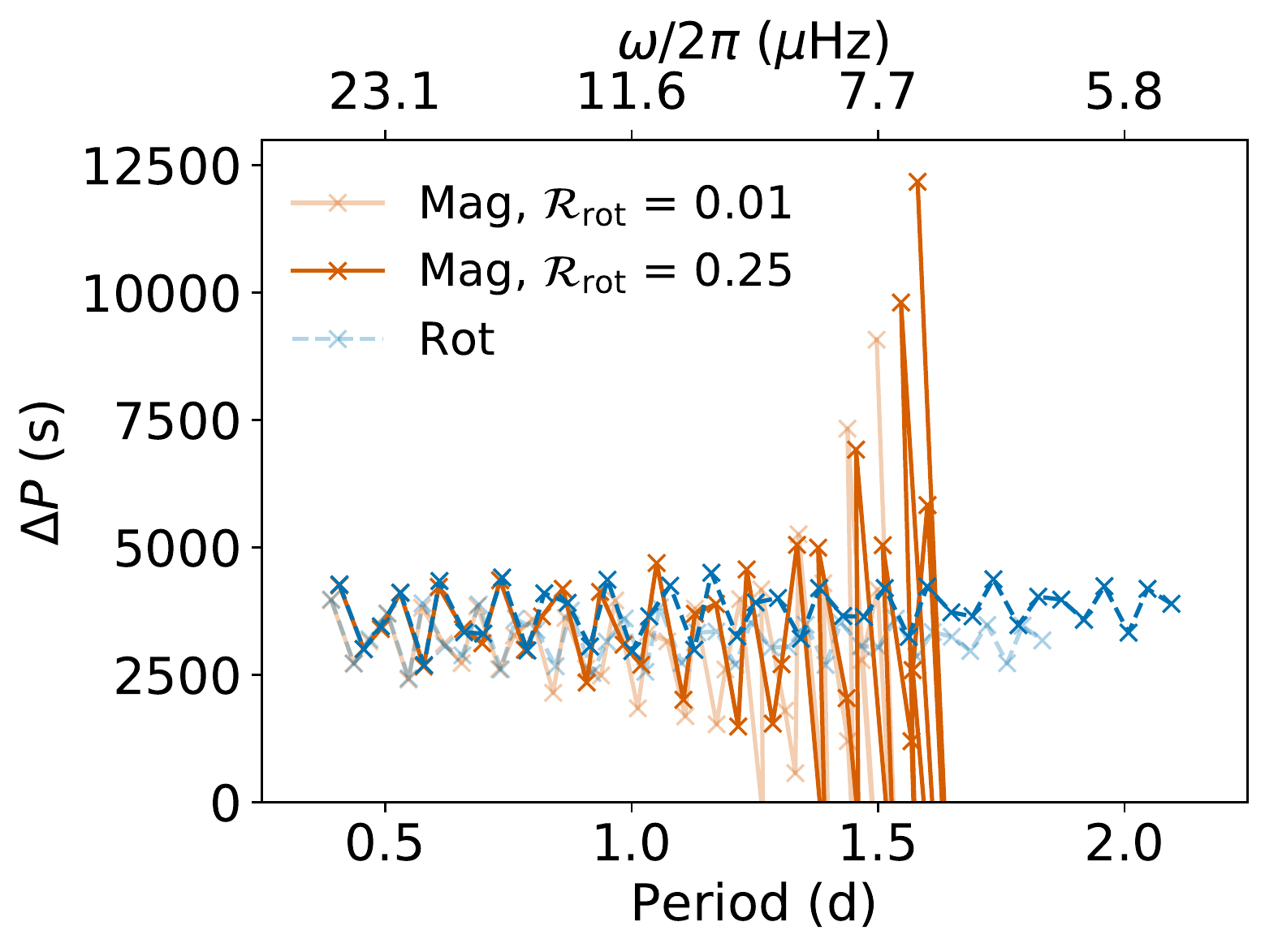}
\caption{Same as Fig. \ref{fig:field_rotation_evolution_prograde}, but for retrograde modes.}
\label{fig:field_rotation_evolution_retrograde}
\end{figure}

\begin{figure*}
\includegraphics[width=6cm,height=4.5cm]{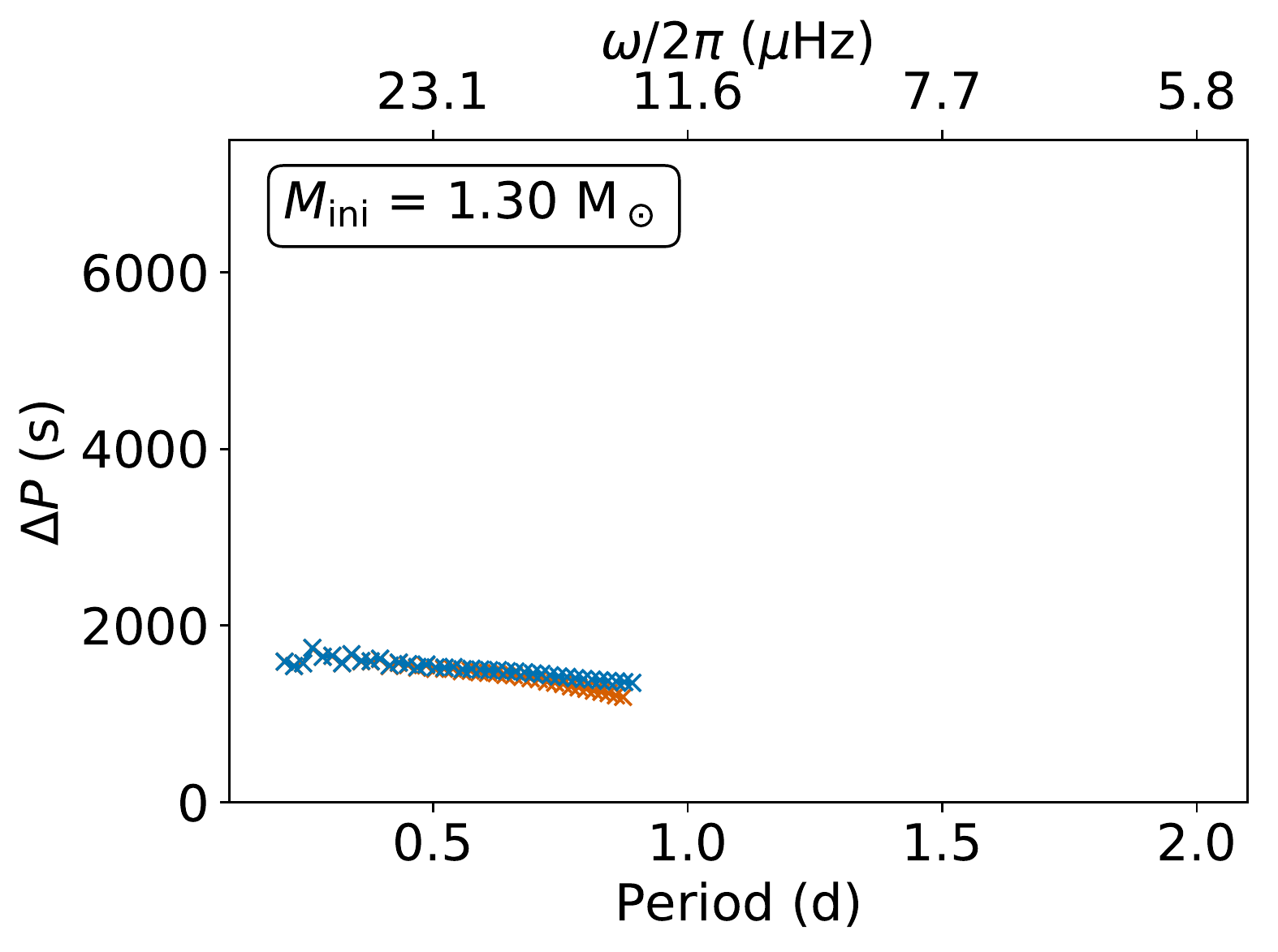}\includegraphics[width=6cm,height=4.5cm]{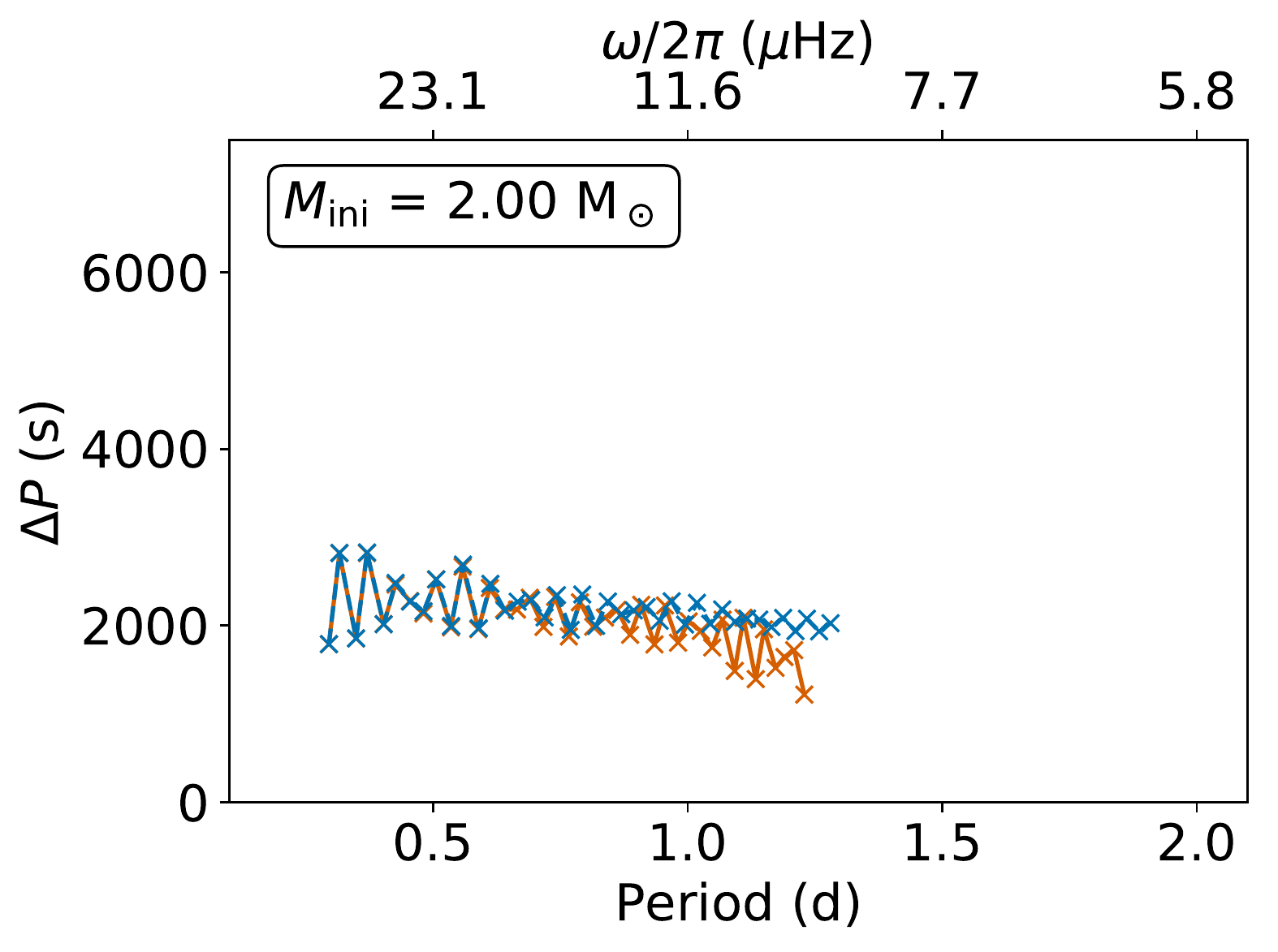}\includegraphics[width=6cm,height=4.5cm]{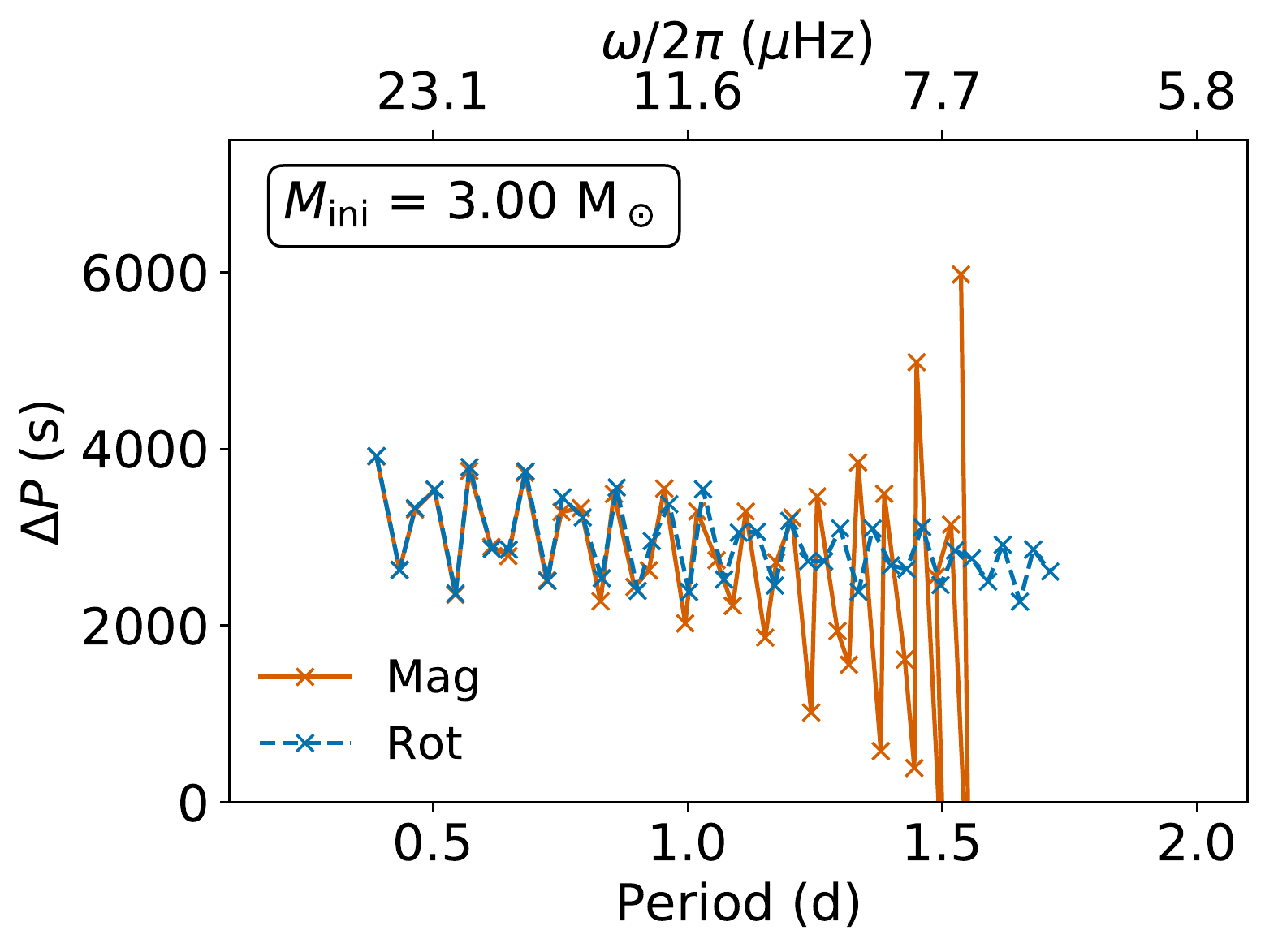}
\caption{Same as Fig. \ref{fig:B0_differences}, but varying $M_{\rm ini}$.}
\label{fig:mini_influence}
\end{figure*}

\begin{figure*}
\includegraphics[width=8.5cm]{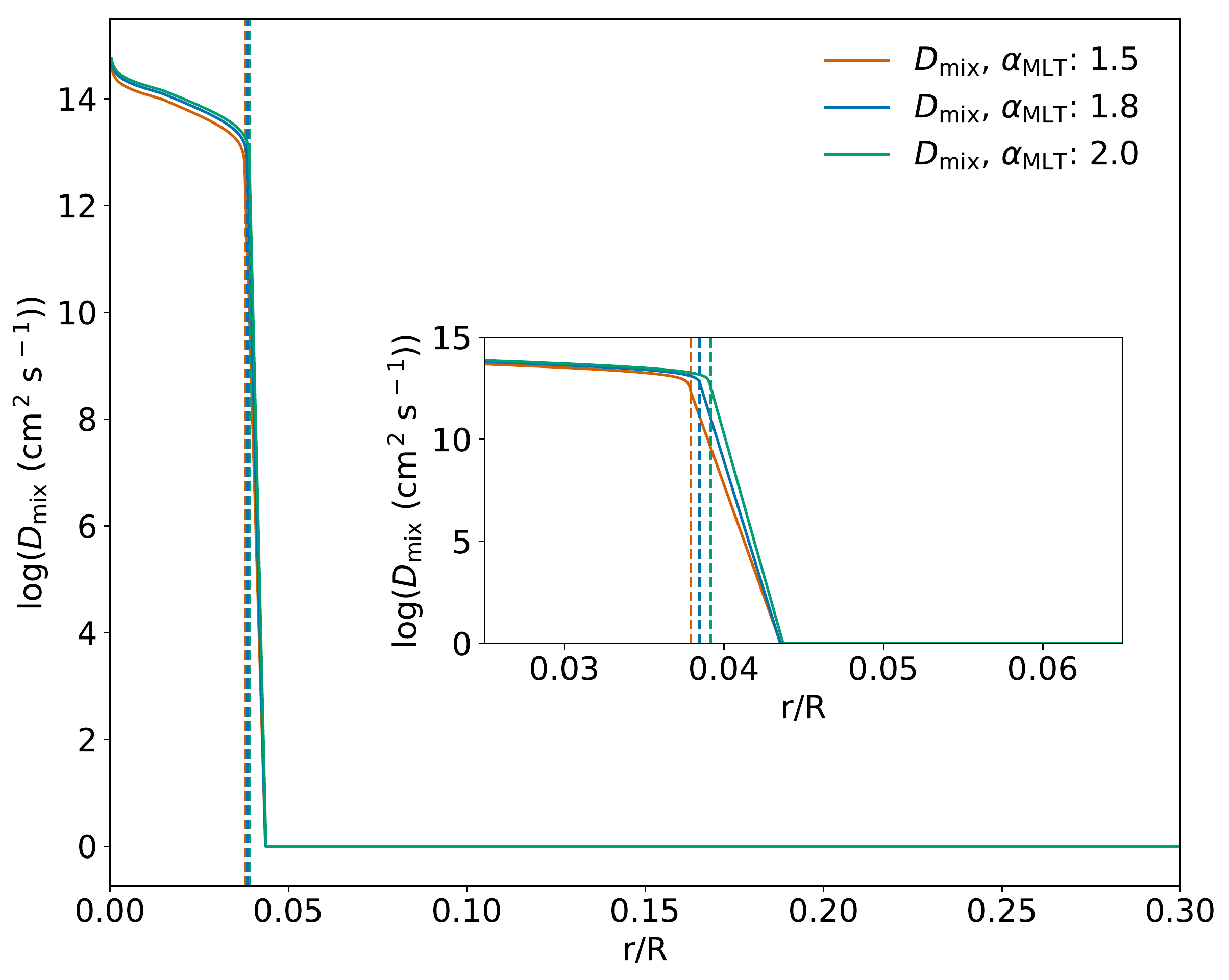}\hfill\includegraphics[width=8.5cm]{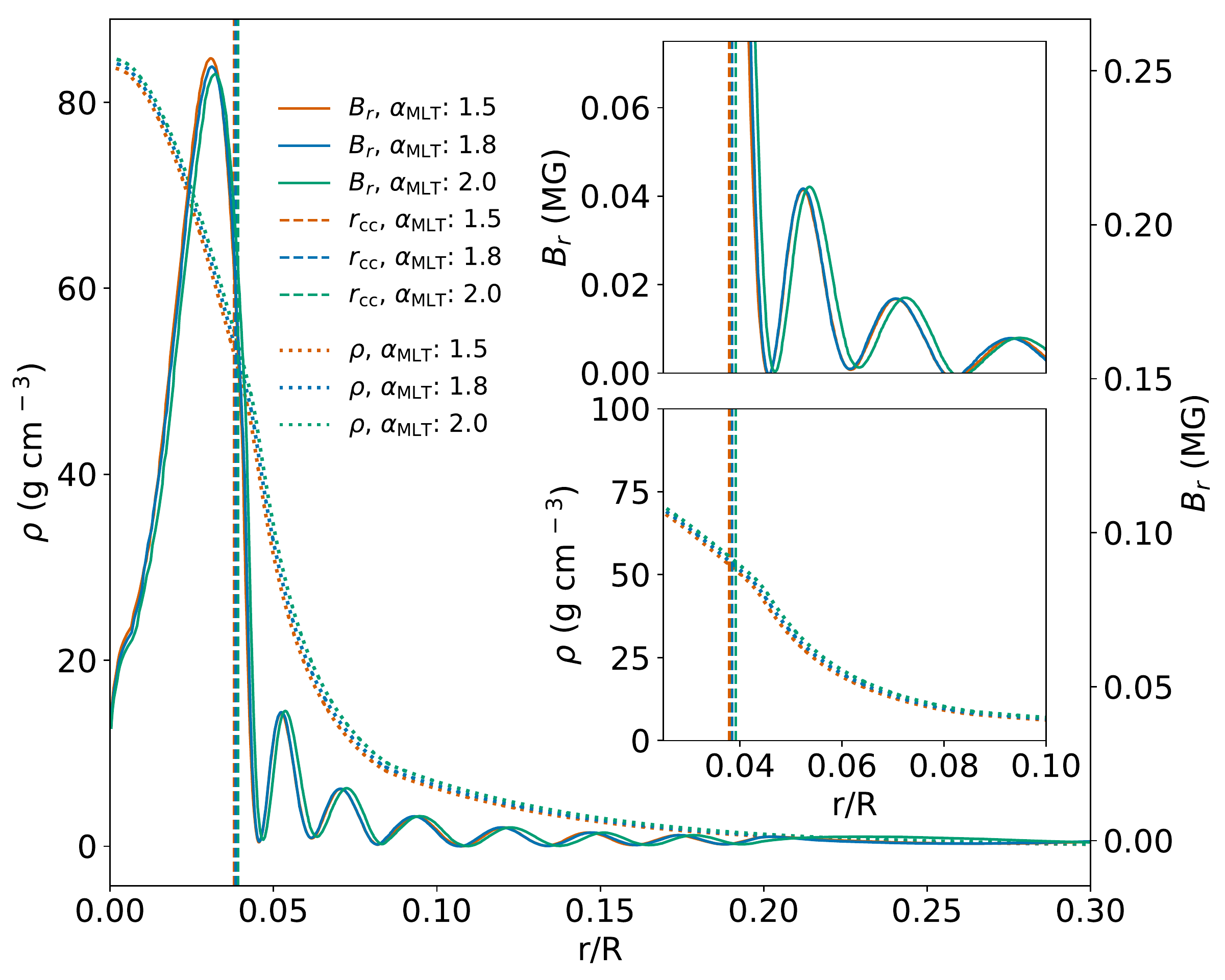}\\
\includegraphics[width=6cm,height=4.5cm]{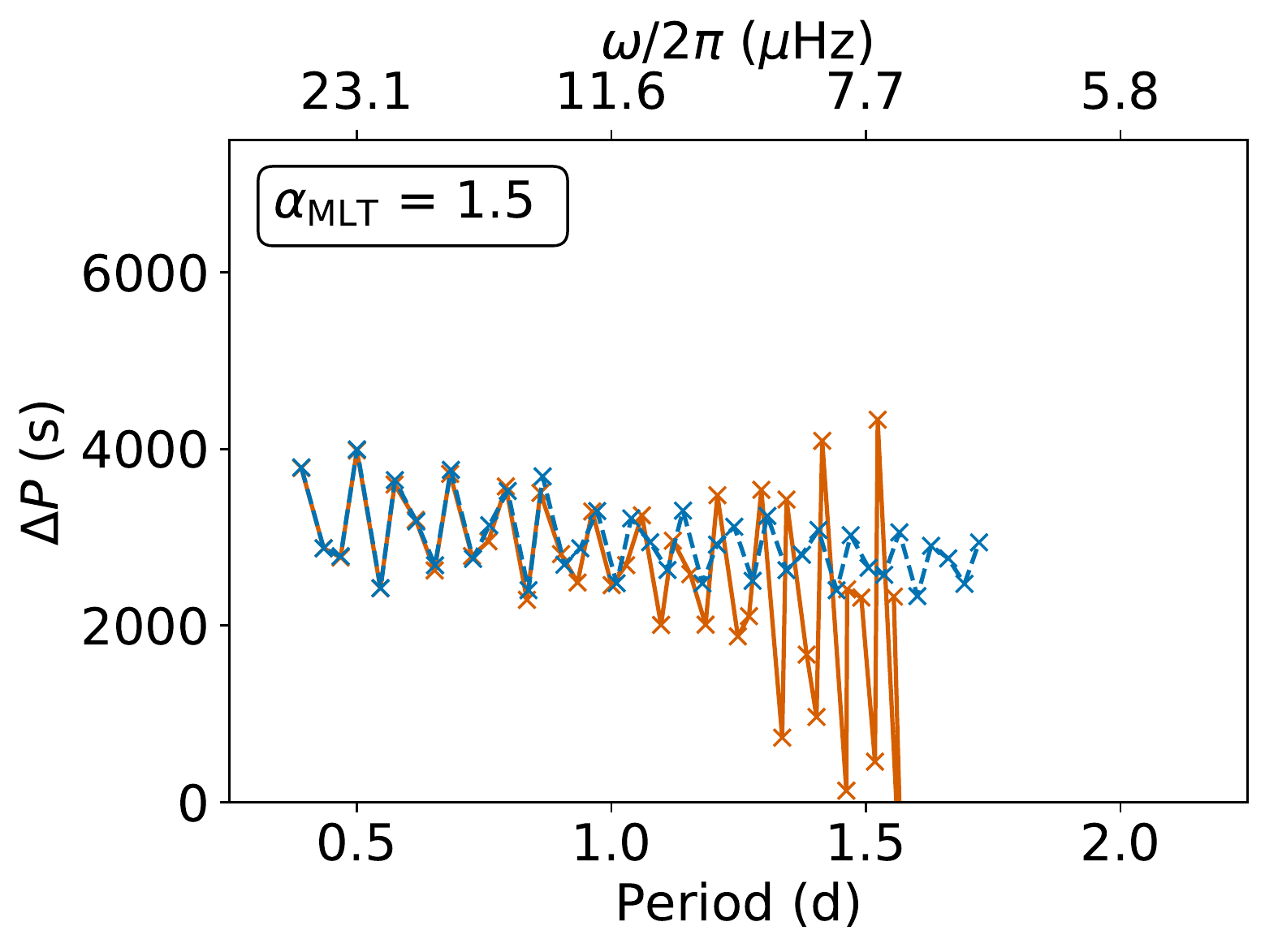}\includegraphics[width=6cm,height=4.5cm]{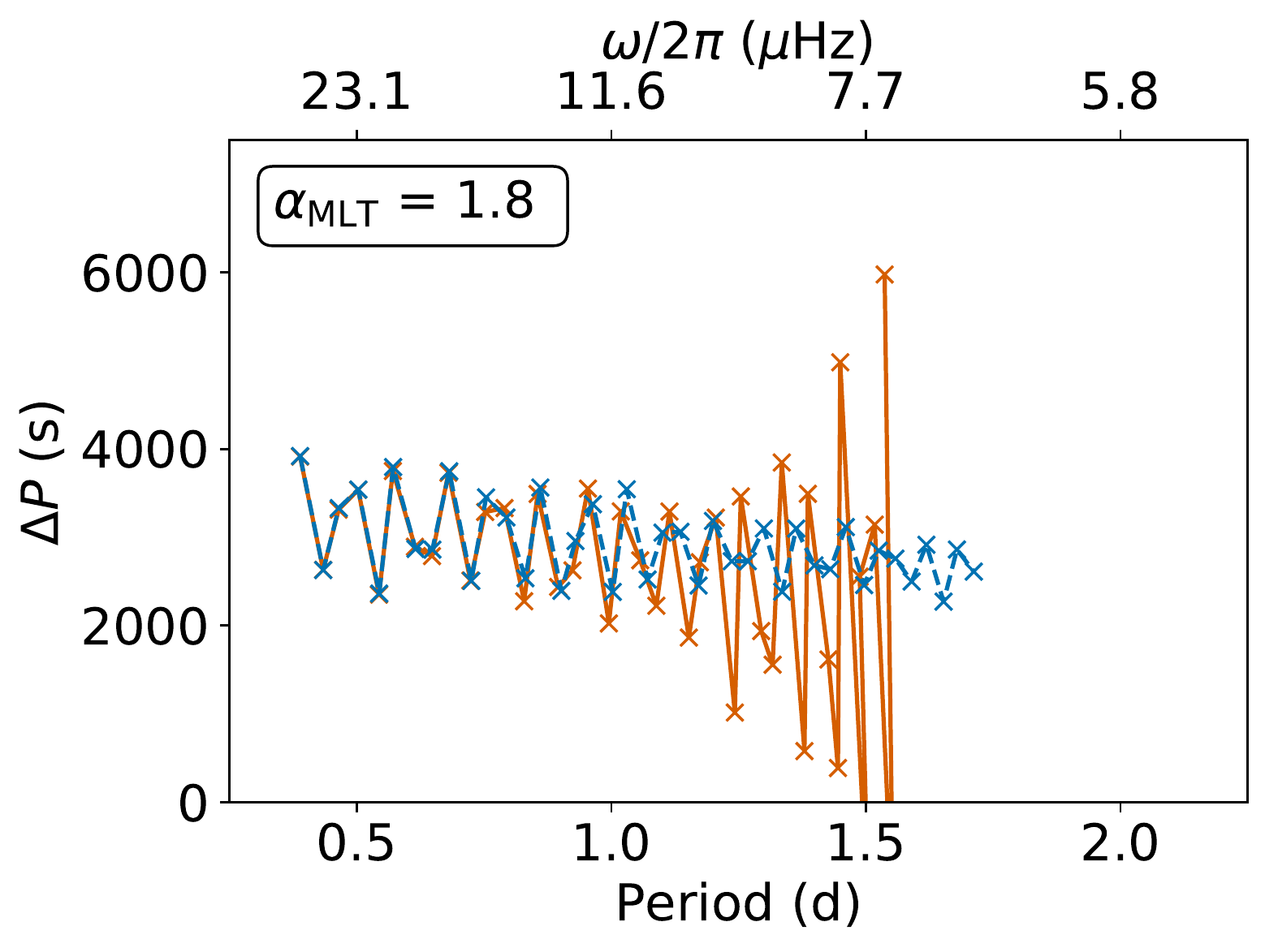}\includegraphics[width=6cm,height=4.5cm]{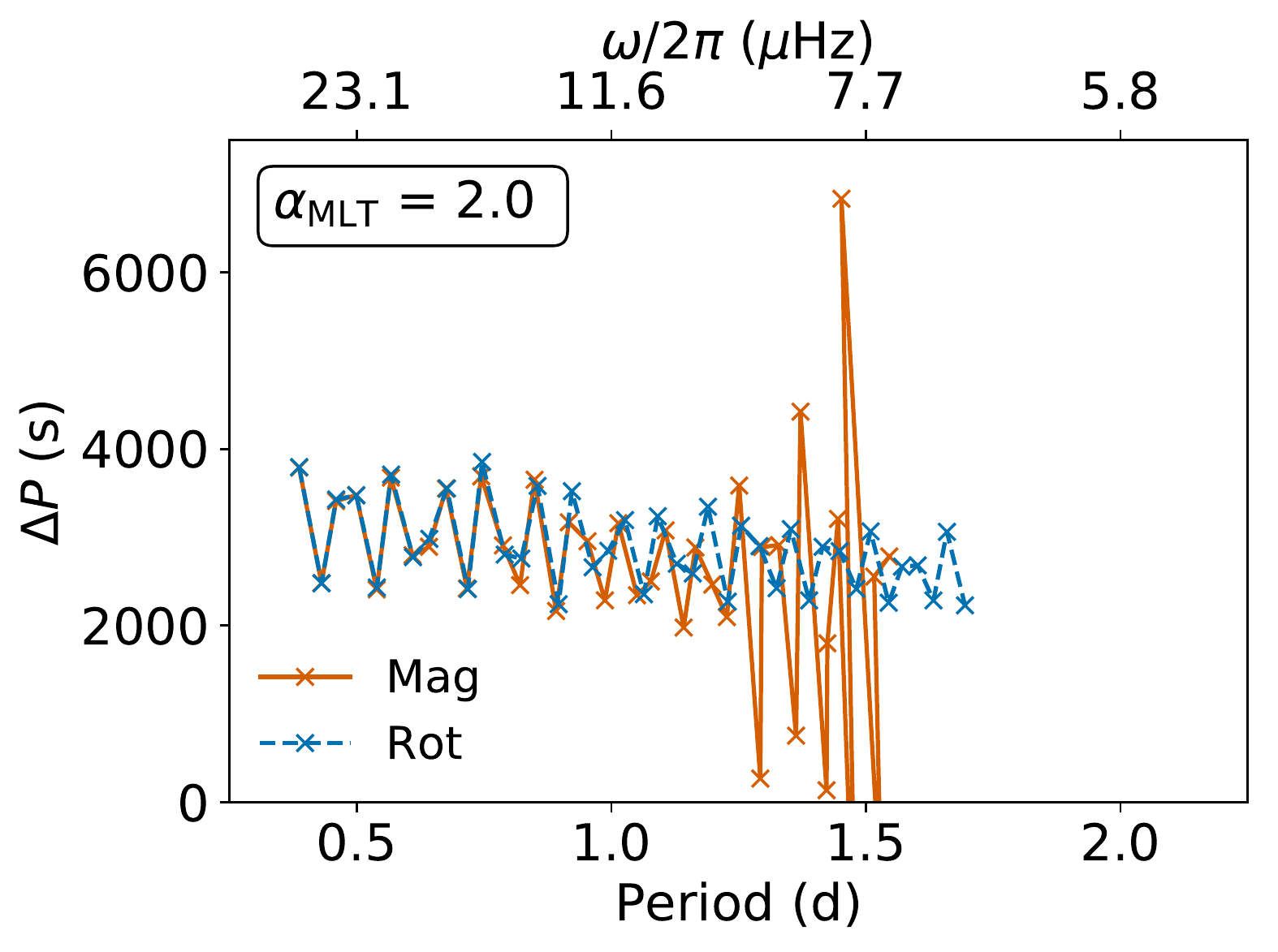}
\caption{Same as Fig. \ref{fig:fov_influence}, but varying $\alpha_{\rm MLT}$.}
\label{fig:AMLT_influence}
\end{figure*}

\begin{figure*}
\includegraphics[width=88mm, height=6.25cm]{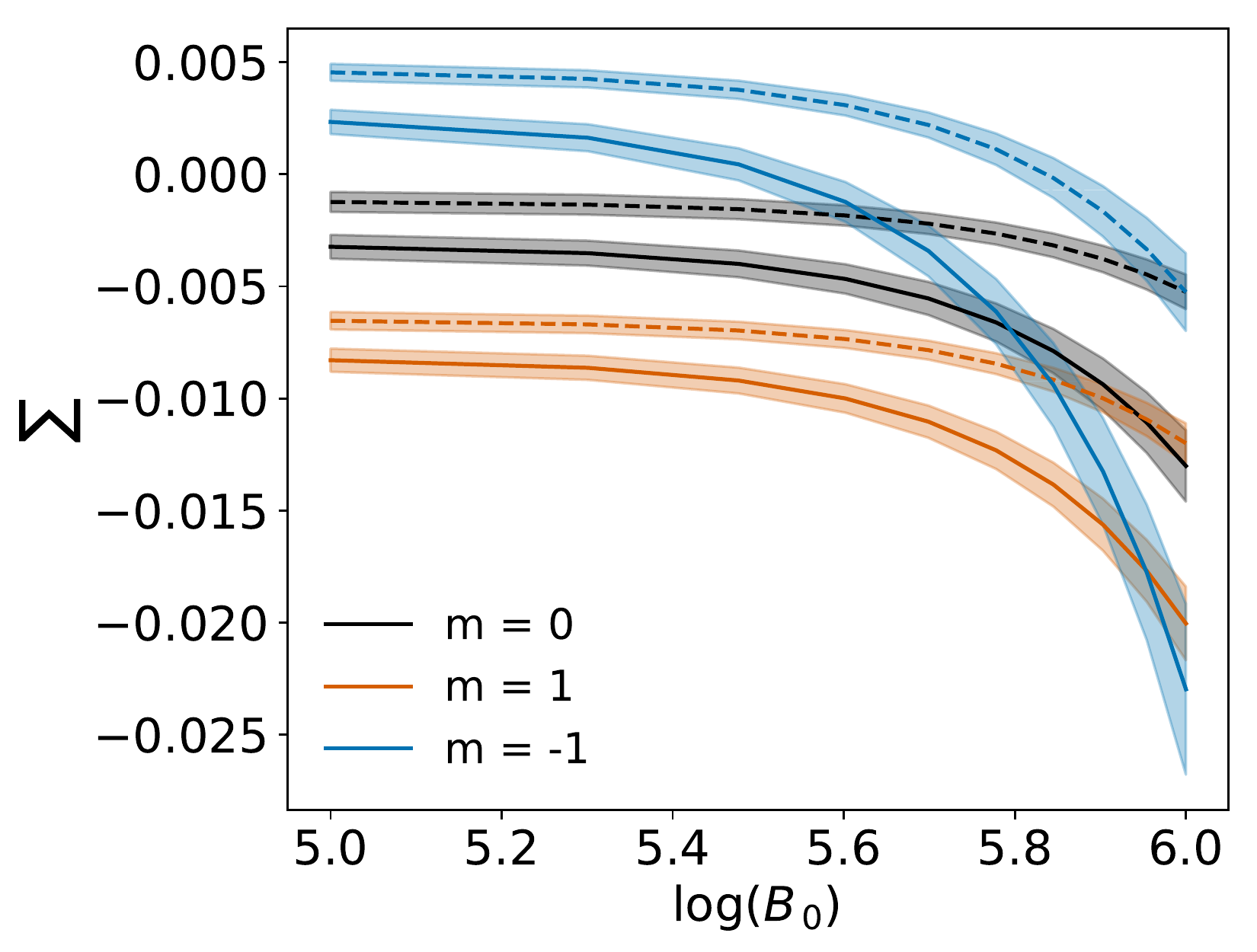}
\caption{Same as Fig. \ref{fig:slopes_fov}, but for the $2$-M$_\sun$ reference model.}
\label{fig:slopes_fov_M2}
\end{figure*}

\begin{figure*}
    \includegraphics[width=8.5cm]{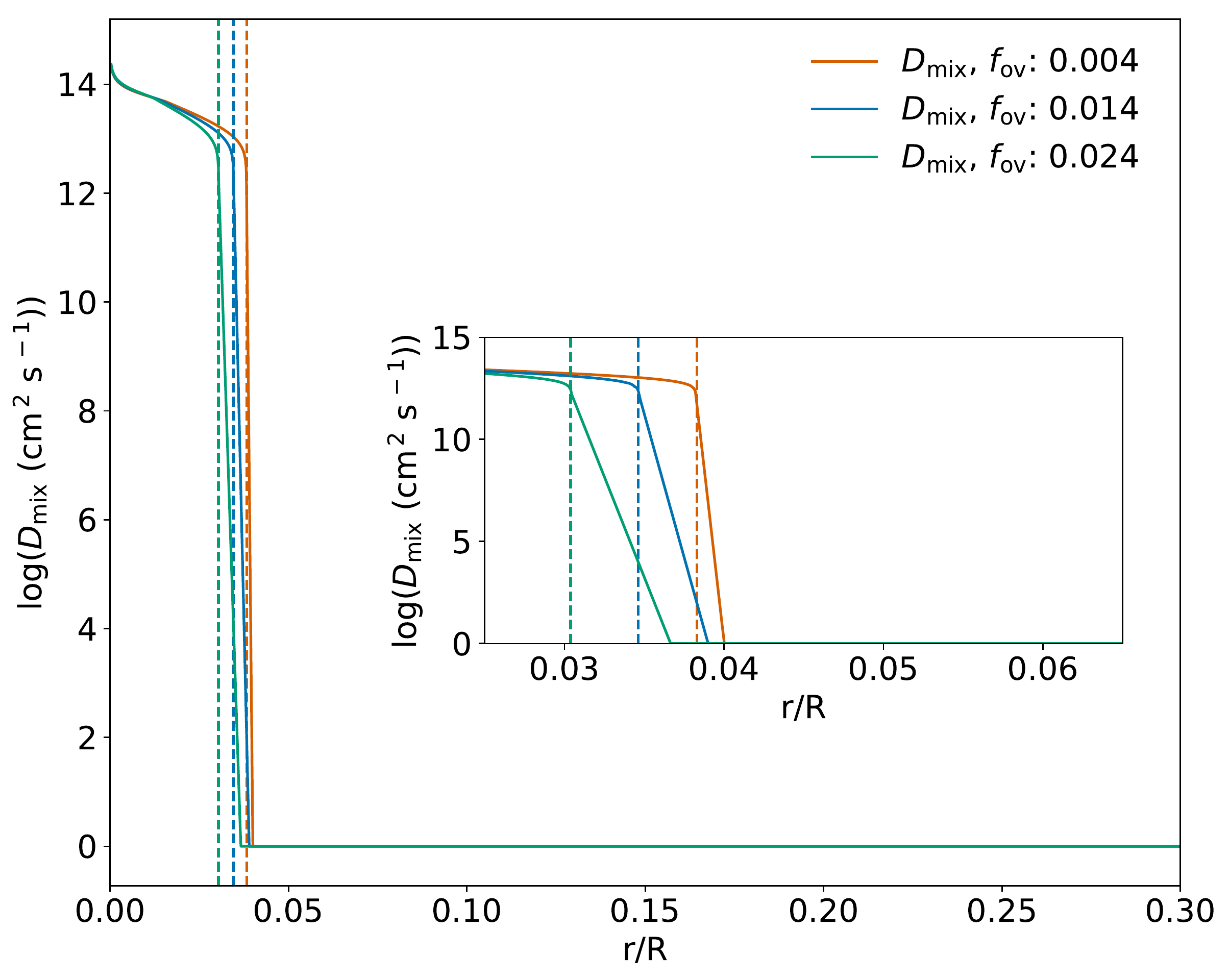}\hfill\includegraphics[width=8.5cm]{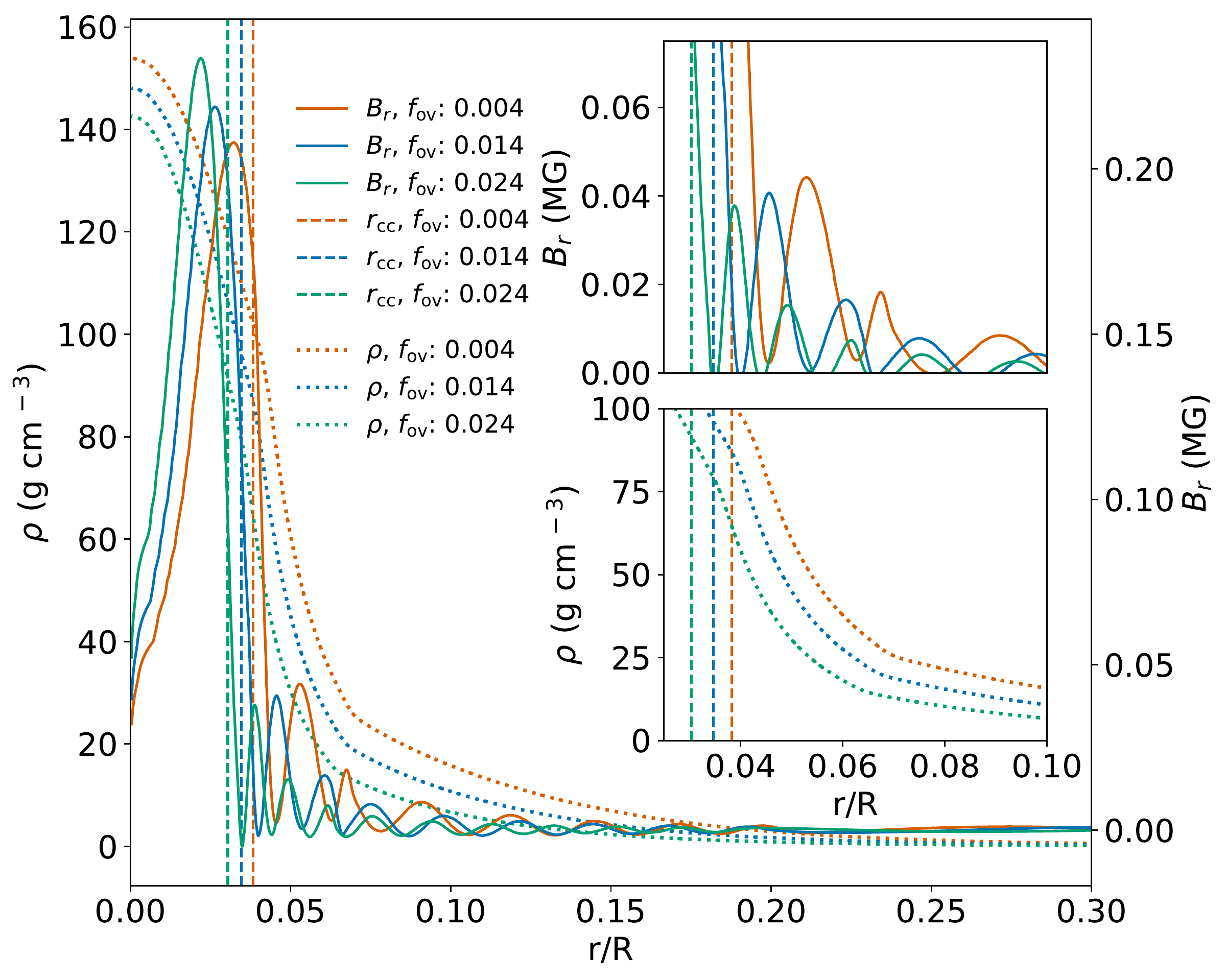}\\
    \includegraphics[width=6cm,height=4.5cm]{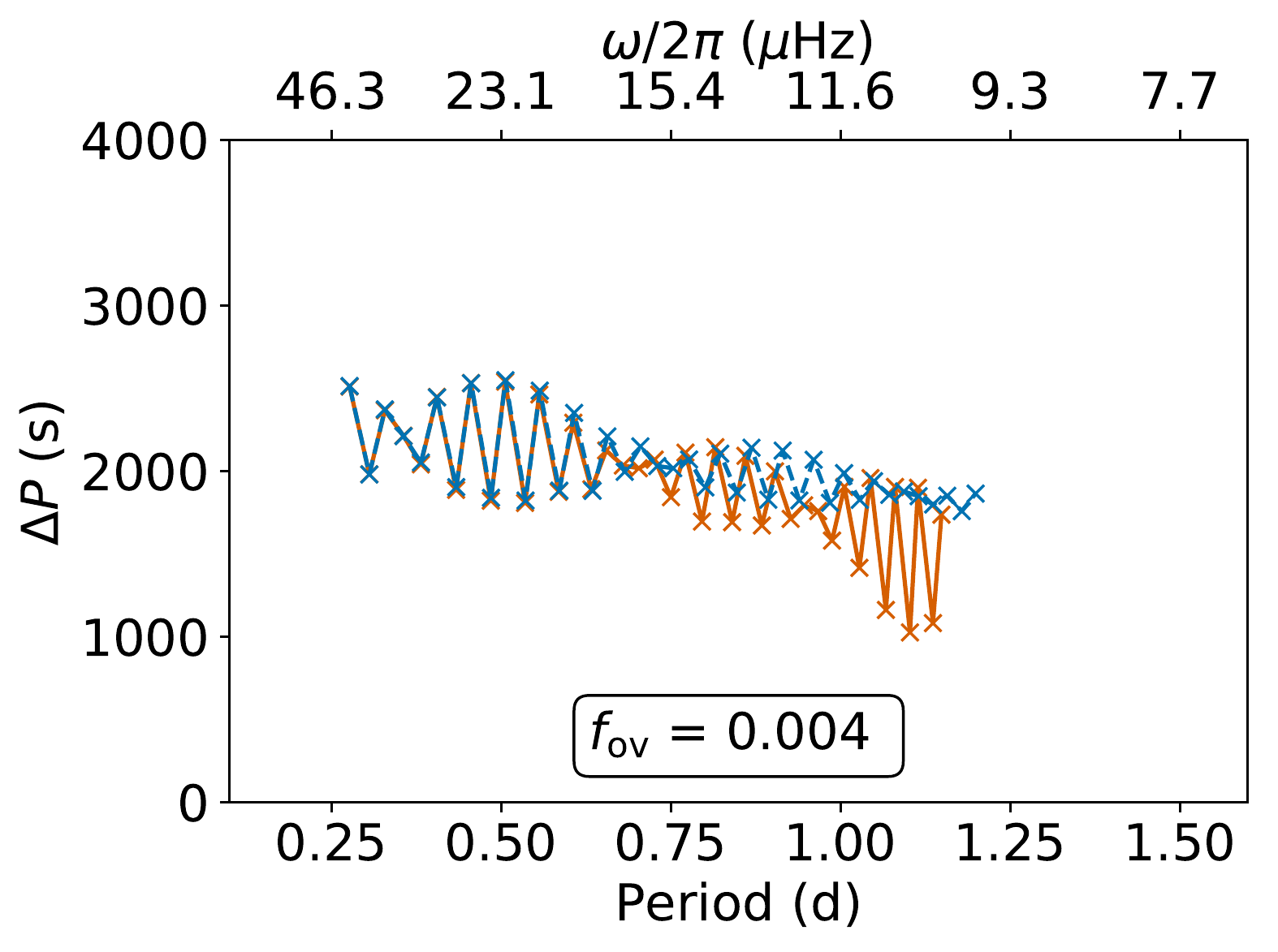}\includegraphics[width=6cm,height=4.5cm]{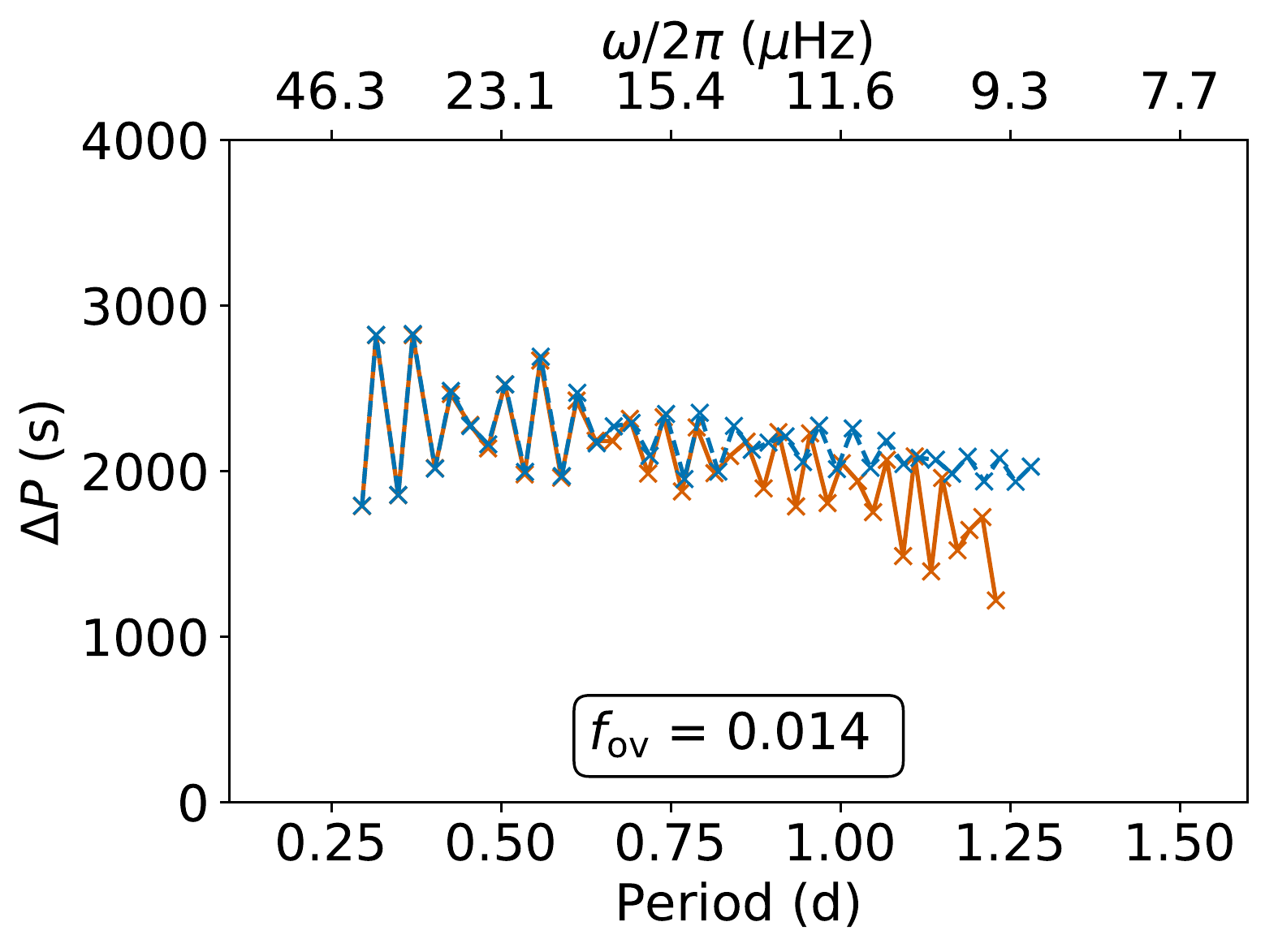}\includegraphics[width=6cm,height=4.5cm]{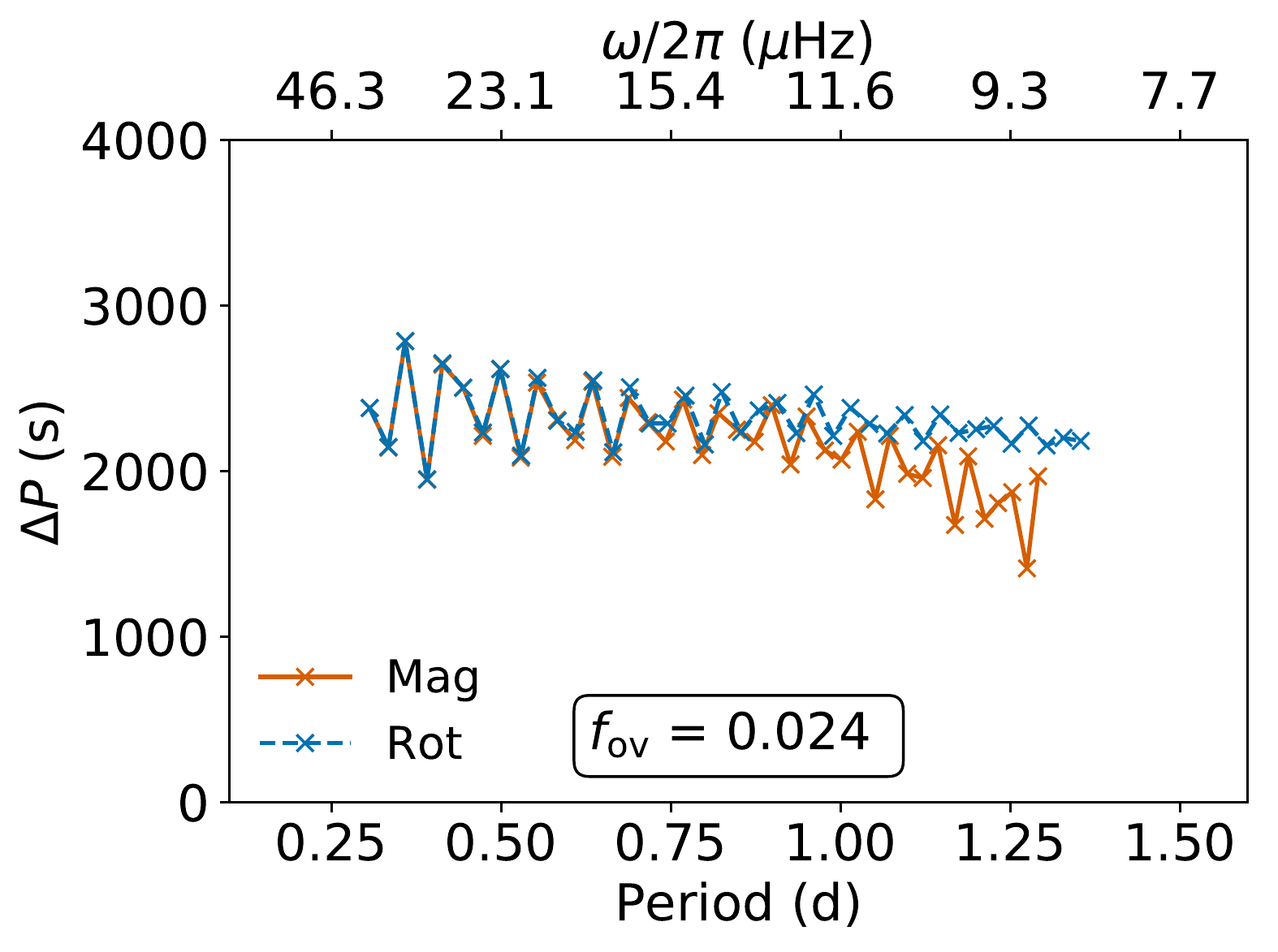}
    \caption{Same as Fig. \ref{fig:fov_influence}, but for the $2$-M$_\sun$ reference model.}
    \label{fig:fov_influence_M2}
\end{figure*}

\begin{figure*}
    \includegraphics[width=8.5cm]{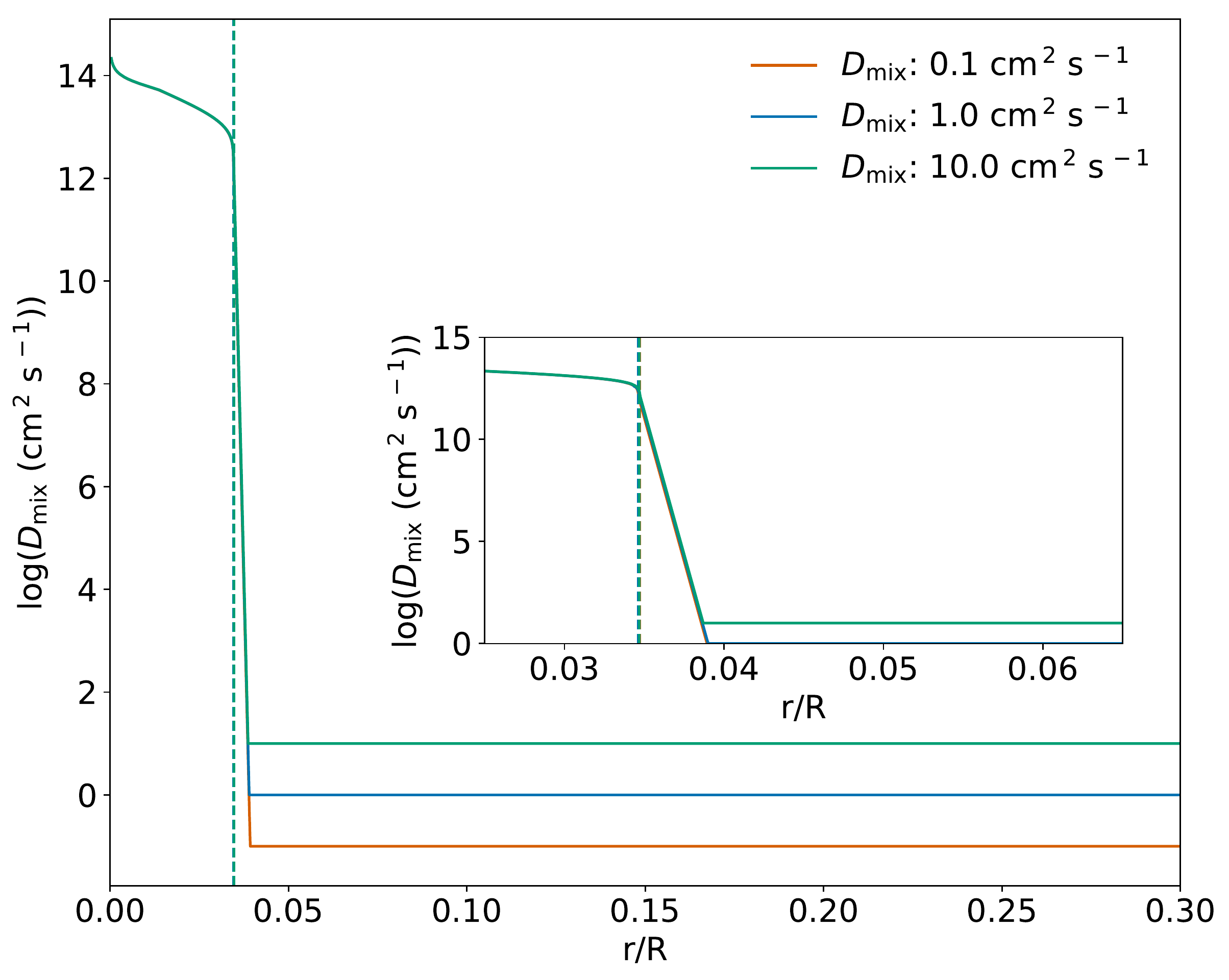}\hfill\includegraphics[width=8.5cm]{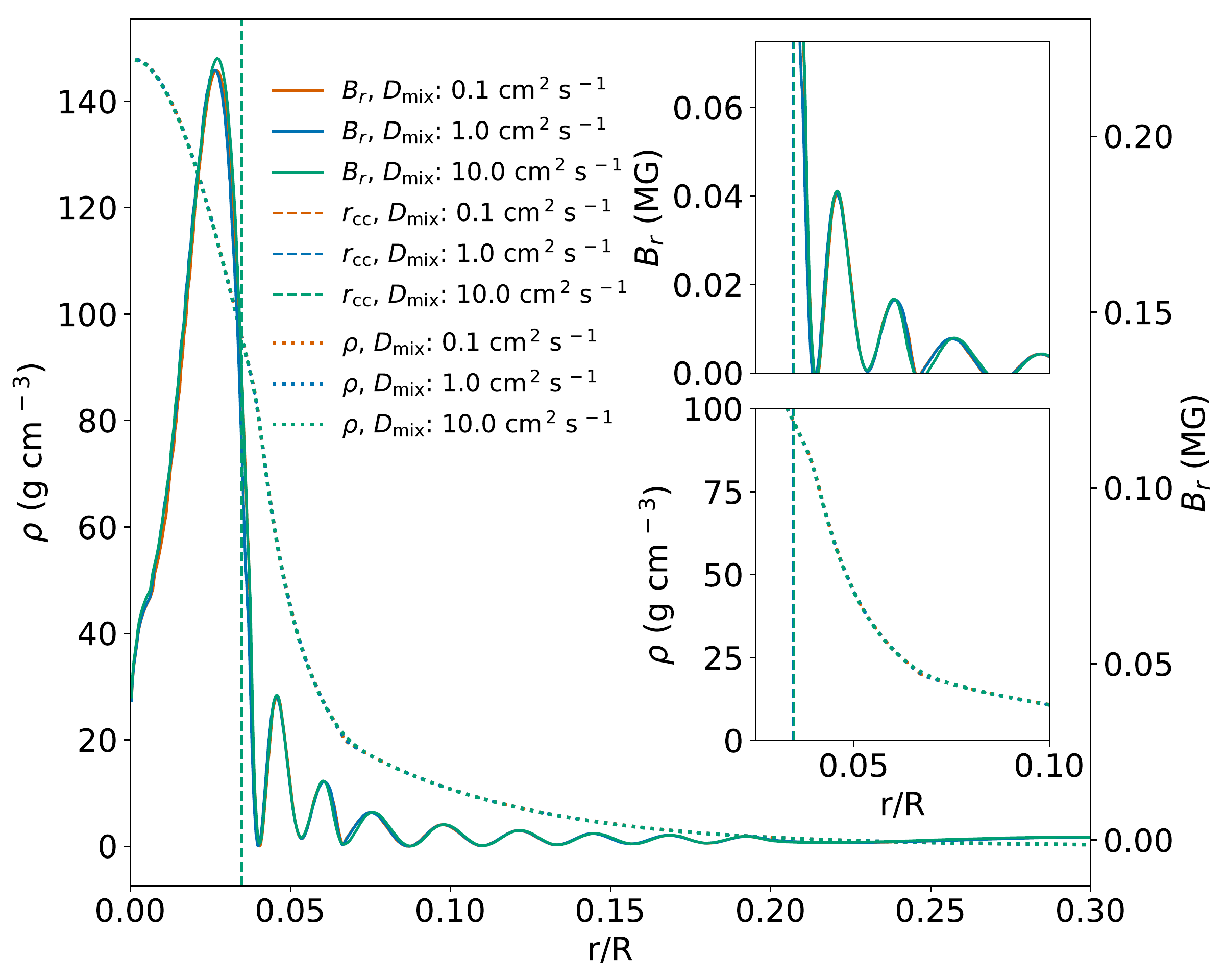}\\    \includegraphics[width=6cm,height=4.5cm]{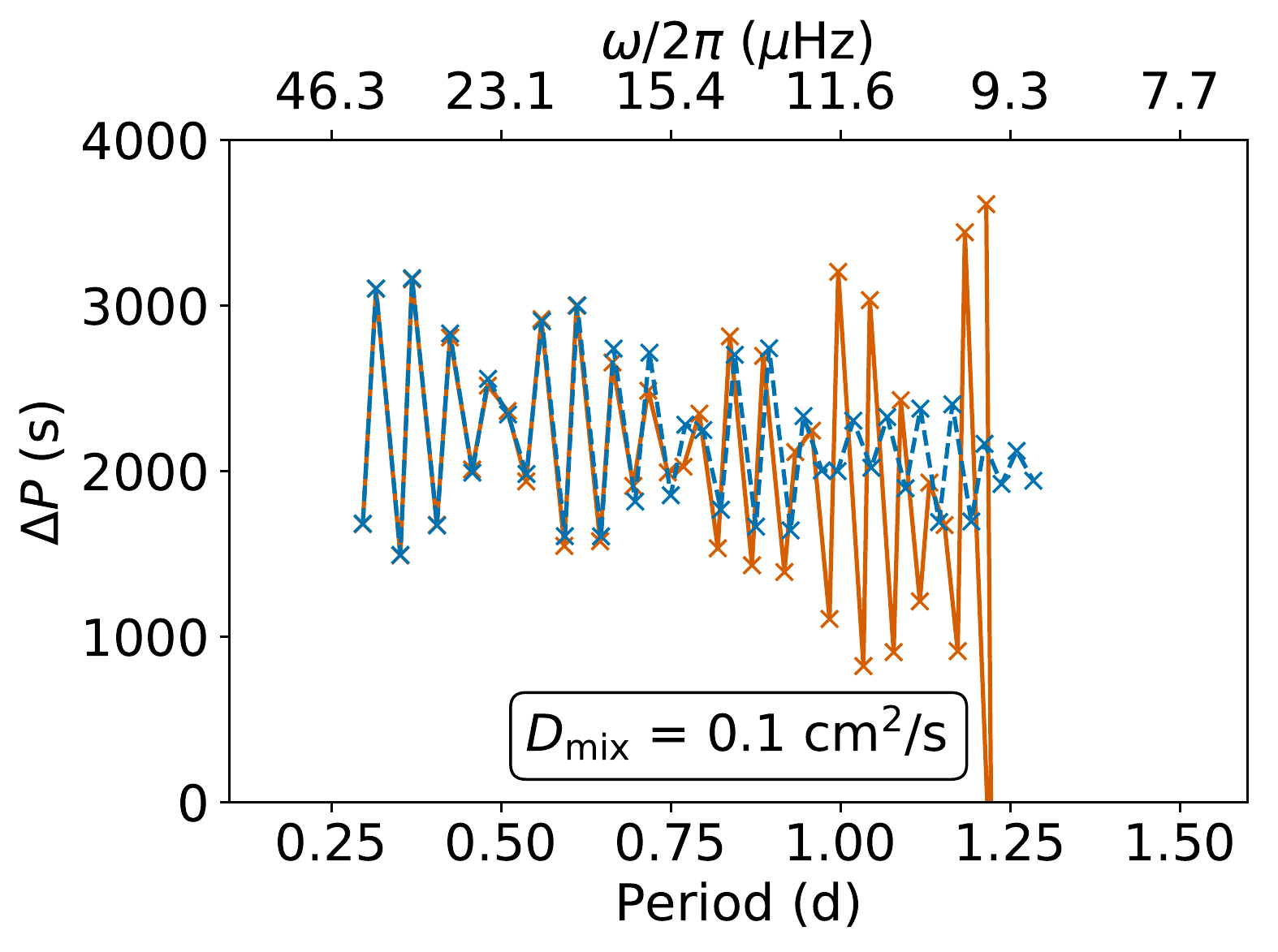}\includegraphics[width=6cm,height=4.5cm]{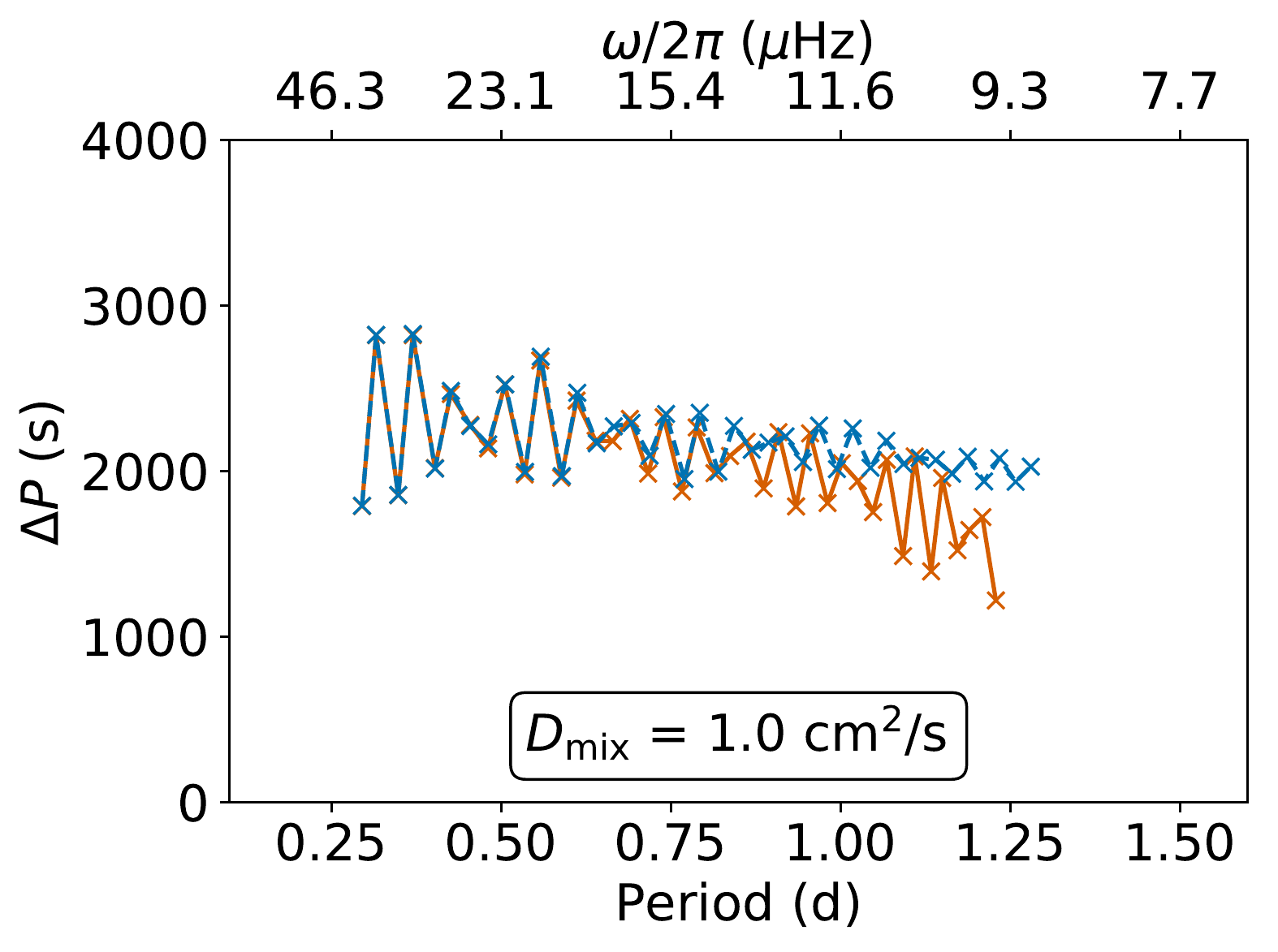}\includegraphics[width=6cm,height=4.5cm]{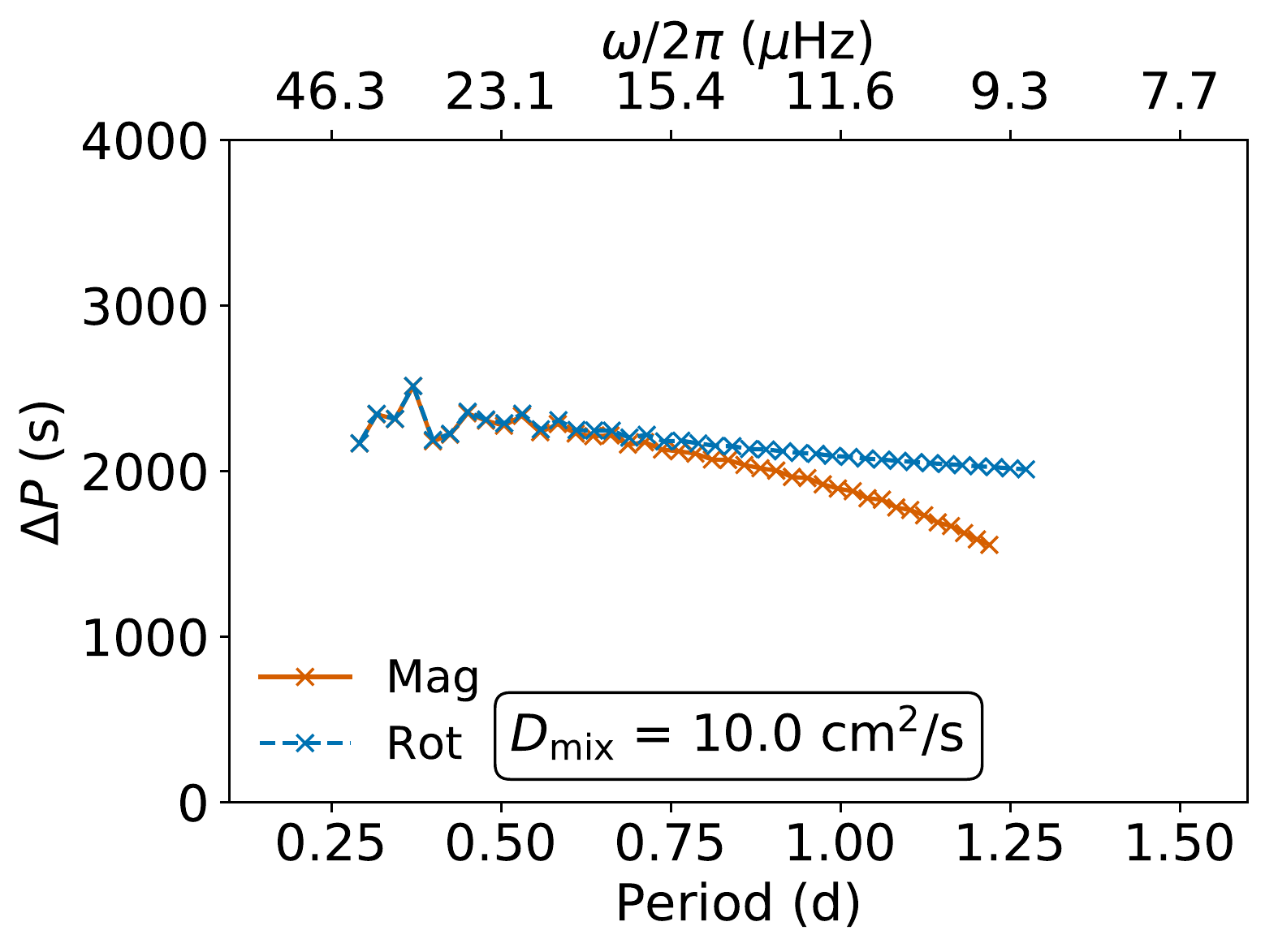}
    \caption{Same as Fig. \ref{fig:Dmix_influence}, but for the $2$-M$_\sun$ reference model.}
    \label{fig:D_mix_influence_M2}
\end{figure*}

\begin{figure*}
    \includegraphics[width=8.5cm]{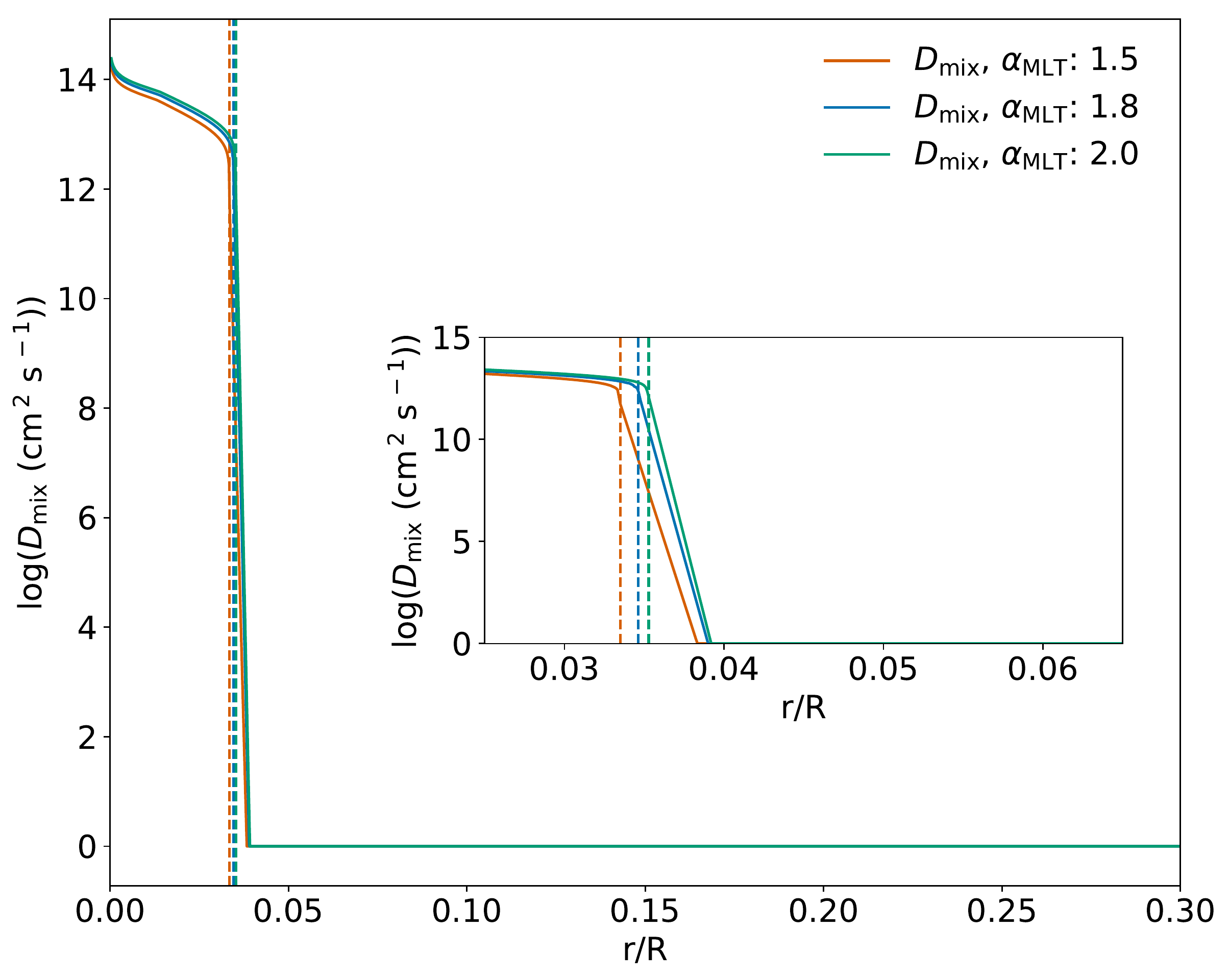}\hfill\includegraphics[width=8.5cm]{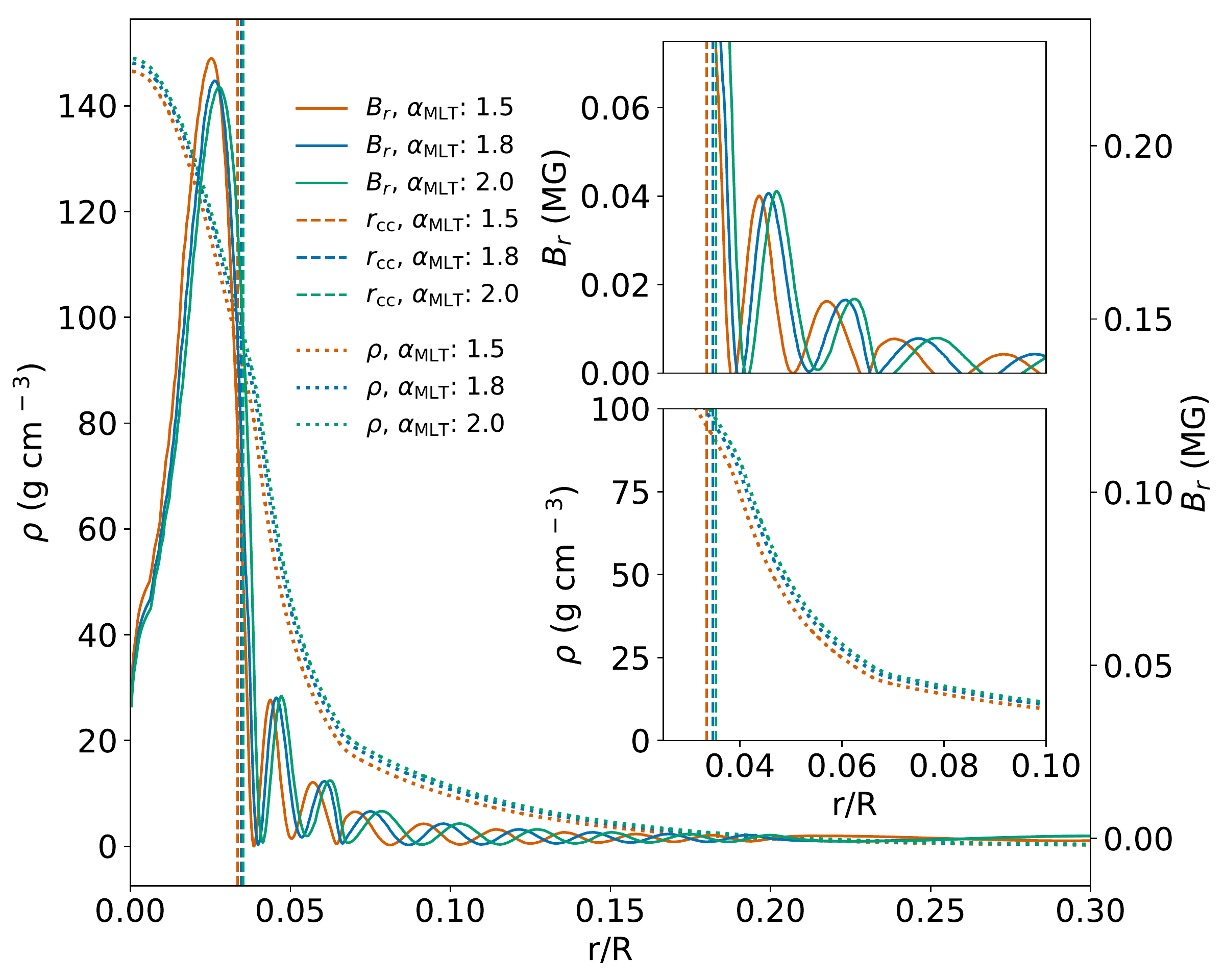}\\    \includegraphics[width=6cm,height=4.5cm]{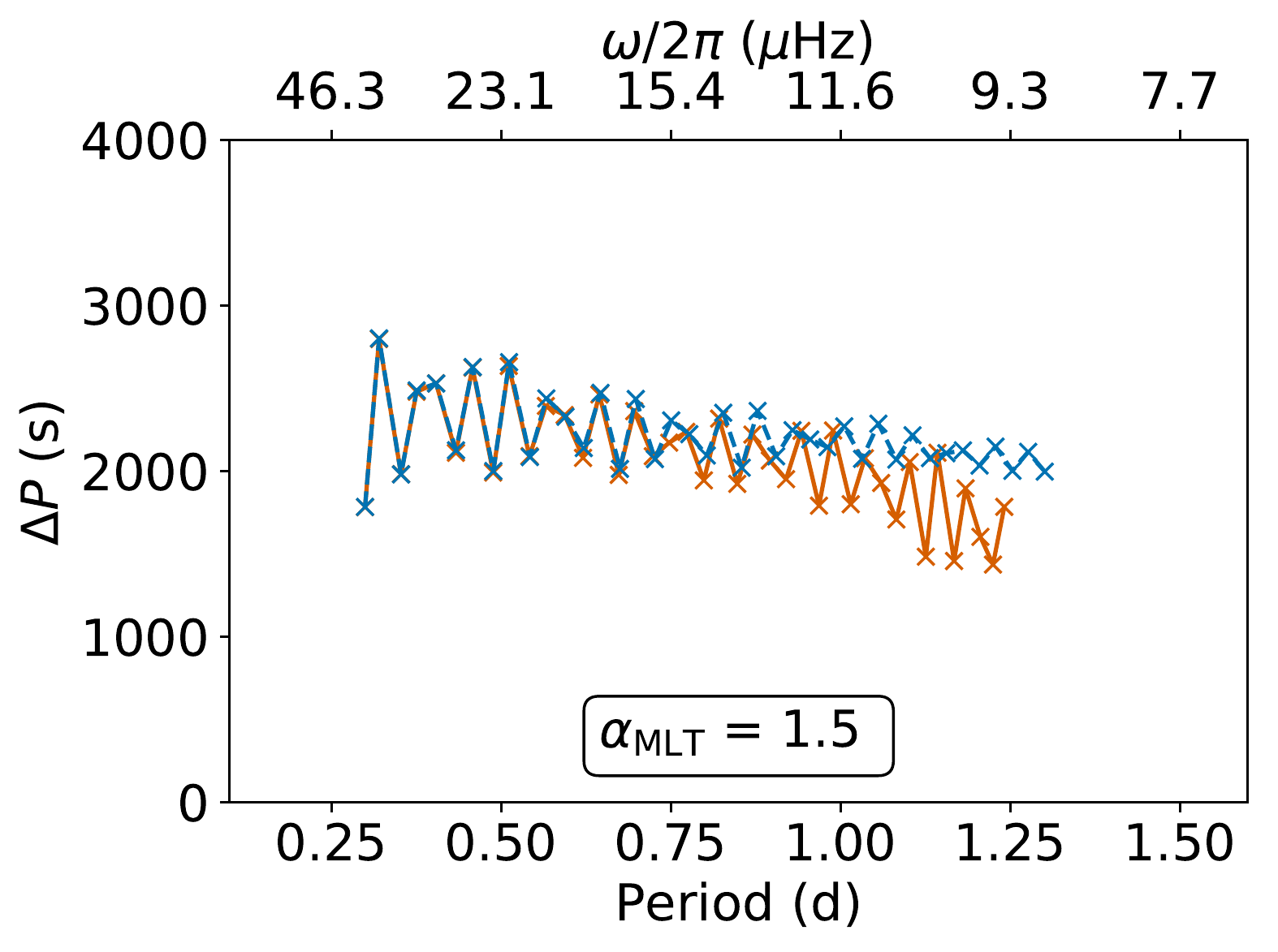}\includegraphics[width=6cm,height=4.5cm]{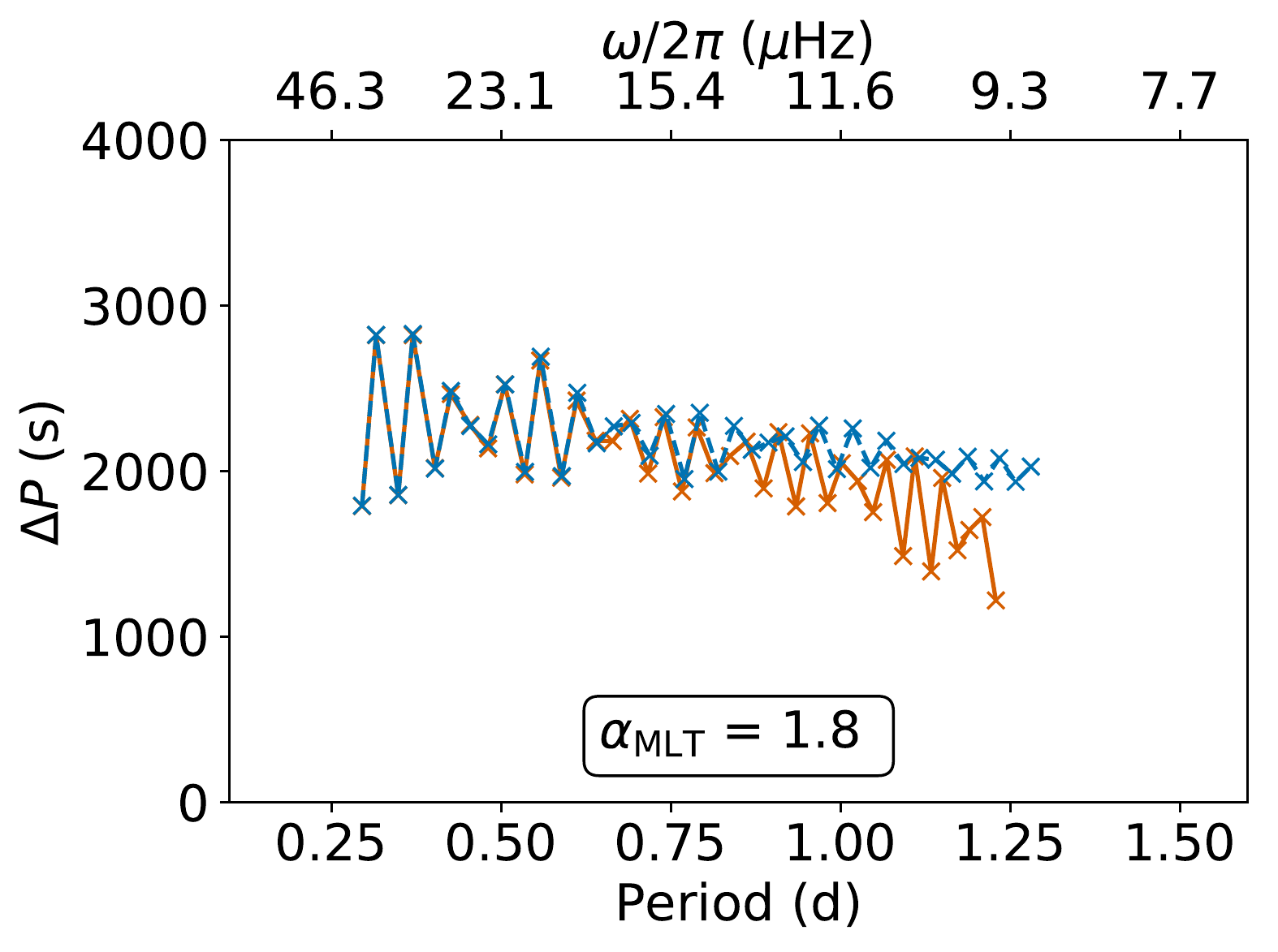}\includegraphics[width=6cm,height=4.5cm]{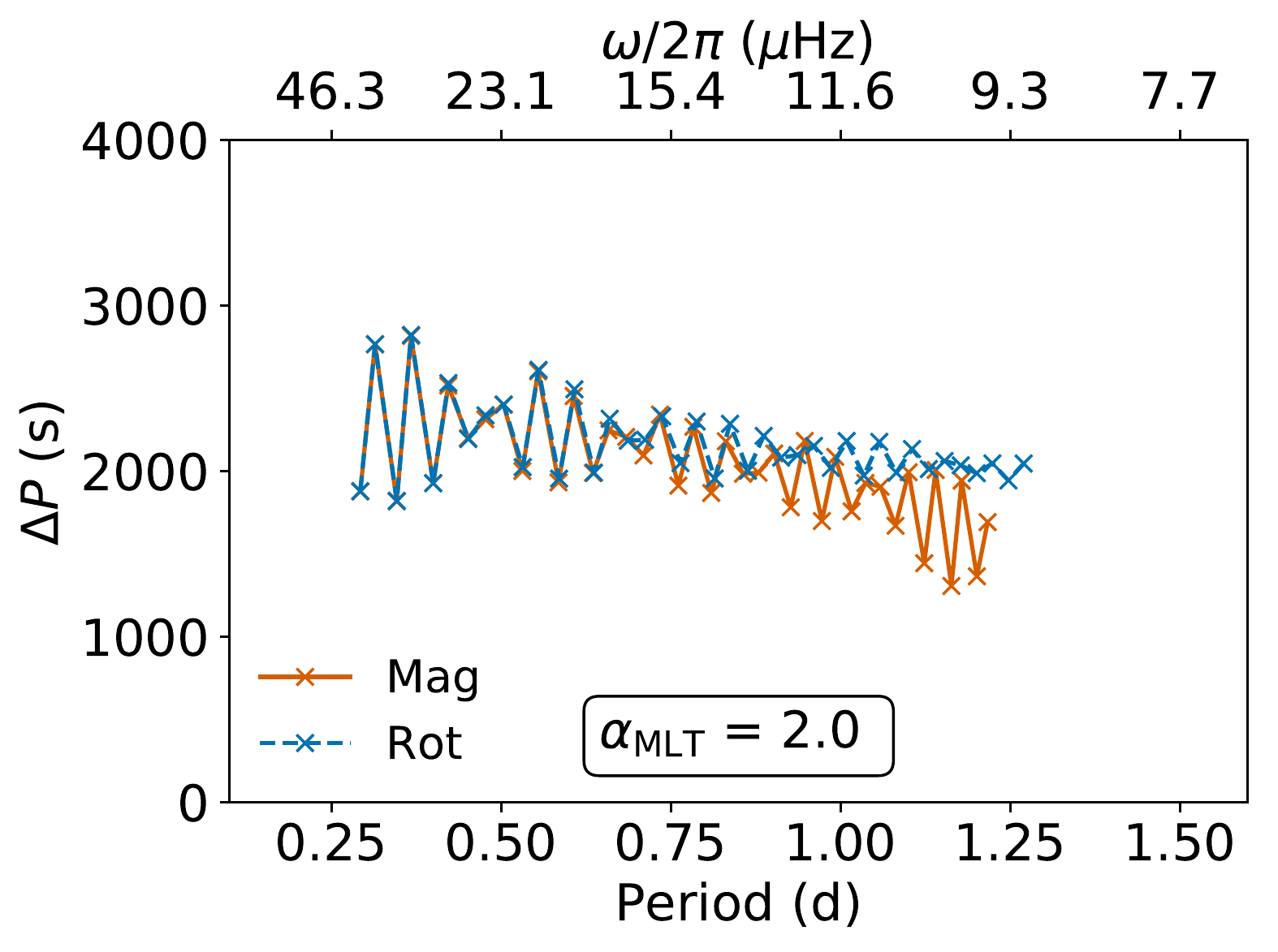}
    \caption{Same as Fig. \ref{fig:AMLT_influence}, but for the $2$-M$_\sun$ reference model.}
    \label{fig:AMLT_influence_M2}
\end{figure*}

\begin{figure*}
    \includegraphics[width=8.5cm]{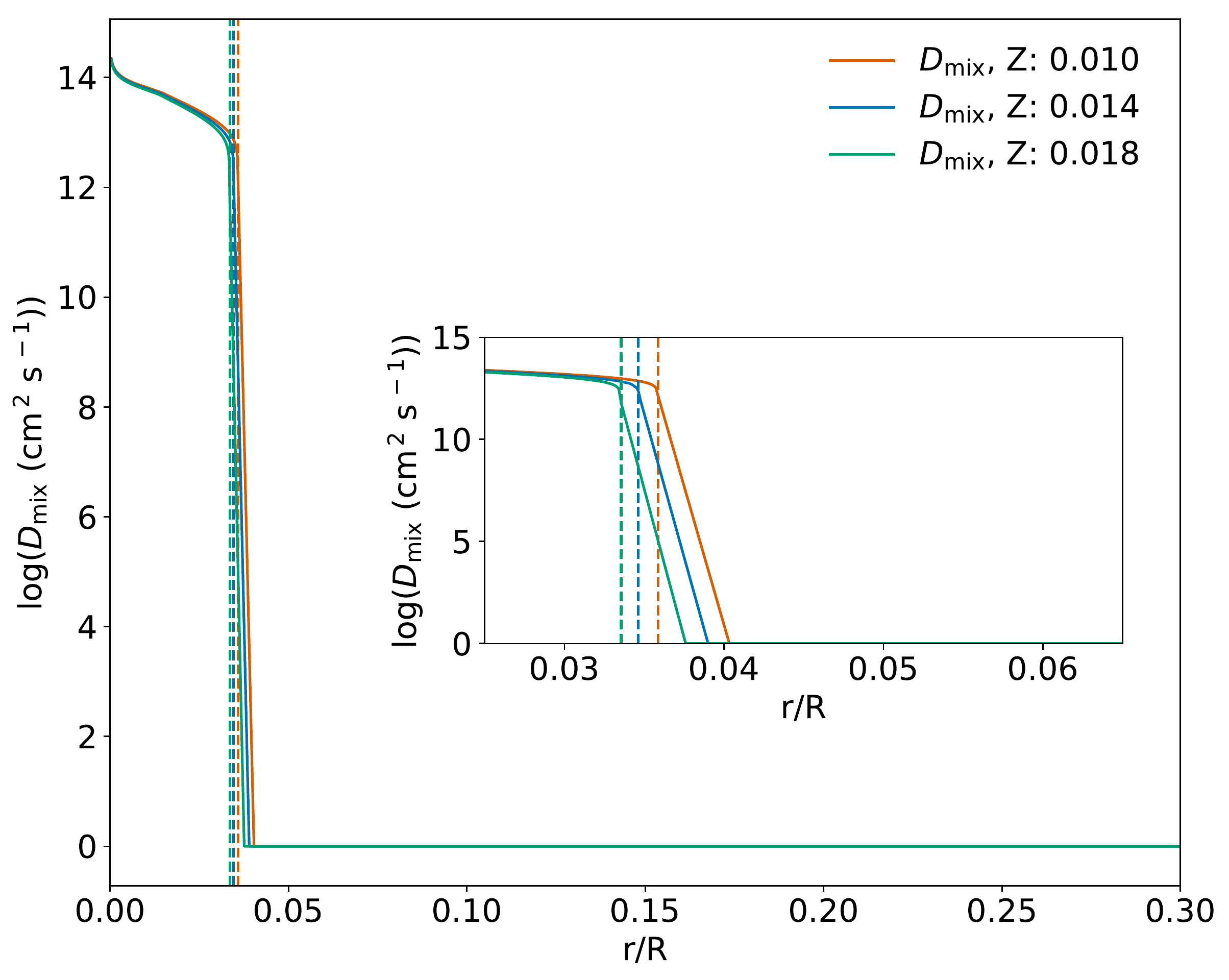}\hfill\includegraphics[width=8.5cm]{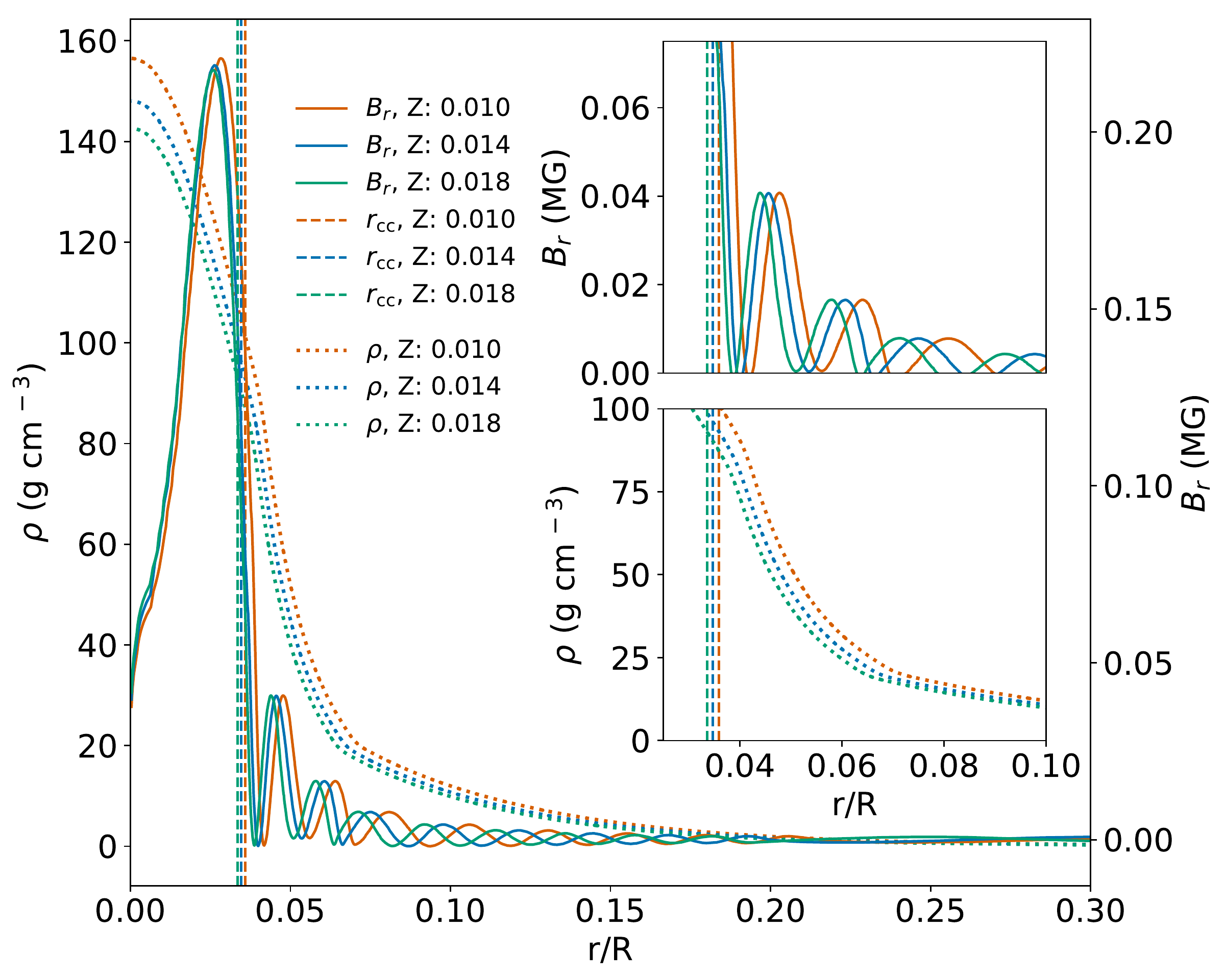}\\    \includegraphics[width=6cm,height=4.5cm]{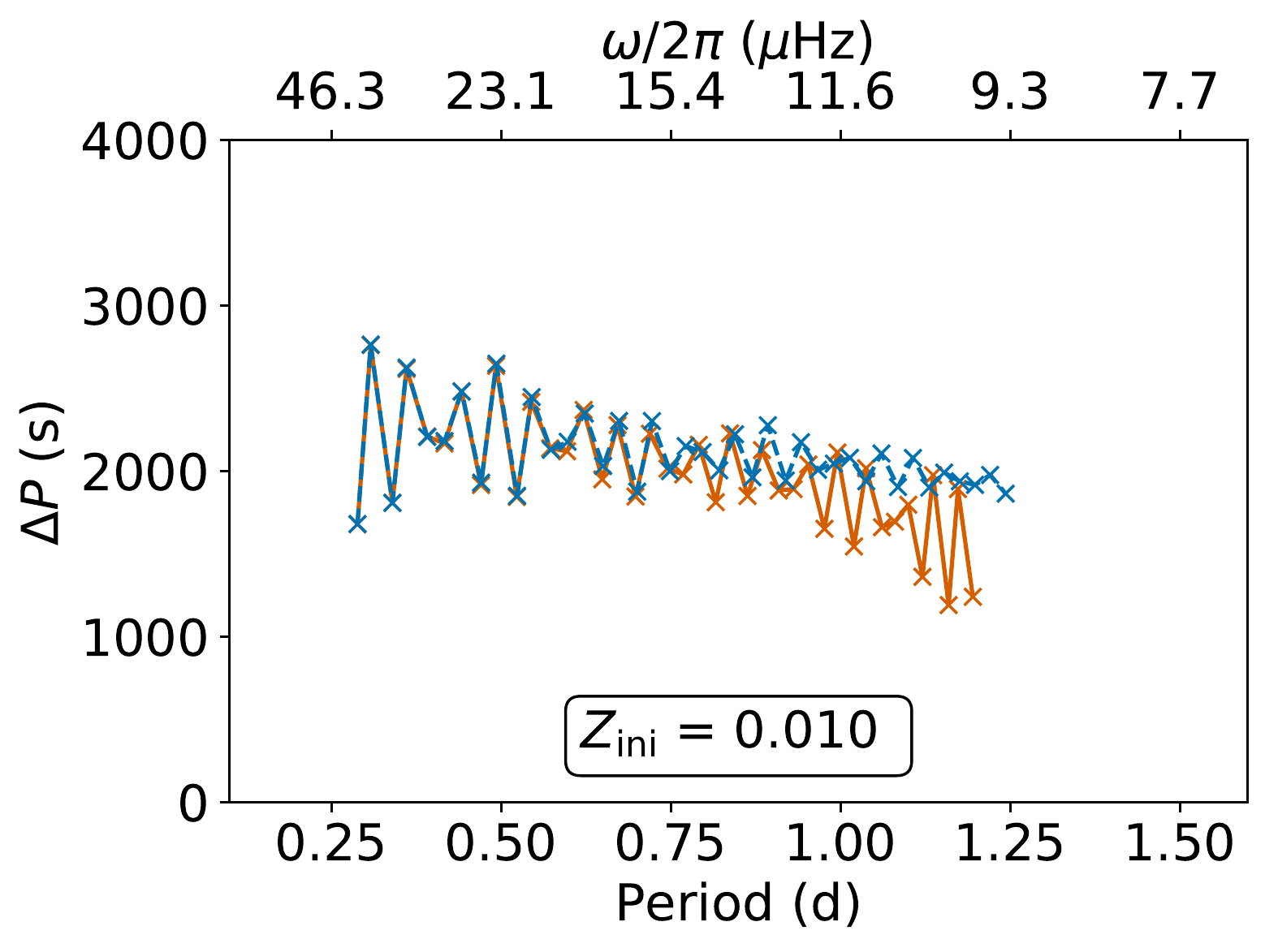}\includegraphics[width=6cm,height=4.5cm]{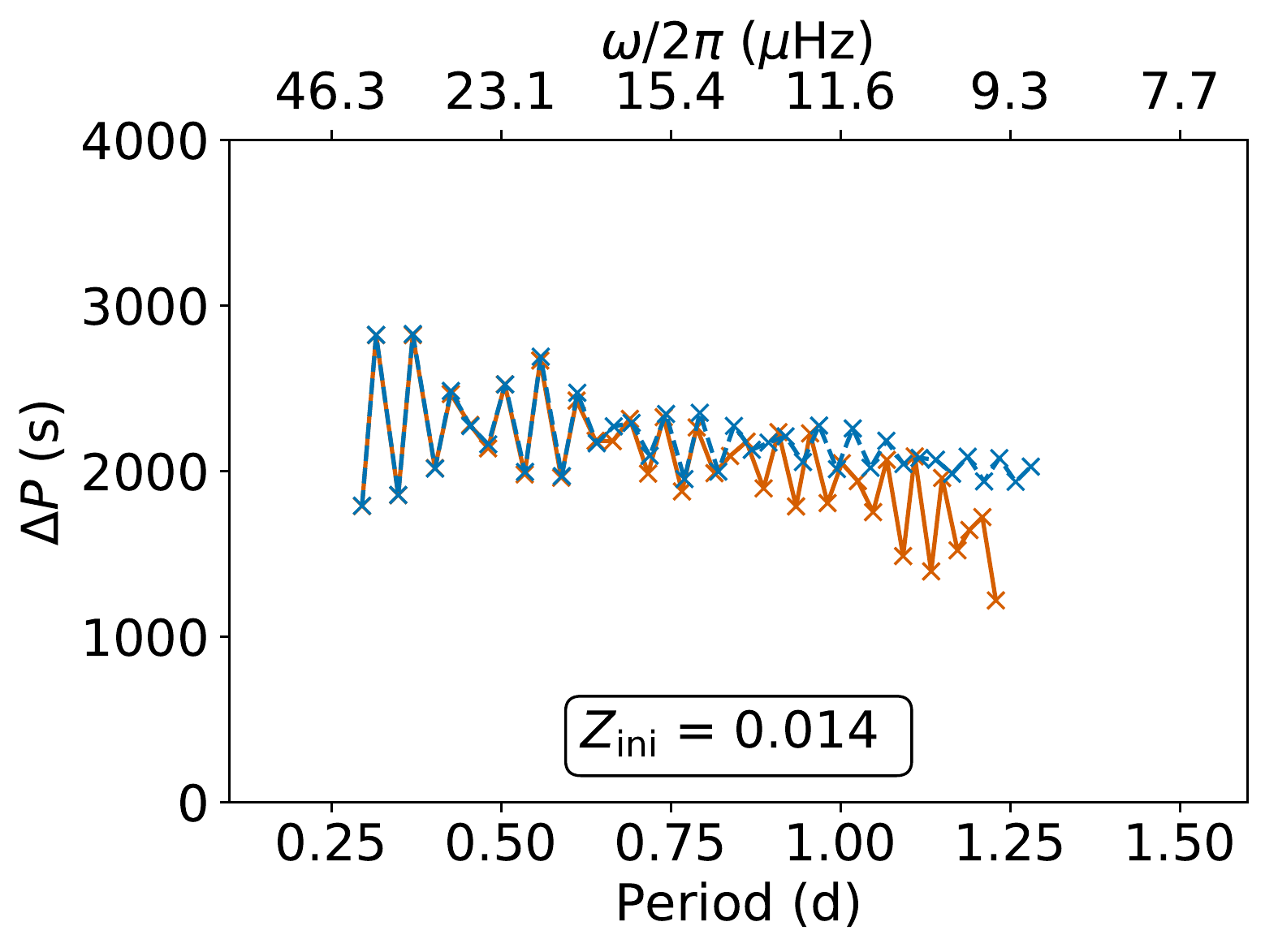}\includegraphics[width=6cm,height=4.5cm]{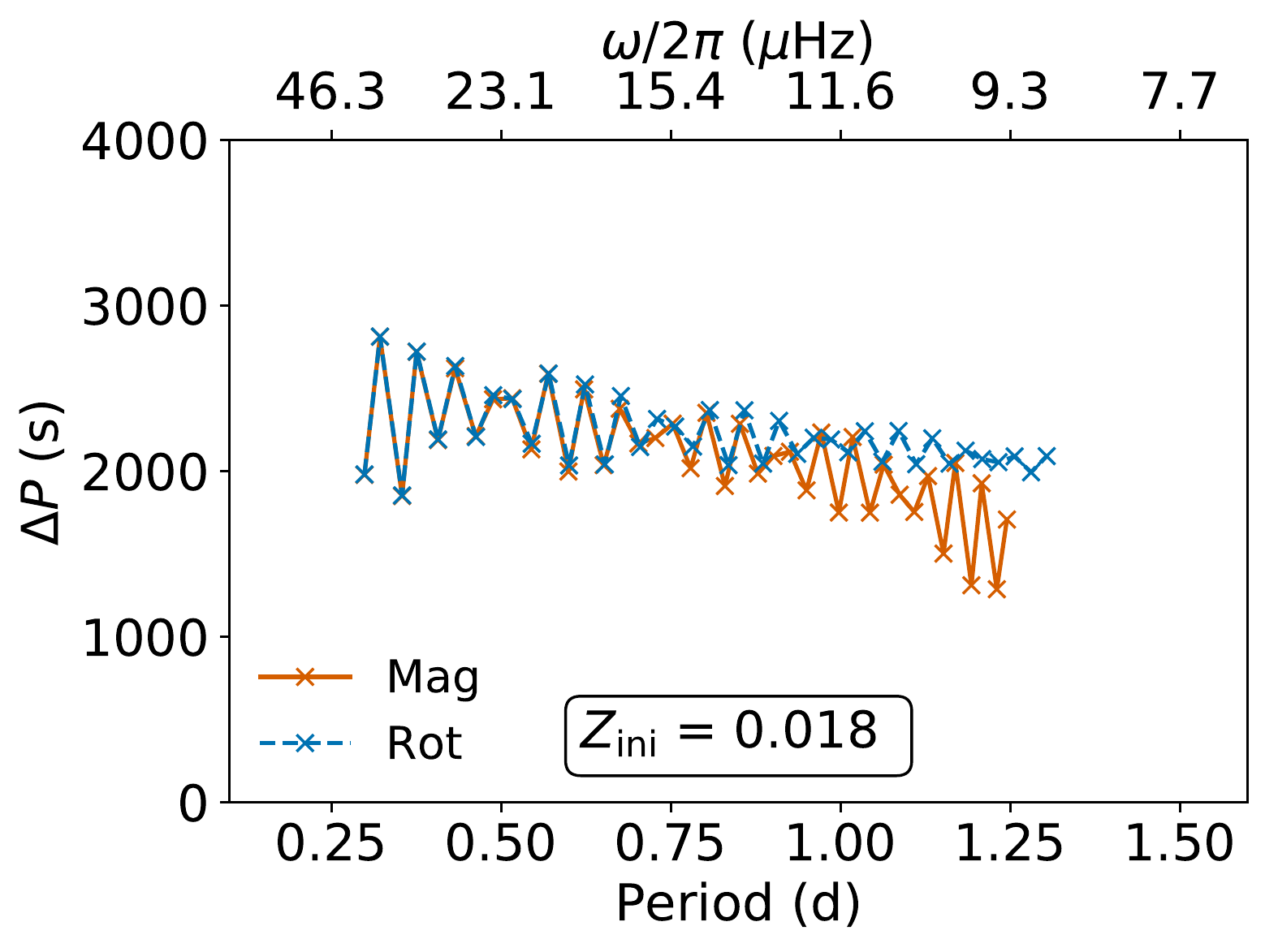}
    \caption{Same as Fig. \ref{fig:Z_influence}, but for the $2$-M$_\sun$ reference model.}
    \label{fig:Z_influence_M2}
\end{figure*}

\end{appendix}

\end{document}